\documentclass[12pt,a4paper]{article}

\usepackage[T1]{fontenc}
\usepackage[dvipsnames]{xcolor}
\usepackage{jheppub}

\usepackage{amsmath}
\usepackage{amssymb}
\usepackage{amsfonts}
\usepackage{amsthm}
\usepackage{mathtools}
\usepackage{mathrsfs}
\usepackage{dsfont}
\usepackage{braket}

\usepackage{tikz}
\usetikzlibrary{decorations.pathreplacing, decorations.markings, calc, shapes.misc, decorations.pathmorphing, patterns.meta, math, shapes.geometric}
\tikzset{snake it/.style={decorate, decoration=snake}}
\tikzset{cross/.style={cross out, draw=black, minimum size=2*(#1-\pgflinewidth), inner sep=0pt, outer sep=0pt},
cross/.default={1pt}}
\tikzset{
    partial ellipse/.style args={#1:#2:#3}{
        insert path={+ (#1:#3) arc (#1:#2:#3)}
    }
}
\usepackage{subcaption}
\usepackage[most]{tcolorbox}

\definecolor{bleudefrance}{rgb}{0.19, 0.55, 0.91}
\definecolor{candyapplered}{rgb}{1.0, 0.03, 0.0}
\definecolor{vert}{rgb}{0.1367 0.543 0.1367}

\hypersetup{
    pdfencoding=unicode,
    colorlinks=true,
    urlcolor=Maroon,
    linkcolor=RoyalBlue,
    citecolor=Maroon,
    pdftitle={c=1 strings as a matrix integral},
    pdfauthor={Scott Collier, Lorenz Eberhardt, Victor A. Rodriguez},
    pdfdisplaydoctitle=true,
    pdfstartview=FitH,
    linktocpage=true
}

\newcommand{\ba}{\begin{align}}

\newcommand{\be}{\begin{equation}}
\newcommand{\ee}{\end{equation}}
\def\bd{\begin{tikzpicture}}
\def\ed{\end{tikzpicture}}

\DeclareMathOperator\tr{tr}
\newcommand\e{\mathop{\text{e}}}
\newcommand\ex{{\mathrm e}}

\renewcommand\Im{\mathop{\text{Im}}}

\renewcommand\Re{\mathop{\text{Re}}}
\newcommand\Res{\mathop{\text{Res}}}
\DeclareMathOperator{\Aut}{Aut}

\DeclareMathOperator{\sgn}{\text{sgn}}

\renewcommand\d{\text{d}}
\newcommand\half{\frac{1}{2}}

\allowdisplaybreaks[1]

\newcommand\SL{\text{SL}}

\newcommand\CC{\mathbb{C}}
\newcommand\ZZ{\mathbb{Z}}
\newcommand\RR{\mathbb{R}}
\newcommand\NN{\mathbb{N}}

\newcommand\TT{\mathbb{T}}
\newcommand\QQ{\mathbb{Q}}
\newcommand\LL{\mathbb{L}}
\newcommand{\id}{\mathds{1}}

\newcommand{\bM}{\overline{\mathcal{M}}}

\newcommand{\xx}{\mathsf{x}}
\newcommand{\yy}{\mathsf{y}}

\title{$\boldsymbol{c=1}$ strings as a matrix integral}

\author[1,2]{Scott Collier}\emailAdd{scolli32@syr.edu}
\author[3]{\!\!, Lorenz Eberhardt}\emailAdd{l.eberhardt@uva.nl}
\author[4]{\!\!, Victor A. Rodriguez}\emailAdd{varodriguez@ucsb.edu}

\affiliation[1]{Department of Physics, Syracuse University, Syracuse, NY, 13244, USA}
\affiliation[2]{Institute for Quantum \& Information Sciences, Syracuse University, Syracuse, NY, 13244, USA}
\affiliation[3]{Institute for Theoretical Physics, University of Amsterdam, Amsterdam, 1098XH, NL}
\affiliation[4]{Department of Physics, University of California, Santa Barbara, CA 93106, USA}

\abstract{
We study the perturbative $S$-matrix of the $c=1$ string and show that it admits a description in terms of a double-scaled (0+0)-dimensional matrix integral based on the spectral curve $\xx(z) = 2\sqrt{2}\cos(z)$, $\yy(z)=\sin(z)$. Combined with the famous duality to matrix quantum mechanics, this establishes a triality between three formulations of the theory: the worldsheet description, matrix quantum mechanics, and a matrix integral.

Starting from the intersection number expressions for the complex Liouville string, we derive closed-form Feynman rule expressions for the $c=1$ amplitudes as intersection numbers on the moduli space of Riemann surfaces. The intersection theory naturally computes amplitudes corresponding to a discretized target space where momentum is conserved only modulo an integer. The physical $S$-matrix elements are recovered by restriction to the first `Brillouin zone' and analytic continuation to Lorentzian kinematics. We prove that these amplitudes satisfy perturbative spacetime unitarity directly from the intersection theory expressions, and show that they satisfy a Mirzakhani-type recursion relation. We show detailed agreement with the known matrix quantum mechanics results, providing strong evidence for the triality.
}

\begin{document}

\maketitle

\makeatletter
\g@addto@macro\bfseries{\boldmath}
\makeatother

\section{Introduction}
The $c=1$ string is one of the simplest and perhaps most thoroughly studied models of string theory with a dynamical target space. It has provided a rich and tractable testing ground for basic ideas relating to quantum gravity and duality for over 30 years \cite{Klebanov:1991qa,Ginsparg:1993is,Jevicki:1993qn,Polchinski:1994mb}. Its worldsheet theory is defined by coupling a timelike free boson to $c=25$ Liouville CFT (together with the usual $bc$-ghost system), and it describes strings propagating in a (1+1)-dimensional target space with a Liouville wall and a weakly-coupled region where asymptotic $S$-matrix elements may be defined.

The theory only has a single physical field, a massless boson (the `tachyon'), but its perturbative $S$-matrix is highly nontrivial, encoding the quantum gravitational interactions between strings and scattering of strings off the Liouville wall. Despite the conceptual simplicity of the worldsheet definition, the direct evaluation of the moduli space integrals that define the string scattering amplitudes is technically very challenging and obscures many physical features of the model, such as unitarity and polynomiality of the amplitudes.

Part of the reason for the intense interest in the $c=1$ string is that it is conjectured to admit an exact dual description in terms of matrix quantum mechanics (MQM): the quantum mechanics of a single $N \times N$ Hermitian matrix in an inverted harmonic oscillator potential, in the double-scaling limit $N \to \infty$. Evidence for this duality was accumulated over the course of the early 1990s through the work of many authors \cite{Brezin:1990rb,Douglas:1989ve,Gross:1989vs,Gross:1989aw,Gross:1990ay,Brezin:1989ss,Ginsparg:1990as,Gross:1990st,Das:1990kaa,Polchinski:1991uq,Moore:1991sf,Mandal:1991ua,Demeterfi:1991tz,DiFrancesco:1991ocm,Moore:1992gb,Natsuume:1994sp,Moore:1991zv}. 
The original route proceeded through random triangulations of the worldsheet, which provide a lattice regularization of the path integral. In the double-scaling limit, the resulting matrix model reduces to a free-fermion system. Moore, Plesser and Ramgoolam \cite{Moore:1991zv} developed a diagrammatic formalism that extracts the \emph{exact} non-perturbative $S$-matrix from this system via an LSZ-type prescription, and agreement was verified with worldsheet computations to the extent that the latter existed at the time.

Interest in the $c=1$ string was rekindled in the mid-2000s via work on its D-brane spectrum, non-perturbative effects and as a laboratory for time-dependent backgrounds in string theory \cite{McGreevy:2003kb,Klebanov:2003km,Douglas:2003up,Takayanagi:2003sm,Alexandrov:2003nn,Alexandrov:2003un,Alexandrov:2004ks,Alexandrov:2004ip,Maldacena:2005hi,Fidkowski:2005ck,Martinec:2004td,Karczmarek:2003pv,Karczmarek:2004ph}. 
In particular, Alexandrov \cite{Alexandrov:2004ks} identified a spectral curve for the $c=1$ string from the disk amplitude with FZZT boundary conditions, foreshadowing a connection to the matrix integral paradigm that we will discuss below.
The topic was further revitalized in 2017 by \cite{Balthazar:2017mxh}, who revisited the $c=1$ $S$-matrix from the worldsheet using modern CFT techniques, and performed the first precision tests of the MQM/worldsheet duality beyond tree level. This work culminated in striking confirmations of the duality at the non-perturbative level through the study of ZZ-instanton effects \cite{Balthazar:2019rnh,Balthazar:2019ypi,Sen:2019qqg}, and stimulated a broader program aimed at understanding D-instanton corrections in string theory \cite{Sen:2020cef,Sen:2020eck, Eniceicu:2022xvk}.

An independent perspective on the $c=1$ string arose from the study of topological strings on Calabi-Yau threefolds. In particular, \cite{Ghoshal:1995wm} showed that the compactified $c=1$ string at the self-dual radius (in Euclidean target-space signature) is equivalent to the topological B-model on the deformed conifold, whose perturbative genus expansion is computed by Kodaira-Spencer theory \cite{Bershadsky:1993cx}. The open-closed duality of  \cite{Gopakumar:1998ki} then identifies the topological string free energy with that of the Gaussian one-matrix model in the 't Hooft limit, thereby relating the $c=1$ string at the self-dual radius to a matrix model. Dijkgraaf and Vafa \cite{Dijkgraaf:2003xk} gave this duality an explicit realization as a $\widehat{A}_1$ quiver matrix model, and extended it to integer multiples $k$ of the self-dual radius, which were shown to be governed by $\widehat{A}_{2k-1}$ quiver matrix models. In particular, the decompactified limit $k\to\infty$ yields a $\widehat{A}_\infty$ quiver matrix model, with the infinite array of nodes playing the role of a discretized target space. While the resulting matrix models are structurally distinct from the matrix integrals that are derived directly from the worldsheet in this paper, it is intriguing that both descriptions involve a discretization of the $c=1$ target space (as we will see shortly). 

A complementary approach to two-dimensional string theories has emerged from the study of matrix integrals and topological recursion. A well-established but a priori distinct paradigm in two-dimensional string theory is the duality between minimal string theories and double-scaled matrix integrals \cite{Gross:1989vs, Douglas:1989ve, Brezin:1990rb}, in which the perturbative genus expansion of the string theory is governed by the loop equations of a matrix integral and the geometry of its spectral curve. In recent years this paradigm has been sharpened through the development of topological recursion \cite{Eynard:2007kz}, and extended to several additional models of two-dimensional gravity, most notably Jackiw-Teitelboim dilaton gravity (JT gravity) \cite{Saad:2019lba}.

Another important example is the Virasoro minimal string (VMS) \cite{Collier:2023cyw}, which is defined by coupling $c\geq 25$ Liouville CFT to timelike Liouville CFT with central charge $\hat{c} =26-c$ on the worldsheet. It admits a matrix integral description governed by topological recursion on a specific spectral curve, and its string amplitudes --- the ``quantum volumes'' $\mathsf{V}_{g,n}^{(b)}$ --- are expressible as certain intersection numbers on the Deligne-Mumford compactification $\overline{\mathcal{M}}_{g,n}$ of the moduli space of the worldsheet Riemann surface.

These ideas were further developed with the construction of the complex Liouville string ($\CC$LS) \cite{Collier:2024kmo}, defined on the worldsheet by coupling two copies of Liouville CFT with complex conjugate central charges $c = 13 \pm i \mathbb{R}$.
The $\CC$LS string amplitudes $\mathsf{A}^{(b)}_{g,n}$ were bootstrapped from the analytic structure of the Liouville CFT correlators in \cite{Collier:2024kwt} (see also \cite{Khromov:2025awh} for a similar analysis of VMS amplitudes). In \cite{Collier:2024lys} it was argued that the $\CC$LS amplitudes are captured by a double-scaled two-matrix integral characterized by a specific spectral curve. The string amplitudes were expressed as sums over `colored' stable graphs, with each graph contributing an intersection number on the product of moduli spaces associated to its vertices. Remarkably, the intersection theory data for each graph reassembled into a product of the VMS quantum volumes of the vertices, together with vertex, edge, and leg factors determined by the spectral curve.

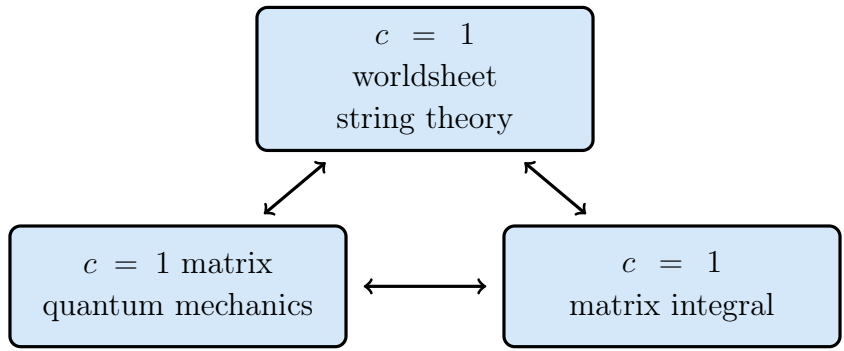
\begin{figure}[h]
\centering
\begin{tikzpicture}[ thick]

\tikzset{
theory/.style={
draw,very thick,
rounded corners,
align=center,
text width=4cm,
minimum height=1.6cm,
inner sep=6pt,
fill=bleudefrance,
fill opacity=0.2,
text opacity=1,
draw opacity=1
}
}

\node[theory] (str) at (0,2.75)
{$c=1$\\worldsheet string theory};

\node[theory] (mqm) at (-3.25,0)
{$c=1$ matrix\\quantum mechanics};

\node[theory] (mat) at (3.25,0)
{$c=1$\\matrix integral};

\draw[<->, very thick, shorten >=6pt, shorten <=6pt] (str) -- (mqm);
\draw[<->, very thick, shorten >=6pt, shorten <=6pt] (str) -- (mat);
\draw[<->, very thick, shorten >=6pt, shorten <=6pt] (mqm) -- (mat);

\node[white] at (0,-0.75){};

\end{tikzpicture}
\caption{Dual descriptions of $c=1$ strings.}\label{fig:triality}
\end{figure}

The main purpose of \textbf{this paper} is to demonstrate that the $c=1$ string also fits squarely in this paradigm and admits a description in terms of topological recursion on a spectral curve, and to elucidate the matrix integral that captures the $c=1$ amplitudes.
The matrix integral description presented in this paper thus completes a \emph{triality} that has no known analogue among the other examples discussed above: see figure \ref{fig:triality}.

The first leg of the triality (worldsheet~$\leftrightarrow$~MQM) is the well-established conjectural duality from the early 90s. The second leg (MQM~$\leftrightarrow$~matrix integral) was partially anticipated by Alexandrov \cite{Alexandrov:2004ks}, who identified a spectral curve from the FZZT disk amplitude, but was not developed into a full computational framework. What is new in this paper is (i) a self-contained derivation of the matrix integral and its topological recursion starting from the worldsheet, (ii) the identification of a Mirzakhani-type recursion relation for the $c=1$ amplitudes, and (iii) closed-form intersection number expressions that are analogous to Feynman rules for the amplitudes. We verify that the resulting amplitudes reproduce the known MQM results for the $S$-matrix elements, providing nontrivial evidence for the triality. Our starting point is the observation that $c=1$ string amplitudes arise as residues of (the analytic continuation of) $\CC$LS amplitudes at specific poles in the external momenta, and we leverage the intersection number description of \cite{Collier:2024lys} to derive the results.

We now summarize the main results of the paper.

\paragraph{$c=1$ amplitudes as intersection numbers.}
We derive new closed-form expressions for the $c=1$ string amplitudes as intersection numbers on the moduli space of Riemann surfaces. Our route exploits the observation that the $c=1$ string amplitudes can be extracted from the complex Liouville string by taking a specific limit. The $\CC$LS amplitudes, viewed as functions of the external momenta $p_j$, exhibit an infinite series of resonance poles whenever the background charge of either of the two Liouville factors is saturated \cite{Goulian:1990qr,Teschner:2001rv,Collier:2024kwt}. The residue of the leading pole reduces that Liouville factor to a free boson (or linear dilaton) correlator with background charge $Q = i(b^{-1}-b)$. Taking the subsequent limit $b\to 1$ (such that the background charge of the free boson factor vanishes) thus recovers the worldsheet correlator of the $c=1$ string.

This procedure can be directly carried out at the level of the intersection number expressions \cite{Collier:2024lys} for the $\CC$LS amplitudes, which are naturally given as a sum over so-called colored stable graphs. The residue and $b\to 1$ limit can then be carried out graph by graph, yielding a closed-form expression for the $c=1$ amplitudes as a sum over stable graphs (graphs labeling possible degenerations of Riemann surfaces), with each graph contributing an intersection number on the product of moduli spaces associated to its vertices. To our knowledge, these are the first intersection number expressions for the $c=1$ string amplitudes.

The result can be viewed as position-space Feynman rules for the $c=1$ string amplitudes: each vertex $v$ carries an integer `color' $m_v\in\mathbb{Z}$, defined up to overall translations $\boldsymbol{m}\sim\boldsymbol{m}+1$, playing the role of a lattice position in a discretized target space. The propagator depends only on the color difference $m_\circ - m_\bullet$ between adjacent vertices. Passing to momentum space by discrete Fourier transform in the vertex colors leads to a more compact expression. The sum over colors is performed in closed form and the propagator becomes a Bernoulli polynomial $B_{2d+2}(\{p_e\})$ depending on the fractional parts of the edge momenta $p_e$, and the loop momenta are integrated over the compact space $\mathbb{R}/\mathbb{Z}$. The intersection numbers can be reduced to the quantum volumes of the Virasoro minimal string \cite{Collier:2023cyw} dressed with simple Feynman propagators, see \eqref{eq:Feynman rules integer shifts}.

\paragraph{Discretized amplitudes and the first Brillouin zone.}
The intersection number formula naturally computes a version of the $c=1$ amplitudes in which total (Euclidean) momentum is only conserved modulo an integer, reflecting a discrete rather than continuous translation symmetry in target space. In this paper we refer to the discretized $c=1$ amplitudes as $\mathsf{s}_{g,n}(p_1,\ldots,p_n)$, where $p_j$ are the momenta in Euclidean signature. The full amplitude can thus be expressed in terms of a Dirac comb $\delta_\ZZ(x)=\sum_{m \in \ZZ} \delta(x-m)$
\be
    \mathsf{s}_{g,n}(p_1,\ldots,p_n) = \delta_\ZZ(p_1+\cdots+p_n) \, \hat{\mathsf{s}}_{g,n}(p_1,\ldots,p_n)\,. \label{eq:full s Dirac comb}
\ee
It is only the sector where momentum is exactly conserved that corresponds to the $S$-matrix elements of the physical $c=1$ string (in Euclidean signature). Restricting to the physical sector amounts to evaluating $\hat{\mathsf{s}}_{g,n}$ in the ``first Brillouin zone'' --- the region $|p_{I}|<1$ for all partial sums of momenta $p_{I}$ --- and then analytically continuing to Lorentzian kinematics via an $i\varepsilon$ prescription \cite{Witten:2013pra,Eberhardt:2022zay}. We refer to $\mathsf{S}_{g,n}(\omega_1,\ldots,\omega_n)$ as the physical $c=1$ amplitudes obtained in this way, where $\omega_j$ are the momenta in Lorentzian signature. Left implicit in this notation is the choice of ingoing and outgoing momenta, which affects the analytic continuation.
More precisely, the matrix integral and intersection number expressions compute the discretized amplitudes, which contain the physical $c=1$ amplitudes computed by the worldsheet and MQM upon restriction to the first Brillouin zone. 
The discretized amplitudes themselves cannot be directly recovered from the physical $c=1$ amplitudes, except in the first Brillouin zone.

The discretized amplitudes enjoy several attractive features. Loop momenta range over the compact space $\mathbb{R}/\mathbb{Z}$, rendering all loop integrals manifestly UV-finite. The amplitudes are piecewise polynomial: on each cell of the Brillouin zone decomposition the amplitude $\hat{\mathsf{s}}_{g,n}$ is a polynomial in the external momenta, with the different cells related by non-analyticities encoded by Bernoulli polynomials depending on the fractional parts of partial sums of the momenta.

Within the first Brillouin zone, the non-analyticities have a direct worldsheet origin: the moduli space integral (\ref{eq:c=1 S-matrix worldsheet definition}) diverges near the boundary of moduli space for physical kinematics (real Lorentzian energies), and one must choose an $i\varepsilon$ prescription (amounting to a Wick rotation of worldsheet proper time) to define the amplitude. This choice selects a branch of $\sqrt{-\omega_e^2}$ for each intermediate energy $\omega_e$. The non-analyticities at the higher Brillouin-zone boundaries $p_{I}\in\mathbb{Z}$ reflect the discrete translation symmetry of the colored stable graphs inherited from the $\CC$LS amplitudes. The full discretized amplitude $\hat{\mathsf{s}}_{g,n}$ packages these features together into a single piecewise polynomial object.

The situation is loosely analogous to phonon scattering in a one-dimensional lattice. The lattice provides a natural UV cutoff, momentum is conserved only modulo a reciprocal lattice vector, and the physical S-matrix elements $\mathsf{S}_{g,n}$ emerge in the long-wavelength (first Brillouin zone) limit where the lattice structure becomes invisible. The non-analyticities at the Brillouin zone boundaries are the origin of the branch cuts of the physical amplitudes and play a central role in the proof of perturbative spacetime unitarity that we turn to next.

\paragraph{Perturbative unitarity.}
We give a direct proof that the $c=1$ string amplitudes as defined by the intersection theory formula satisfy spacetime unitarity, using the holomorphic cutting rules of \cite{Hannesdottir:2022bmo}. The main observation is that the branch cuts of the physical amplitudes originate entirely from the edge factors in the stable graph expansion: the Bernoulli polynomials $B_{2d+2}(\{p_e\})$ are piecewise smooth and their discontinuities at integer momenta factorize into products of lower-point on-shell amplitudes weighted by the phase space measure. Summing over all possible cuts verifies the optical theorem order-by-order in the genus expansion. The proof works entirely from the intersection theory expressions and does not use the MQM description.

\paragraph{Spectral curve and topological recursion.}
We identify a double-scaled matrix integral whose perturbative expansion captures the $c=1$ string amplitudes. As in the other examples discussed above, the matrix integral is fully determined at the perturbative level by its spectral curve characterizing the large-$N$ distribution of eigenvalues, which we find to be
\be\label{eq:c1 spectral curve}
    \xx(z) = 2\sqrt{2}\cos(z), \quad \yy(z) = \sin(z)\, ,
\ee
parameterizing the ellipse $\xx^2/8+\yy^2 = 1$. The $2\pi$ periodicity of $\xx(z)$ realizes this as an infinite-sheeted covering of the plane, with branch points at $z_m^* = \pi m$, for $m\in\mathbb{Z}$. Topological recursion \cite{Eynard:2002kg,Chekhov:2006vd} on this curve, summing over residues at the branch points, generates the perturbative resolvents $\omega_{g,n}$ of the matrix integral. As in $\CC$LS the dictionary between the resolvents and the $S$-matrix elements $\hat{\mathsf{s}}_{g,n}$ likewise involves a sum over branch points. The residue at $z_m^*$ picks up the contribution from the $m^{\text{th}}$ sheet, and summing over $m$ with the phase $\e^{2\pi i m p}$ implements the discrete Fourier transform from the sheet index (lattice position) to momentum space.

We translate the topological recursion to a recursion relation for the discrete string amplitudes $\hat{\mathsf{s}}_{g,n}$, which has the form of a \emph{Mirzakhani-type recursion relation}. Like Mirzakhani's original recursion for the Weil-Petersson volumes \cite{Mirzakhani:2006fta}, it is a three-term recursion corresponding to the three distinct ways of embedding a pair of pants with a distinguished external leg into the worldsheet surface: creating a handle, splitting the surface, or joining the distinguished leg with another external leg. We simplify the recursion into a form that is efficient for explicit computation. The recursion kernel separates cleanly into the recursion kernel of the VMS quantum volumes at $c=25$, plus corrections involving the fractional parts $\{p_{I}\}$ of partial sums of the external momenta. The latter may be viewed as boundary corrections from the compactification of moduli space that are specific to the $c=1$ string. Importantly, the recursion involves a sum over sub-amplitudes where momentum is conserved only modulo an integer. Thus the recursion does not close on physical $c=1$ string amplitudes where total momentum is strictly conserved; the sectors describing `Umklapp scattering' where momentum is only conserved up to an integer are necessary for the recursive structure.

\paragraph{Polynomial structure and zeros on the physical sheet.}
The physical $c=1$ amplitudes exhibit a striking polynomial structure. The non-discretized physical amplitude $\mathsf{S}_{g,n}$, evaluated on the physical sheet, is a polynomial in the external momenta of degree $4g-3+n$. This reduced degree is well-known from MQM \cite{Moore:1991zv}, but is highly nontrivial from the worldsheet perspective: it is considerably lower than the naive upper bound of $6g-6+2n$ suggested by the dimension of the worldsheet moduli space $\mathcal{M}_{g,n}$, indicating large cancellations within the intersection theory expressions. We verify these cancellations in several examples. For instance, in the $n-1\to 1$ scattering channel the physical amplitudes take the remarkably simple form
\be
    \mathsf{S}_{g,n}(\boldsymbol{\omega}) = (1+ie_1)_{2g-3+n}P_{g}(e_1,\ldots,e_{2g})\, ,
\ee
where $e_i$ is the $i^\text{th}$ symmetric polynomial in the incoming momenta $\omega_1,\ldots,\omega_{n-1}$ ($e_i=0$ for $i>n-1$), $(a)_n$ is the rising Pochhammer symbol, and $P_{g}$ is a polynomial that we determine explicitly for $g\leq 4$ and low-lying $n$ (expressed in terms of the momenta, it is a polynomial of degree $2g$). Strikingly the $n$-dependence enters only through the Pochhammer factor, which produces a pattern of zeros in the total outgoing momentum that is natural from the MQM free fermion description (they arise from the perturbative expansion of the MPR formula \eqref{eq:MPR n to 1}) but that is totally obscured in the worldsheet or matrix integral formulations.

At genus zero, we are able to prove the reduced polynomial degree by rewriting the sum over stable graphs as a single intersection number on $\overline{\mathcal{M}}_{0,n}$ involving pushforwards of boundary classes. In this form, the degree bound $n-3$ is manifest. For higher genus, a similar rewriting is possible, but only after discretization. It requires us to pass to a larger moduli space, the so-called moduli space of $r$-spin curves, which parametrizes stable surfaces together with certain line bundles. 

We also show that the integrand of the intersection theory expression for the amplitudes factorizes appropriately under pinching of the surface. These factorization properties underlie the topological recursion of the dual matrix integral.

\paragraph{Relation to MQM and the chain of matrices.}
We verify that the physical amplitudes extracted from the $c=1$ spectral curve and topological recursion agree with the known results from the dual MQM, as computed via the free-fermion formalism of Moore, Plesser and Ramgoolam \cite{Moore:1991zv}. The agreement is checked to genus 5 for the partition function, and for higher multiplicities at lower genera, see eq.~\eqref{eq:Sgn direct computation}. In the \hyperref[sec:discussion]{discussion} we speculate about a more direct connection between the matrix integral and discretized matrix quantum mechanics descriptions through Eynard's ``chain of matrices'' model \cite{Eynard:2003kf, Eynard:2009zz}.

\paragraph{Outline.}
This paper is organized as follows. In section \ref{sec:CLS to c1} we derive the intersection theory expressions for the $c=1$ string amplitudes, starting from the resonance residues of the $\CC$LS amplitudes. We present the position and momentum space Feynman rules, discuss the restriction to the first Brillouin zone and the analytic continuation to physical kinematics, and give a direct proof of perturbative unitarity. We also discuss the rewriting of the amplitudes as a single intersection number involving boundary classes, and characterize the higher-genus amplitudes in terms of cohomological field theory. In section \ref{sec:matrix integral} we elucidate the $c=1$ matrix integral that governs the discretized amplitudes: we present the spectral curve, the topological recursion, the dictionary relating resolvents and string amplitudes, and the Mirzakhani-type recursion relation for the discretized amplitudes. We conclude in the \hyperref[sec:discussion]{discussion} with a discussion of generalizations --- including the extension to $b\ne 1$ --- and future directions, such as the chain-of-matrices interpretation of the matrix integral and non-perturbative effects. Appendix \ref{app:MPR comparison} contains the comparison with the MQM free-fermion formalism, appendix \ref{app:algebraic geometry} reviews the algebraic geometry of the stable graph expansion, and appendix \ref{app:Bernoulli integration} provides a recursion for integrating products of Bernoulli polynomials involving fractional parts of momenta.

\section{\texorpdfstring{$c=1$}{c=1} amplitudes from geometry} \label{sec:CLS to c1}

In this section, we show that $c=1$ string amplitudes can be directly computed as intersection numbers on the moduli space $\bM_{g,n}$, see eqs.~\eqref{eq:c=1 momentum space intersection numbers} and \eqref{eq:Feynman rules integer shifts}. We derive such a representation of the amplitudes from the complex Liouville string which enjoys a similar intersection number expression. One can collapse one of the worldsheet Liouville factors of the complex Liouville string to a free boson factor by taking a resonance residue. One then considers the limit where its central charge becomes 1, which recovers the worldsheet theory of $c=1$ string theory. The resulting amplitudes admit an exact stable-graph/intersection-number expansion, which can be viewed as Feynman diagrams for the theory. Physical $c=1$ S‑matrix elements are obtained by evaluating this expansion in its region of absolute-convergence. We discuss various properties of the resulting amplitudes.

\subsection{Setup and conventions}

\paragraph{$c=1$ string.} The worldsheet theory of the $c=1$ string consists of $c=25$ Liouville theory, coupled to a timelike free boson. Physical vertex operators of the theory consist of tachyon vertex operators in the free boson theory, dressed with an appropriate Liouville vertex operator (as well as $c$-ghosts). We can parametrize the Liouville conformal weight by $h_\text{Liouville}=\frac{c-1}{24}-p^2=1-p^2$, so that the mass-shell condition constrains the free boson conformal weight to be $h_\text{boson}=p^2$. For physical choices of kinematics, the timelike free boson conformal weight should be negative, so that $p$ is purely imaginary.
To connect with the convention used in the earlier literature \cite{Klebanov:1991qa, Polchinski:1991uq, Balthazar:2017mxh}, we should set
\be
p=\frac{i\omega}{2}\,, \label{eq:p omega relation}
\ee
where we think of $\omega$ as the energy of the string state.
We will use the convention where $\omega>0$ for ingoing states and $\omega<0$ for outgoing states.
The $c=1$ string S-matrix elements are obtained by integrating the product of the Liouville correlator and the free boson correlator over moduli space:
\begin{align}
\mathsf{S}_{g,n}(\omega_1,\dots,\omega_n)&=C^{(b)}_{\Sigma_g} \int_{\mathcal{M}_{g,n}} \Big\langle \prod_{k=1}^{3g-3+n} \mathcal{B}_k \widetilde{\mathcal{B}}_k \prod_{j=1}^{n} \mathcal{T}_{p_j=\frac{i}{2} \omega_j} \Big\rangle_{\Sigma_{g,n}} \,, \label{eq:c=1 S-matrix worldsheet definition}
\end{align}
where $C^{(b)}_{\Sigma_g}$ is a normalization factor and $\mathcal{B}_k$ are the $b$-ghosts, paired with a basis of Beltrami differentials.
Here, $\mathcal{T}_{p_j=\frac{i}{2} \omega_j}\simeq \ex^{i\omega_j X^0}V_{p_j=\frac{i}{2} \omega_j}$ are the on-shell `tachyon' vertex operators, constructed from the timelike free boson $X^0$ and a Liouville operator $V_{p_j}$. 
After stripping off the momentum conserving delta functions, we shall denote the scattering amplitude by $\hat{\mathsf{S}}_{g,n}(\omega_1,\dots,\omega_n)$. The precise relation involves a proportionality constant $\mathcal{N}$,
\be
\mathsf{S}_{g,n}(\omega_1,\dots,\omega_n)=\mathcal{N} \, \delta\Big(\sum\nolimits_i \omega_i\Big) \hat{\mathsf{S}}_{g,n}(\omega_1,\dots,\omega_n)\,, \label{eq:S delta function strip off normalization}
\ee
which we will fix once we analyze spacetime unitarity and fix all other normalization constants, see \eqref{eq:N value}.

The integral \eqref{eq:c=1 S-matrix worldsheet definition} converges only for a certain region of parameter space. In particular, it converges for \emph{real} $p$, i.e. \emph{imaginary} energies, whereas we are ultimately interested in real energies for the physical amplitudes. This situation is completely analogous to the situation for type II superstring scattering amplitudes, see e.g.\ \cite{Witten:2013pra, DHoker:1994gnm}. Nonetheless, this fact motivates us to study \eqref{eq:c=1 S-matrix worldsheet definition} for complex values of $\omega_j$.
We can define the physical amplitudes either by analytic continuation, suitable regularization \cite{Balthazar:2017mxh}, or the use of the $i \varepsilon$-prescription \cite{Witten:2013pra}, which we shall discuss further in section~\ref{subsec:remarks}.
It is customary to also include so-called leg factors in the definition of the amplitudes owing to the arbitrary normalization of vertex operators. We will make a convenient choice for those factors later. Let us also remark the obvious property
\be
\hat{\mathsf{S}}_{g,n}(\omega_1,\dots,\omega_n)=\hat{\mathsf{S}}_{g,n}(-\omega_1,\dots,-\omega_n)\,, \label{eq:flipping all momenta}
\ee
which follows from the fact that the free boson correlator only depends quadratically on the momenta. We will in the following often denote a collection of quantities by boldface letters, such as $\hat{\mathsf{S}}_{g,n}(\boldsymbol{\omega})=\hat{\mathsf{S}}_{g,n}(\omega_1,\dots,\omega_n)$.

\paragraph{Complex Liouville string.} We will leverage insights from the study of another string theory with a 2d target space to study $c=1$ string theory --- the so-called Complex Liouville String ($\CC$LS) \cite{Collier:2024kwt}. Let us also review its basic definition from the worldsheet.
Its worldsheet theory consists of two Liouville theories of central charge $c=1+6(b+b^{-1})^2$ and central charge $26-c$. In \cite{Collier:2024kwt}, the theory was mostly studied in the regime where $b^2 \in i \RR$ (i.e.\ $c \in 13+i \RR$), since the physical Hilbert space admits an inner product in that case. However, observables can be defined for any $c \in \CC \setminus \{(-\infty,1] \cup [25,\infty)\}$. Physical vertex operators can be constructed as products of primary vertex operators in the two Liouville theories of momentum $p$ and $i p$ (so that the mass-shell condition is satisfied). It is again possible to define the observables of the $\CC$LS for any choice of complex momentum, although the string correlators only converge absolutely in a certain region in parameter space and have to be defined by analytic continuation in the rest of parameter space. We will denote the $\CC$LS correlation functions by $\mathsf{A}^{(b)}_{g,n}(p_1,\dots,p_n)$.

It is also worth pointing out that in the $\CC$LS, the states with momenta $p$ and $-p$ are identical (reflection symmetry), while in the $c=1$ string, they correspond to in- and out-states, respectively. In particular, there is no form of reflection symmetry in $c=1$ string theory, since the momenta $p$ and $-p$ are inequivalent for the free boson factor. The $c=1$ S-matrix is only invariant under a joint reflection in all vertex operators, see \eqref{eq:flipping all momenta}.

\subsection{Degenerating \texorpdfstring{$\CC$LS}{CLS} amplitudes} \label{subsec:degenerating CLS amplitudes}
We shall now explain how the $c=1$ S-matrix is contained in the $\CC$LS correlation functions by taking a certain degeneration limit.

\paragraph{Resonance poles.} Let us recall the following property of Liouville correlators. Liouville correlators have a series of poles as a function of the external momenta. In particular, they have poles whenever the background charge is saturated, meaning that \cite{Goulian:1990qr, Teschner:2001rv}\footnote{There are also corresponding poles for $p_i \to -p_i$, but we will focus on these.}
\be
\sum_{i=1}^n p_i=\frac{b+b^{-1}}{2}
(2g-2+n)+r b+ s b^{-1}\,, \qquad r,\, s \in \ZZ_{\ge 0}\,. \label{eq:background charge poles}
\ee
The existence of these poles can be derived axiomatically from the OPE expansion. More conceptually, they arise from constructing Liouville theory as a deformation of a free boson with a background charge by the non-normalizable operator $\mu \, \ex^{2b \phi}$. In particular, the residue of the first pole for $r=s=0$ equals the free boson correlator with background charge $Q=b+b^{-1}$. Residues of higher poles can in principle be computed in the Coulomb gas formalism, but we will not make use of them.

Thus, the path from $\mathbb{C}$LS to the $c=1$ string is clear: we will extract the leading residue of the $\mathbb{C}$LS correlators when the second Liouville theory (for which $p_j \to i p_j$ and $b \to i b$ compared to the first Liouville factors) saturates the background charge, i.e.\ for
\be
\sum_j p_j=\frac{b-b^{-1}}{2}(2g-2+n)\,. \label{eq:anomalous momentum conservation}
\ee
Notice the minus sign on the RHS of \eqref{eq:anomalous momentum conservation} compared to \eqref{eq:background charge poles}. Afterwards, we can take the limit $b \to 1$. This switches off the background charge in the free boson factor and turns the anomalous momentum conservation \eqref{eq:anomalous momentum conservation} into ordinary momentum conservation. We should note that the order of limits is important, since the limit $b \to 1$ is only well-defined after extracting the resonance pole \cite{Ribault:2015sxa, Collier:2024kwt}.
We should also note that many of the results of this paper hold before taking $b \to 1$ and we will mention some of the modifications that have to be performed in the \hyperref[sec:discussion]{discussion}.

\paragraph{Branch cuts and analytic structure.} At the level of the integrand, this procedure clearly reduces the $\CC$LS integrands to the $c=1$ string integrands. Thus, as long as the moduli space integral converges absolutely, the same holds for the integrated amplitudes. Beyond the region of absolute convergence, however, the situation is more subtle. Both the $\CC$LS correlators and the $c=1$ $S$-matrix have branch cuts, and some branches present in $\CC$LS drop out once the correlator is degenerated to that of the $c=1$ string. Taking the $b \to 1$ limit on a sheet that survives this degeneration is expected to reproduce the $c=1$ correlator by analytic continuation, but taking it on a sheet that does not survive need not yield a result related to $c=1$ string theory.

Thus, we see that beyond the region of absolute convergence of the moduli space integral, the resulting limit depends on a branch choice. In
\cite{Collier:2024kwt, Collier:2024lys}, explicit formulas were proposed for the extension of the $\CC$LS correlators outside of this region. These formulas necessarily make a branch choice. It will turn out that unless the external momenta are chosen to be sufficiently small to lie in the region of absolute convergence, this branch choice \emph{does not} correspond to a branch of the $c=1$ amplitudes. When taking the limit, we find versions of $c=1$ amplitudes that \emph{differ} from the usual $c=1$ amplitudes when we go outside the region of absolute convergence. This modified $c=1$ string will have the interpretation of a theory with a discrete target space as we will discuss in section~\ref{subsec:remarks}.
We emphasize that this is not a fatal problem since we can always evaluate an amplitude in a region where the moduli space integral converges absolutely and analytically continue to the desired kinematics, recovering the physical $c=1$ amplitudes.

\begin{figure}[ht]
    \centering
    \begin{subfigure}[b]{.45\textwidth}
        \includegraphics[width=\textwidth]{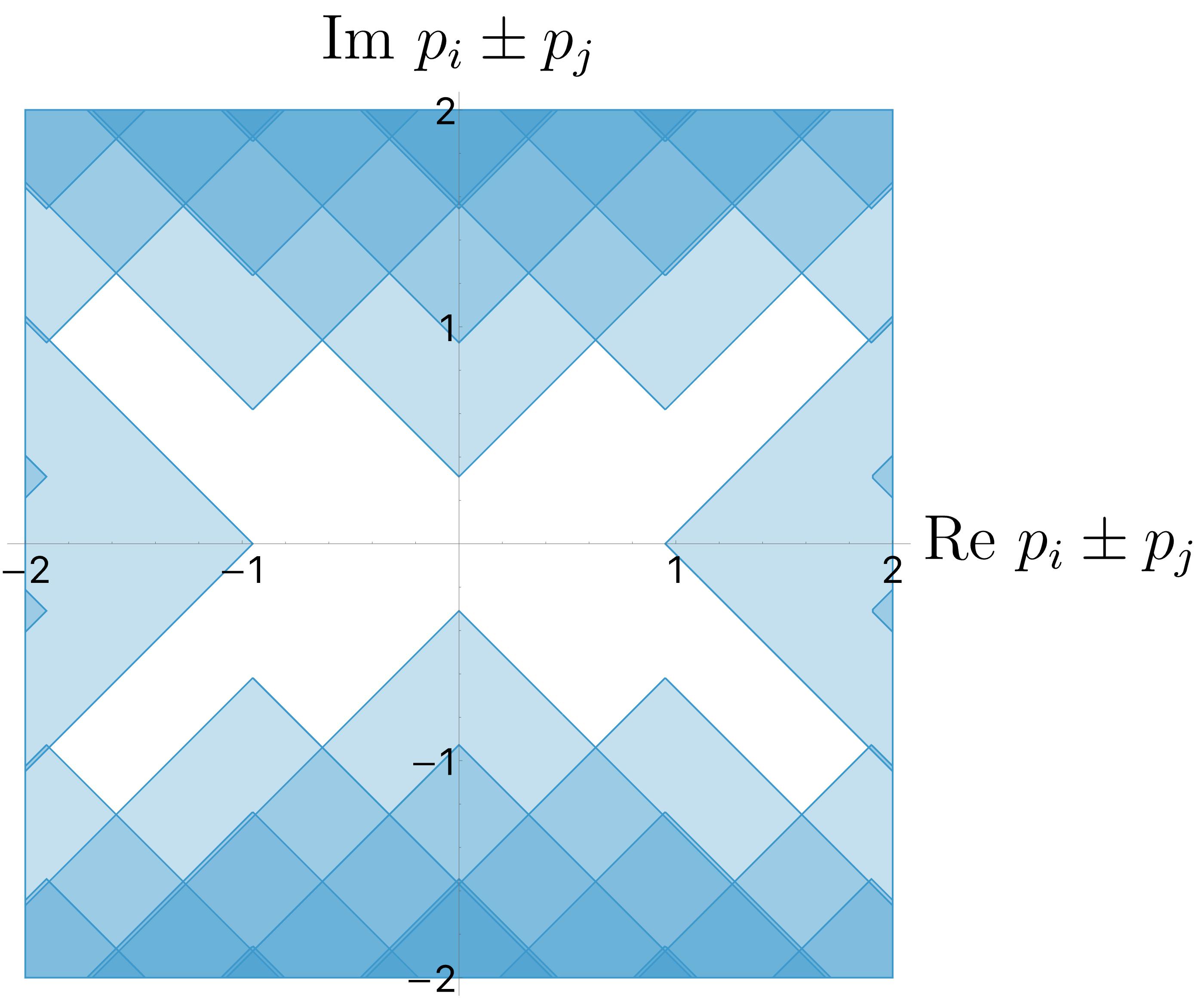}
        \caption{$b = \mathrm{e}^{\frac{\pi i}{10}}$}
    \end{subfigure}
    ~
    \begin{subfigure}[b]{.45\textwidth}
        \includegraphics[width=\textwidth]{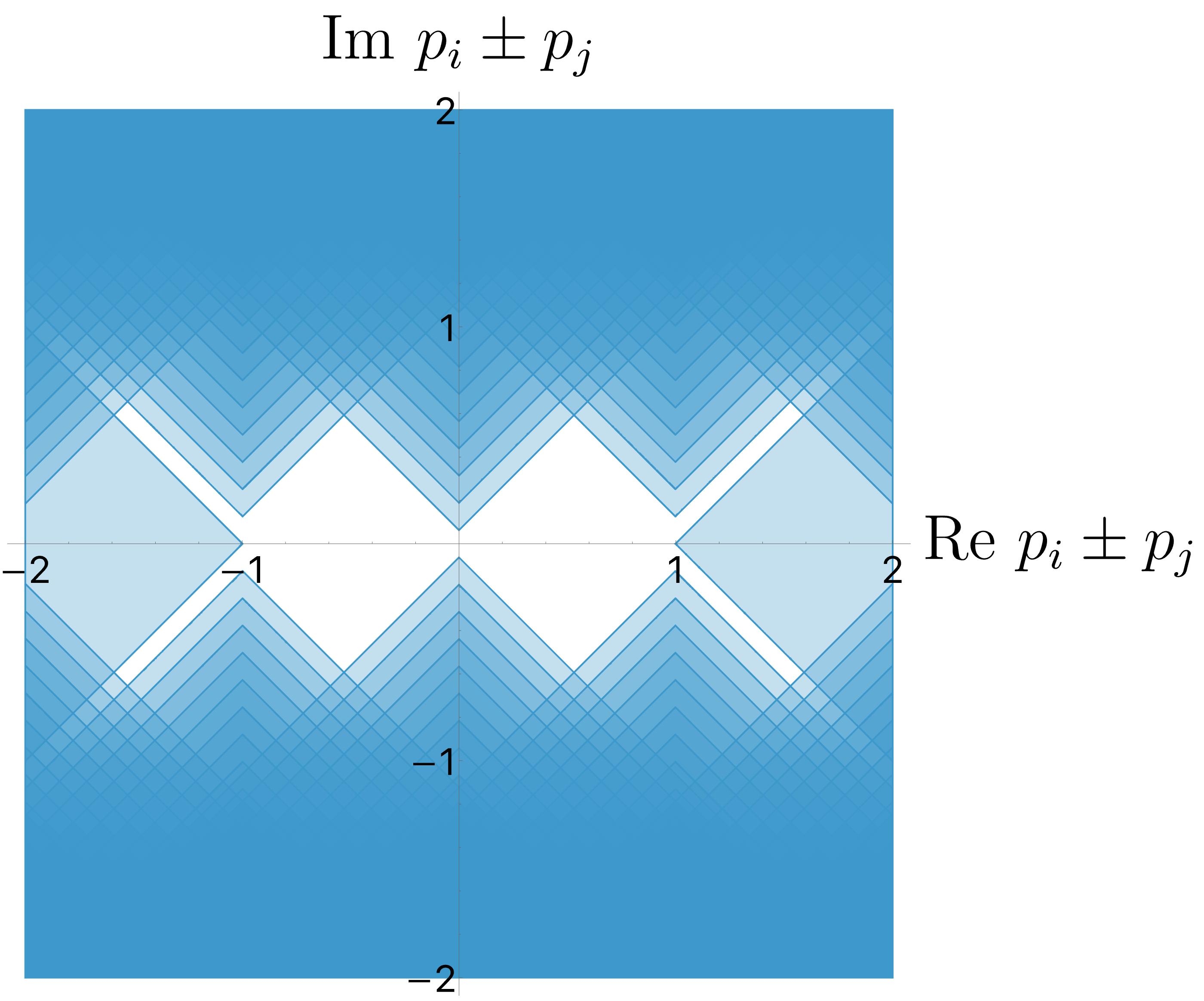}
        \caption{$b = \mathrm{e}^{\frac{\pi i}{50}}$}
    \end{subfigure}
    \caption{The unshaded area represents the region in the external Liouville momenta where the moduli space integral that defines the $\CC$LS four-point function converges. 
    The shaded regions correspond to the 90-degree wedges of divergence emanating from the branch points of the amplitude.
    In the limit that $b$ is taken to one, the region of convergence pinches off and infinitely many branch points collide at $p_i\pm p_j = -1,0,1$, corresponding to the Brillouin zone boundaries described in the introduction.}\label{fig:region of analyticity}
\end{figure}

Let us illustrate this phenomenon in more detail for the sphere four-point function. The worldsheet integral of the $\CC$LS correlator $\mathsf{A}_{0,4}^{(b)}(p_1,p_2,p_3,p_4)$ has branch points for $p_i \pm p_j \in (\ZZ+\frac{1}{2}) b \oplus (\ZZ+\frac{1}{2}) b^{-1}$.
The region of absolute convergence is the complement of all `wedges' emanating away from the origin from these branch points. We refer to \cite[Section 3.1]{Collier:2024kwt} for a more precise discussion. See also Figure~\ref{fig:region of analyticity} for an illustration for various values of $b$.
In particular, the branch points closest to the origin in $p_i \pm p_j$ are located at $\frac{1}{2}(b+b^{-1})$, $\frac{1}{2}(b-b^{-1})$, $\frac{1}{2}(-b-b^{-1})$ and $\frac{1}{2}(-b+b^{-1})$. Let us now choose all $p_j$'s real. Then the region of absolute convergence is the interval $p_i\pm p_j \in [-\frac{1}{2}\Re(b+b^{-1}),\frac{1}{2}\Re(b+b^{-1})]$. Upon taking the limit $b \to 1$, absolute convergence forces $p_j \pm p_i \in [-1,1]$, where we will get agreement of the degenerated $\CC$LS correlator and the $c=1$ correlator. See also Figure~\ref{fig:region of analyticity} for how the region of analyticity degenerates in the limit $b \to 1$. This is the origin of the Brillouin zone boundaries $|p_I|=1$ encountered in the introduction. We also see that the region of analyticity becomes pinched at the origin, leading to a discontinuity of the $c=1$ amplitudes. 

\paragraph{Three-point function.} Let us illustrate the procedure with the simple example of the three-point function. We will substantially generalize this computation below. In $\mathbb{C}$LS, it takes the form \cite{Collier:2024kwt},
\be
\mathsf{A}_{0,3}^{(b)}(\boldsymbol{p})=\sum_{m=1}^\infty \frac{2b (-1)^m \sin(2\pi m b p_1)\sin(2\pi m b p_2)\sin(2\pi m b p_3)}{\sin(\pi m b^2)}\,. \label{eq:CLS three-point function}
\ee
This representation converges for small enough $p_i$. It starts to diverge for $p_1+p_2+p_3 \to \frac{b-b^{-1}}{2}$, which is precisely the limit that we need to analyze in order to compute the residue, see \eqref{eq:anomalous momentum conservation}. The sum over $m=1,\dots,M$ for any finite $M$ would not have residues and thus only the asymptotic tail as $m \to \infty$ is important for the residue. Thus, we can replace $\sin(\pi m b^2) \rightsquigarrow \frac{i}{2} \ex^{-\pi i m b^2}$, since we assumed that $\Im b^2 \to 0$ from above. We can also expand the sines in the numerator in terms of exponentials. Only one combination leads to the desired pole, the others lead to the reflected poles. This leads to the sum
\begin{align}
\mathsf{A}_{0,3}^{(b)}(\boldsymbol{p})&=-\frac{b}{2}\sum_{m=1}^\infty \ex^{-2\pi i m b(p_1+p_2+p_3-\frac{b-b^{-1}}{2})}+\text{regular} \\
&=\frac{i}{4\pi (p_1+p_2+p_3-\frac{b-b^{-1}}{2})}+\text{regular}\,.
\end{align}
Thus extracting the residue leads to a constant $c=1$ three-point function. We will fix the normalization conventions below.
\subsection{\texorpdfstring{$c=1$}{c=1} amplitudes as intersection numbers} \label{subsec:c=1 amplitudes as intersection numbers}
\paragraph{Intersection number expression.} We can apply a similar manipulation to a general $\CC$LS correlation function. These admit an expansion in terms of so-called stable graphs labelling all possible degenerations of Riemann surfaces.  The formula is \cite[eq.~(B.19)]{Collier:2024lys}
\begin{align}
    \mathsf{A}_{g,n}^{(b)}(\boldsymbol{p})&= \sum_{\Gamma \in \mathcal{G}_{g,n}^\infty}\frac{1}{|\text{Aut}(\Gamma)|}\prod_{v \in \mathcal{V}_\Gamma} \left(\frac{b (-1)^{m_v}}{\sqrt{2}\sin(\pi m_v b^2)}\right)^{2g_v-2+n_v} \!\! \int_{\bM_\Gamma} \prod_{v \in \mathcal{V}_\Gamma}\ex^{\frac{b^2+b^{-2}}{4} \kappa_1-\sum_k\frac{B_{2k}\kappa_{2k}}{(2k)(2k)!}} \nonumber\\
    &\qquad\times\!\prod_{(\bullet,\circ) \in \mathcal{E}_\Gamma} \sum_{d=0}^\infty \frac{\Gamma(d+\tfrac{3}{2})}{\sqrt{\pi}(\pi b)^{2d+2}}\left(\frac{\delta_{m_\bullet\ne m_\circ}}{(m_\bullet-m_\circ)^{2d+2}}-\frac{1}{(m_\bullet+m_\circ)^{2d+2}}\right)(\psi_\bullet+\psi_\circ)^d \nonumber\\
    &\qquad\times \prod_{i=1}^n \ex^{-p_i^2 \psi_i} \sqrt{2}\sin(2\pi m_i b p_i)\,.\label{eq:CLS intersection expression}
\end{align}
Here we used the abbreviation $\boldsymbol{p}=\{p_1,\dots,p_n\}$.
Let us review the different ingredients going into this formula. To begin with, the objects $\Gamma \in \mathcal{G}_{g,n}^\infty$ that we are summing over are unoriented connected graphs (self-loops and multi-edges allowed) where every vertex $v \in \mathcal{V}_\Gamma$ is assigned a genus $g_v$ as well as a `color' $m_v\in \ZZ_{\ge 1}$. The total genus is fixed to be $g=\sum_v g_v+L$, where $L$ is the number of loops in the graph ($L=\dim \mathrm{H}^1(\Gamma,\RR)$). The number of external legs $n$ is also fixed.
For a vertex $v \in \mathcal{V}_\Gamma$, $n_v$ denotes the number of emanating internal or external edges. The graph is moreover required to be stable, meaning that $n_v \ge 1$ if $g_v=1$ and $n_v \ge 3$ if $g_v=0$. From Euler's formula, we have the simple relations
\be
2g-2+n=\sum_{v \in \mathcal{V}_\Gamma} (2g_v-2+n_v)\,.
\ee
The automorphism group $\text{Aut}(\Gamma)$ is defined to be the automorphism group of the underlying graph where we forget about the genus and color assignments. For each graph, we integrate over the compactified moduli space $\overline{\mathcal{M}}_\Gamma=\prod_{v \in \mathcal{V}_\Gamma} \overline{\mathcal{M}}_{g_v,n_v}$.\footnote{In the mathematical literature, it is more common to define $\overline{\mathcal{M}}_\Gamma=\prod_{v \in \mathcal{V}_\Gamma} \overline{\mathcal{M}}_{g_v,n_v}/\text{Aut}(\Gamma)$, which absorbs the automorphism factor into the integral. We choose not to do so, since \eqref{eq:CLS intersection expression} has the flavour of a Feynman diagram expression.} Passing to the compactification is necessary in order to define the intersection pairing (i.e.\ have a non-trivial top cohomology class). This formula expresses this integral in terms of the standard $\psi$- and $\kappa$-classes on moduli space. These are degree-2 cohomology classes. There is one $\psi$-class associated to every marked point (external or internal) as well as one $\kappa_m$ class on the moduli space at every vertex. In the second line \eqref{eq:CLS intersection expression}, we take the product over all edges in the graph. We take $\bullet$ and $\circ$ as a shorthand for the adjacent vertices. In particular, $m_\bullet$ and $m_\circ$ are the color assignments of the adjacent vertices and $\psi_\bullet$ and $\psi_\circ$ are the $\psi$-classes of the corresponding marked point on the two vertex moduli spaces. Finally, in the last line, we have the $\psi$-classes of the external marked points appearing, which belong to one of the vertex moduli spaces.

The formula \eqref{eq:CLS intersection expression} was derived in \cite{Collier:2024kwt, Collier:2024lys} via an analytic bootstrap: rather than evaluating the moduli space integral over Liouville correlators directly (which appears prohibitively difficult), one deduces the answer by constraining its analytic properties. Under favorable assumptions, these constraints fix the result uniquely and lead to \eqref{eq:CLS intersection expression}. We should note, however, that while a lot of evidence for \eqref{eq:CLS intersection expression} was assembled in \cite{Collier:2024kwt, Collier:2024lys} (including extensive numerical verification and consistency with analytic constraints on the correlators), a complete proof from the worldsheet does not yet exist. All the results derived in this section (the extraction of $c=1$ amplitudes, and the unitarity proof) are conditional on the validity of this formula and provide further nontrivial consistency checks.

\paragraph{Extracting the residue.} Even though the expression \eqref{eq:CLS intersection expression} looks perhaps somewhat complicated, the required steps to extract the residue are essentially the same as for the three-point function \eqref{eq:CLS three-point function}. The sum over the colors $m_v$ only converges in a region of the parameter space of external momenta $p_i$. We precisely get the first divergence when the background charge is saturated as in \eqref{eq:anomalous momentum conservation}. It arises again from a geometric series. This geometric series corresponds to the diagonal subsum over the vertex colors $m_v$. Thus we will in the following replace $m_v \rightsquigarrow m_v+m$ and perform the sum over $m$. This subsum is the only one leading to a pole in the amplitude. One can see this for example from observing that the sine function in the third line of \eqref{eq:CLS intersection expression} when rewritten in terms of exponentials becomes $\ex^{2\pi i b\sum_i m_i p_i}$ and thus only diagonal translations in the space of colors are multiplied by the sum of momenta.

We will be slightly more general than it is needed to extract the $c=1$ amplitudes and extract the residue for
\be
\sum_{j=1}^n p_j=\frac{b-b^{-1}}{2}(2g-2+n)+b^{-1}r \label{eq:shifted anomalous momentum conservation}
\ee
with $r \in \ZZ_{\ge 0}$. We can see that this generalization is trivial in \eqref{eq:CLS intersection expression}, since the expression is invariant under $p_j \to p_j+b^{-1}$, except for the external factor $\ex^{-p_j^2 \psi_j}$, which does not involve any of the colors.

To obtain a geometric series, we only have to keep terms that are not polynomially suppressed in $m$. The second term in the parenthesis of the second line in \eqref{eq:CLS intersection expression} becomes $(m_\bullet+m_\circ+2m)^{-2d-2}$ upon performing this replacement. It thus will not lead to poles and we can drop it. Similar to the three-point function, we can also replace $\sin(\pi (m_v+m) b^2)$ by $\frac{i}{2}\ex^{-\pi i (m+m_v)b^2}$. Thus the vertex factor becomes
\be
\prod_{v \in \mathcal{V}_\Gamma} \left(\frac{b (-1)^{m_v+m}}{\sqrt{2}\sin(\pi (m_v+m) b^2)}\right)^{2g_v-2+n_v} \rightsquigarrow \prod_{v \in \mathcal{V}_\Gamma} \left(-\sqrt{2}bi\, \ex^{\pi i (m+m_v)(b^2-1)}\right)^{2g_v-2+n_v}\,.
\ee
We convert the trigonometric factors in the last line of \eqref{eq:CLS intersection expression} to exponentials and only keep the exponential term that leads to the correct pole, namely
\be
\sqrt{2}\sin(2\pi (m_i+m) b p_i) \rightsquigarrow \frac{i}{\sqrt{2}} \, \ex^{-2\pi i (m_i+m) b p_i}\,.
\ee
One can then perform the sum over $m$, which leads to the sought-for pole. We obtain
\begin{align}
    \Res_{\sum_i p_i=(\frac{b}{2}-\frac{1}{2b})(2g-2+n)+\frac{r}{b}} \!\!\!\mathsf{A}^{(b)}_{g,n}(\boldsymbol{p})&=\frac{ b^{n-1}(-2b^2)^{g-1}}{2\pi i} \!\!\!\!\sum_{\Gamma \in \mathcal{G}_{g,n}^\infty/\ZZ}\frac{1}{|\text{Aut}(\Gamma)|}\prod_{v \in \mathcal{V}_\Gamma} \ex^{\pi i m_v(2g_v-2+n_v)(b^2-1)}\nonumber\\
    &\qquad\times\int_{\bM_\Gamma} \prod_{v \in \mathcal{V}_\Gamma}\ex^{\frac{b^2+b^{-2}}{4} \kappa_1-\sum_k\frac{B_{2k}\kappa_{2k}}{(2k)(2k)!}}\prod_{i=1}^n \ex^{-p_i^2 \psi_i-2\pi i b m_i p_i} \nonumber\\
    &\qquad\times\!\prod_{(\bullet,\circ) \in \mathcal{E}_\Gamma} \sum_{d=0}^\infty \frac{\Gamma(d+\tfrac{3}{2})(\psi_\bullet+\psi_\circ)^d\delta_{m_\bullet\ne m_\circ}}{\sqrt{\pi}(\pi b)^{2d+2}(m_\bullet-m_\circ)^{2d+2}} \,. \label{eq:position space Feynman rules b ne 1}
\end{align}
Here, we are summing over colored stable graphs, where we identify graphs related by overall color shifts, i.e.\ $m_v \to m_v+1$ for all vertices simultaneously. This can be viewed as a translation invariance dual to momentum conservation. Perhaps interestingly, one might have expected that we land on a continuous translation symmetry and not a discrete one. We will have more to say about this below.

At this point it is not necessary that the colors are chosen as positive integers, since we can always take an equivalence class in which they are positive. Thus we think of the vertex colors $m_v \in \ZZ$. As remarked above, one can in principle keep $b$ general in the following, but for the purpose of studying $c=1$ amplitudes, we will now specialize to $b=1$. After taking the residue, this is possible in this expression without problems, provided we choose $p_i \in \RR$, so that the infinite sum over colors converges.

We will also multiply by the factor $(-\frac{1}{2})^{g-1}$. This corresponds to a renormalization of the string coupling. The minus sign is perhaps puzzling. It originates from the fact that an imaginary string coupling was chosen in \cite{Collier:2024kwt}, which was motivated from certain consistency conditions that are absent in the $c=1$ string. We thus want to revert back to a real string coupling. We also remove the prefactor $\frac{1}{2\pi i}$, which we absorb into the momentum-conserving delta-function. This leads to a closed form expression of the $c=1$ string amplitudes:
\begin{multline}
    \hat{\mathsf{s}}_{g,n}(\boldsymbol{p})=\sum_{\Gamma \in \mathcal{G}_{g,n}^\infty/\ZZ}\frac{1}{|\text{Aut}(\Gamma)|}\int_{\bM_\Gamma} \prod_{v \in \mathcal{V}_\Gamma}\ex^{-\sum_k\frac{B_{k}\kappa_{k}}{k \, k!}}\prod_{i=1}^n \ex^{-p_i^2 \psi_i-2\pi i m_i p_i} \\
    \times\!\prod_{(\bullet,\circ) \in \mathcal{E}_\Gamma} \sum_{d=0}^\infty \frac{\Gamma(d+\tfrac{3}{2})(\psi_\bullet+\psi_\circ)^d\delta_{m_\bullet\ne m_\circ}}{\pi^{2d+\frac{5}{2}}(m_\bullet-m_\circ)^{2d+2}} \,. \label{eq:c=1 position space intersection numbers}
\end{multline}
It will turn out that these are not quite the conventional $c=1$ string amplitudes, which is why we use lowercase letters to distinguish them. The momentum is naturally only conserved mod 1 in this expression, originating from the parameter $r$ in \eqref{eq:shifted anomalous momentum conservation}.
We took advantage of the fact that $B_1=-\frac{1}{2}$ to rewrite the sum over kappa-classes as a sum over all Bernoulli numbers.
This can be viewed as the position space Feynman rules for the $c=1$ string. Here $m_i$ plays the role of the (discretized) position, the vertex factor is $\exp\big(-\sum_{k \ge 1} \frac{B_k \kappa_k}{k\, k!}\big)$, the leg factors are $\exp\big(-p_i^2 \psi_i-2\pi i m_i p_i\big)$ and the propagator is given in the second line of \eqref{eq:c=1 position space intersection numbers}. All these three ingredients are still of cohomological nature and need to be integrated over the appropriate moduli space, reflecting the fact that this should be viewed as a gravitational theory.

\paragraph{From position space to momentum space.} The corresponding momentum space Feynman rules express the amplitude in a much more economical and transparent way. We shall now explain their derivation. For this, we only have to assume that momentum is conserved mod 1. One can do this graph by graph and we thus focus on a given stable graph $\Gamma$. In order to talk about momenta, we have to pick an orientation on the edges of the graph. We pick the orientations of the external legs as ingoing. At every vertex, we require momentum conservation to hold mod 1. This leaves over a number $L_\Gamma$ of loop momenta, just like in ordinary Feynman diagrams, which naturally take values in the circle $\RR/\ZZ$.

\begin{figure}[ht]
\centering
\begin{tikzpicture}[baseline={([yshift=10pt]current bounding box.center)}]

        \begin{scope}[shift={(-4,0)}]
        \draw[ultra thick, ->] (11.5,1) to (12.5,1);

        \draw[fill=bleudefrance, draw=bleudefrance, opacity=.2] (6,2.5) to[out=0, in = 180] (8.25,1) to [out=180, in =0] (6,-.5) to [out=20, in = -20, looseness=.95] (6,0.5) to[out= 0, in = 0] (6,1.5) to [out=20, in = -20, looseness=.95] (6,2.5);

        \draw[fill=candyapplered, draw=candyapplered, opacity=.2] (10.5,2.5) to[out=180, in = 0] (8.25,1) to[out=0, in = 180] (10.5,-.5) to[out=160, in = 200, looseness=.95] (10.5,0.5) to[out=180, in = 180] (10.5,1.5) to[out=160, in = 200, looseness=.95] (10.5,2.5);

        \draw[very thick] (6,0) ellipse (0.25 and .5);
        \draw[very thick] (6,2) ellipse (0.25 and .5);

        \draw[very thick] (10.5,0) ellipse (0.25 and .5);

        \draw[very thick] (10.5,2) ellipse (0.25 and .5);

        \draw[very thick] (6,.5) to[out=0, in =0] (6,1.5);
        \draw[very thick] (10.5,.5) to[out=180, in =180] (10.5,1.5);
        \draw[very thick] (6,2.5) to[out=0, in = 180] (8.25,1) to[out=0, in=180] (10.5,2.5);
        \draw[very thick] (6,-.5) to[out=0, in = 180] (8.25,1) to[out=0, in = 180] (10.5,-.5);

        \node[scale=1., inner sep = 1.5pt] (m1right) at (6.9,1) {$m_{\bullet e}$};
        \node[scale=1., inner sep = 1.5pt] (m2right) at (9.6,1) {$m_{e\circ}$};
        \fill (8.25,1.) ellipse (0.1 and 0.1);

        \end{scope}

        \begin{scope}[shift={(3.5,0)}]

        \draw[very thick] (6,0) ellipse (0.25 and .5);
        \draw[very thick] (6,2) ellipse (0.25 and .5);
        \draw[very thick] (11.5,0) ellipse (0.25 and .5);
        \draw[very thick] (11.5,2) ellipse (0.25 and .5);

        \draw[very thick] (6,.5) to[out=0, in =0] (6,1.5);
        \draw[very thick] (11.5,.5) to[out=180, in =180] (11.5,1.5);

        \draw[very thick] (6,2.5) to[out=0, in = 180] (8.25,1.35) to[out=0, in=180] (9.25,1.35) to[out=0, in=180] (11.5,2.5);
        \draw[very thick] (6,-.5) to[out=0, in = 180] (8.25,0.65) to[out=0, in=180] (9.25,0.65) to[out=0, in = 180] (11.5,-.5);

        \draw[fill=green, draw=green, opacity=.2] (8.0,1.35) to[out=0, in = 180] (9.5,1.35) to[out=-40, in=40] (9.5,0.65) to[out=0, in=180] (8.0,0.65) to[out=130, in=-130] (8.0,1.35);

        \draw[very thick, bend right=40] (8.0,1.375) to (8.0,0.625);
        \draw[very thick, bend left=40, dashed] (8.0,1.375) to (8.0,0.625);
        \draw[very thick, bend right=40] (9.5,1.375) to (9.5,0.625);
        \draw[very thick, bend left=40, dashed] (9.5,1.375) to (9.5,0.625);

        \node[scale=1., inner sep = 1.5pt] (m1right) at (8.75,1) {$m_e$};

        \end{scope}

    \end{tikzpicture}
\caption{Passage from vertex colors $m_v$ to edge variables $m_e = m_{e\circ} - m_{\bullet e}$. The left panel shows two vertices connected by an edge, with vertex colors $m_{\bullet e}$ and $m_{e\circ}$. The right panel shows the equivalent description in terms of the edge variable $m_e$, which can be thought of as a discrete difference of positions.}
    \label{fig:colorstoedges}
\end{figure}
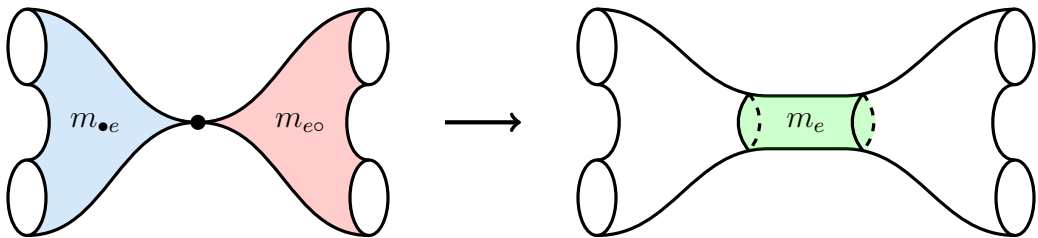

Consider now an edge $e=(\bullet,\circ)$. We will adopt the convention that $\bullet e$ denotes the source and $e\circ$ the target of the edge. We similarly denote by $i \circ$ the vertex to which an external leg $i$ attaches. We now want to think about position differences and define
\be
m_e=m_{e\circ}-m_{\bullet e} \label{eq:me definition}
\ee
instead of the absolute positions $m_v$, see Figure~\ref{fig:colorstoedges}. Since the graph is connected, we can clearly recover the set of $m_v$'s up to overall translation. However, the data is not bijective because for a loop in $\Gamma$, we have $\sum_{e \in \text{loop}} m_e=0$ (assuming that all edges are oriented in the same way).
We also have to account for the leg factors $\ex^{-2\pi i m_i p_i}$ present in \eqref{eq:c=1 position space intersection numbers}. It nicely combines with our previous remark that we want to impose the constraint $\sum_{e \in \text{loop}} m_e=0$ for all loops as follows. Let us compute $\sum_{e \in \Gamma} m_e p_e$ without assuming that the loop constraint is satisfied.
For this, we take a set of edges $\mathcal{L} \subset \mathcal{E}$, that give a basis of the loop momenta $q_1,\dots,q_L$. If we remove those, we end up with a spanning tree, which we denote by $\mathcal{T}$. Its vertex set is still $\mathcal{V}$ and its edge set is $\mathcal{E} \setminus \mathcal{L}$. For the tree $\mathcal{T}$, we can then invert the definition \eqref{eq:me definition}, up to an overall shift. Thus we write
\begin{align}
    \sum_{e \in \mathcal{E}}m_e p_e&=\sum_{e \in \mathcal{L}} m_e p_e+ \sum_{e \in \mathcal{T}} (m_{e\circ}-m_{\bullet e}) p_e \label{eq:loop constraint line 1}\\
    &=\sum_{e \in \mathcal{L}} m_e p_e+ \sum_{v \in \mathcal{V}} m_v \bigg[ \sum_{e \in \mathcal{T},\, e\circ=v} p_e-\sum_{e \in \mathcal{T},\, \bullet e=v} p_e \bigg] \label{eq:loop constraint line 2}\\
    &=\sum_{e \in \mathcal{L}} m_e p_e+ \sum_{v \in \mathcal{V}} m_v \bigg[-\sum_{e \in \mathcal{L},\,e\circ=v}p_e+\sum_{e \in \mathcal{L},\, \bullet e=v}p_e- \sum_{i,\, i\circ=v}  p_i\bigg]  \bmod 1\label{eq:loop constraint line 3}\\
    &=- \sum_i m_i p_i+\sum_{e \in \mathcal{L}} p_e (m_e-m_{e\circ}+m_{\bullet e}) \bmod 1\,. \label{eq:loop constraint line 4}
\end{align}
In \eqref{eq:loop constraint line 1}, we divided the sum into the spanning tree and the basis of loop momenta and used the definition \eqref{eq:me definition}. In \eqref{eq:loop constraint line 2}, we rewrote the sum over the edges of the spanning tree as the sum over vertices. In \eqref{eq:loop constraint line 3}, we used momentum conservation at every vertex mod 1. Finally, in \eqref{eq:loop constraint line 4}, we reorganized the sum as a sum over the loop edges and the external half-edges.
Notice now that $m_e-m_{e\circ}+m_{\bullet e}=0$ is precisely the loop constraint since $m_{e\circ}$ and $m_{\bullet e}$ were obtained by requiring \eqref{eq:me definition} to hold in $\mathcal{T}$.

Thus we can insert the expression $\exp(2\pi i \sum_e m_e p_e)$ and not assume the loop constraint. Integrating out the loop momenta over $\RR/\ZZ$ (one may take the unit interval as a fundamental domain) then imposes the loop constraints and inserts the leg factors. Thus we can rewrite \eqref{eq:c=1 position space intersection numbers} as
\begin{multline}
    \hat{\mathsf{s}}_{g,n}(\boldsymbol{p})=\sum_{\Gamma \in \mathcal{G}_{g,n}}\frac{1}{|\text{Aut}(\Gamma)|}\int_{\RR/\ZZ} \d^{L_\Gamma} \boldsymbol{q}\int_{\bM_\Gamma} \prod_{v \in \mathcal{V}_\Gamma}\ex^{-\sum_k\frac{B_{k}\kappa_{k}}{k \, k!}}\prod_{i=1}^n \ex^{-p_i^2 \psi_i} \\
    \times\!\prod_{e=(\bullet,\circ) \in \mathcal{E}_\Gamma} \sum_{m_e \in \ZZ \setminus \{0\}}\sum_{d=0}^\infty \frac{\Gamma(d+\tfrac{3}{2})(\psi_\bullet+\psi_\circ)^d\ex^{2\pi i m_e p_e}}{\pi^{2d+\frac{5}{2}}m_e^{2d+2}}\,.
\end{multline}
This is a big simplification because we can now perform the sum over $m_e$ in closed form. We have
\be
\sum_{m \ne 0} \frac{\ex^{2\pi i mp}}{m^{2d+2}}=-\frac{(2\pi i)^{2d+2}}{(2d+2)!}\, B_{2d+2}(\{p\})\,, \label{eq:Bernoulli polynomial identity}
\ee
where $B_{2d+2}(x)$ denotes the Bernoulli polynomial and $\{p\}=p-\lfloor p \rfloor$ denotes the fractional part of $p$.
One can prove this by employing the Sommerfeld-Watson trick as follows. First, the answer is clearly periodic in $p$ with period 1. Assuming that $p \in [0,1]$, we can write
\be
\sum_{m \ne 0} \frac{\ex^{2\pi i mp}}{m^{2d+2}}= \oint_\gamma \d z\, \frac{\ex^{2\pi i z p}}{(\ex^{2\pi i z}-1)z^{2d+2}}\,, \label{eq:sum over m Bernoulli polynomial}
\ee
where $\gamma$ encircles all non-zero integers counterclockwise. We can open the contour. Because we chose $p \in [0,1]$, the asymptotic contribution for $|z| \to \infty$ vanishes and we simply pick up the residue at $z=0$. Via the definition of Bernoulli polynomials through their generating function, one concludes \eqref{eq:Bernoulli polynomial identity}. We obtain the momentum space Feynman rules
\begin{multline}
    \hat{\mathsf{s}}_{g,n}(\boldsymbol{p})=\sum_{\Gamma \in \mathcal{G}_{g,n}}\frac{1}{|\text{Aut}(\Gamma)|}\int_{\RR/\ZZ} \d^{L_\Gamma} \boldsymbol{q}\int_{\bM_\Gamma} \prod_{v \in \mathcal{V}_\Gamma}\ex^{-\sum_k\frac{B_{k}\kappa_{k}}{k \, k!}}\prod_{i=1}^n \ex^{-p_i^2 \psi_i} \\
    \times\!\prod_{e=(\bullet,\circ) \in \mathcal{E}_\Gamma} \sum_{d=0}^\infty \frac{B_{2d+2}(\{p_e\})(-\psi_\bullet-\psi_\circ)^d}{(d+1)!}\,. \label{eq:c=1 momentum space intersection numbers}
\end{multline}
The second line now corresponds to the momentum space propagator and the formula includes $L_\Gamma$ loop integrals. This expression holds when momentum conservation holds mod 1 and is a discretized version of the $c=1$ string amplitudes as we shall explain momentarily. Higher-genus amplitudes force us to integrate the edge factors involving Bernoulli polynomials depending on the fractional part of the edge momenta, which can be cumbersome to carry out directly. In appendix \ref{app:Bernoulli integration} we explain how to efficiently carry out these loop integrals, see (\ref{eq:loop momentum integral result}). A structurally similar formula to \eqref{eq:c=1 momentum space intersection numbers} was found independently in \cite{Artemev:2025pvk} for the $(2,p)$ minimal string. 
\paragraph{Simplification.}  One can further rewrite the edge factor in the Feynman rules \eqref{eq:c=1 momentum space intersection numbers} as follows. Similar manipulations were also applied in \cite{Artemev:2025pvk}. We claim that
\be 
\sum_{d=0}^\infty \frac{B_{2d+2}(\{p\}) x^d}{(d+1)!}=-\sum_{\ell \in \ZZ} |\ell+p|^{-s} \, \mathrm{e}^{x(\ell+p)^2}\bigg|_{s=-1}\ . \label{eq:edge factor rewriting asymptotic series}
\ee
Since we want to put $x=-\psi_\bullet-\psi_\circ$ and the moduli space dimension is finite, we treat the LHS as a formal power series in $x$. In particular, the series has vanishing radius of convergence. The meaning of the RHS is as follows. For every term in the formal power series expansion in $x$, the sum over $\ell$ converges for large enough $\Re s$. We define the RHS as the analytic continuation to $s=-1$. This amounts to zeta function regularization of the naive equality with $s=-1$.

Deriving \eqref{eq:edge factor rewriting asymptotic series} is simple. Since both the LHS and the RHS are periodic mod 1, we assume also $p \in [0,1)$. Let us extract the coefficient of $x^d$ on the RHS, which is
\begin{align} 
\frac{1}{d!}\sum_{\ell \in \ZZ} |\ell+p|^{2d-s}\big|_{s=-1}&=\frac{1}{d!}\sum_{\ell=0}^\infty |\ell+p|^{2d-s}+\frac{1}{d!}\sum_{\ell=0}^\infty |\ell+1-p|^{2d-s}\big|_{s=-1} \\
&=\frac{1}{d!}\big(\zeta_\text{H}(-2d-1,p)+\zeta_\text{H}(-2d-1,1-p)\big)\\
&=-\frac{B_{2d+2}(p)}{(d+1)!}\ .
\end{align}
Here $\zeta_\text{H}(s,a)=\sum_{\ell=0}^\infty (\ell+a)^{-s}$ is the Hurwitz zeta function, which defines the analytic continuation away from $\Re s>2d+1$. It satisfies $\zeta_\text{H}(-n,a)=-\frac{B_{n+1}(a)}{n+1}$ for $n \in \ZZ_{\ge 0}$. The last step follows from the symmetry property $B_{2d+2}(p)=B_{2d+2}(1-p)$. Therefore, \eqref{eq:edge factor rewriting asymptotic series} follows.

One can insert \eqref{eq:edge factor rewriting asymptotic series} into the Feynman rules \eqref{eq:c=1 momentum space intersection numbers} with $x=-\psi_\bullet-\psi_\circ$. The integrals over $\bM_\Gamma$ can be done in terms of the quantum volumes of the Virasoro minimal string, defined as\footnote{In \cite{Collier:2023cyw}, this was written in terms of the momenta $P_j=ip_j$. As in \cite{Collier:2024lys}, it will be more convenient for us to use the Wick rotated momenta.}
\be  
\mathsf{V}_{g,n}^{(b)}(\boldsymbol{p})=\int_{\bM_{g,n}} \mathrm{e}^{\frac{b^2+b^{-2}}{4} \kappa_1-\sum_{j=1}^n p_j^2 \psi_j-\sum_{k\ge 1} \frac{B_{2k}\kappa_{2k}}{2k(2k)!}} \, .\label{eq:quantum volume definition}
\ee
This leads to
\begin{tcolorbox}[
  before skip=6pt,
  after skip=6pt,
]
\begingroup
\setlength{\abovedisplayskip}{0pt}
\setlength{\abovedisplayshortskip}{0pt}
\setlength{\belowdisplayskip}{0pt}
\setlength{\belowdisplayshortskip}{0pt}
\be 
\hat{\mathsf{s}}_{g,n}(\boldsymbol{p})=\sum_{\Gamma \in \mathcal{G}_{g,n}} \frac{(-1)^{|\mathcal{E}_\Gamma|}}{|\Aut(\Gamma)|} \int_{\RR/\ZZ} \d^{L_\Gamma} q\, \prod_{e \in \mathcal{E}_\Gamma} \sum_{\ell_e \in \ZZ} |p_e+\ell_e|^{-s} \prod_{v \in \mathcal{V}_\Gamma} \mathsf{V}^{(1)}_{g_v,n_v}(\boldsymbol{p}_v+\boldsymbol{\ell}_v)\bigg|_{s=-1}\ . \label{eq:Feynman rules integer shifts}
\ee
\endgroup
\end{tcolorbox}
\noindent
Let us unpack the notation. For each internal edge, we associate an integer $\ell_e$ and sum over all possible choices. $p_e$ is the momentum associated to an internal edge, which is determined mod 1 by a choice of the loop momenta as well as by imposing momentum conservation mod 1 at every vertex. $\boldsymbol{p}_v+\boldsymbol{\ell}_v$ is the set of momenta emanating from the vertex $v$, shifted by the integers associated to the corresponding edges. By definition, $\ell=0$ for an external leg. As we shall discuss below, the edge factor $-|p+\ell|$ after analytic continuation is the inverse of the two-point function \eqref{eq:worldsheet two-point function} and thus provides the natural propagator.
\subsection{Remarks} \label{subsec:remarks}
A few remarks about \eqref{eq:c=1 momentum space intersection numbers} are in order. We should remark that while \eqref{eq:Feynman rules integer shifts} is a conceptually nice formula, it is computationally simpler to consider \eqref{eq:c=1 momentum space intersection numbers}. Thus we will often work with that version in the following.

\paragraph{Discrete target space.} It is at first glance surprising that the loop momenta are restricted to the region $q \in \RR/\ZZ \cong [0,1)$. This feature has many different avatars in these formulas, such as the edge momenta appearing through their fractional parts in \eqref{eq:c=1 momentum space intersection numbers} and the positions $m_v$ taking only integer values in \eqref{eq:c=1 position space intersection numbers}. Thus, the formulas essentially look like a discrete version of target space, where a natural UV cutoff is built into the loop momenta, which renders all loop integral UV-finite.

In fact, we saw that this formula naturally also holds when momentum conservation is only satisfied mod 1. We thus define the full discrete amplitude to also include a Dirac comb imposing this, see eq.~\eqref{eq:full s Dirac comb} in the introduction.

As an analogy, these formulas look like phonon scattering in a 1D lattice, where the timelike boson is Wick rotated to a spacelike boson and discretized. 
Specifically, the Feynman rules describe the interactions of infinitely many phonons labelled by the integer $\ell$ whose momentum takes values in $p \in \RR/\ZZ$. The vertex factors are simply $\mathsf{V}_{g,n}^{(1)}(\boldsymbol{p}+\boldsymbol{\ell})$. 
The discretization of target space is invisible as long as we take the external momenta small enough so that the sums of momenta are only allowed to explore the first Brillouin zone of the lattice. This condition coincides with the condition that the initial moduli space integral defining the $\mathbb{C}$LS and $c=1$ amplitudes (for purely imaginary external closed string energies, or real-valued $p_j$) is absolutely convergent. Concretely, the momenta are in the first Brillouin zone when $|p_I| < 1$ for all subsets $I \subset [n]$. Thus, we should only compare \eqref{eq:c=1 momentum space intersection numbers} and \eqref{eq:Feynman rules integer shifts} with the $c=1$ amplitudes in the first Brillouin zone. In fact, \eqref{eq:c=1 momentum space intersection numbers} features non-analyticities that are not present in the worldsheet integrals. Thus the correct prescription to recover the conventional $c=1$ amplitudes is to evaluate \eqref{eq:c=1 momentum space intersection numbers} or \eqref{eq:Feynman rules integer shifts} for small momenta and then analytically continue it to the desired momentum.

In order to distinguish the two prescriptions, we will denote by $\mathsf{S}_{g,n}(\boldsymbol{\omega})$, the continuous $c=1$ amplitudes, i.e.\ the result of restricting \eqref{eq:c=1 momentum space intersection numbers} to the first Brillouin zone and Wick rotating the momenta, while we use lowercase letters for the discrete amplitudes as in \eqref{eq:c=1 momentum space intersection numbers}.

For tree-level amplitudes it is simple to give a closed form expression also for the continuous amplitudes $\mathsf{S}_{0,n}$. In that case, we do not have loop integrals and the edge momenta $p_e$ are fully determined through the external momenta. Restricting to the first Brillouin zone means that we assume that $|p_e| < 1$ for all edges $e$. In that case, we can replace $B_{2d+2}(\{p_e\})$ with $B_{2d+2}(\sqrt{p_e^2})$ (using that for $p_e \in (-1,0]$, $B_{2d+2}(\{p_e\})=B_{2d+2}(p_e+1)=B_{2d+2}(-p_e)$). We write $\sqrt{p_e^2}$ instead of $|p_e|$ to emphasize the result is piecewise analytic. Thus we find that
\begin{multline}
    \mathsf{S}_{0,n}(\boldsymbol{\omega})=\sum_{\Gamma \in \mathcal{G}_{0,n}}\frac{1}{|\text{Aut}(\Gamma)|}\int_{\bM_\Gamma} \prod_{v \in \mathcal{V}_\Gamma}\ex^{-\sum_k\frac{B_{k}\kappa_{k}}{k \, k!}}\prod_{i=1}^n \ex^{\frac{1}{4}\omega_i^2 \psi_i} \\
    \times\!\prod_{e=(\bullet,\circ) \in \mathcal{E}_\Gamma} \sum_{d=0}^\infty \frac{B_{2d+2}(\frac{1}{2}\sqrt{-\omega_e^2})(-\psi_\bullet-\psi_\circ)^d}{(d+1)!}\,. \label{eq:c=1 continuous version intersection numbers genus 0}
\end{multline}
The choice of sign for $\sqrt{-\omega_e^2}$ is dictated by causality as we will discuss next.
Thus the actual amplitude is a multi-valued function of the momenta $\omega_i$, although the different sheets are not directly related by analytic continuation. Thus crossing symmetry doesn't hold in this simple sense for $c=1$ strings as was observed previously \cite{Balthazar:2017mxh}.
However, once a physical process is specified, i.e.\ a choice of sign for each external energy $\omega_i$ and each intermediate energy $\omega_I=\sum_{i \in I} \omega_i$ flowing through a cut (labelled by $I \subset [n]$ as in section~\ref{subsec:perturbative unitarity}), the $i\varepsilon$ prescription selects a definite sheet. Since each sheet is polynomial in the momenta, the analytic continuation from the convergence region to physical kinematics is unique.
It appears complicated to analytically continue the higher genus intersection number in closed form in the same way.

\paragraph{Physical sheet and $i \varepsilon$-prescription.} Let us now discuss how these amplitudes can be analytically continued to the physical sheet, i.e.\ to purely imaginary $p=\frac{i \omega}{2}$. Clearly, there is a choice related to whether we approach the imaginary axis from the left or from the right. Which one is the right one is determined by the $i \varepsilon$ prescription \cite{Witten:2013pra}, see also \cite{Eberhardt:2022zay, Eberhardt:2023xck, Eberhardt:2024twy, Figueiredo:2025fnr, Baccianti:2025whd} for the recent implementation in 10D superstring theory.

Let us first explain the procedure from the worldsheet. For real choices of $\omega_i$, we can make sense of the moduli space integral by deforming the integration contour into the complexification of moduli space, $\mathcal{M}_{g,n}^\CC$. Consider what happens near a degeneration of moduli space. There is a plumbing parameter $q$ describing the deformation away from this degeneration. The local behavior of the worldsheet integral is
\be
|q|^{2 p_e^2-2} (-\log |q|)^{-\frac{3}{2}}\,,
\ee
where $p_e$ is the momentum flowing through the tube. This follows from the OPE expansion. The term $(-\log |q|)^{-\frac{3}{2}}$ is produced from the fact that the OPE integral in Liouville theory behaves as $\int_0^\infty \d P\, P^2 \, |q|^{2P^2+\Delta}$ for small $q$ (with fixed $\Delta$ that cancels once the mass shell condition is imposed). This can be evaluated via saddle point approximation in the small-$q$ limit, see e.g.\ \cite{Ribault:2015sxa}. Clearly, the integral is only convergent provided that $\Re p_e^2 \ge 0$, which is violated on the physical spectrum where all momenta are purely imaginary.
It is convergent when all momenta are real, which is what we exploited in our analysis so far.

To define the physical integral for imaginary choices of $p_e$, we deform the integral over $|q|=\ex^{-t}$ into the complex plane. Here, $t$ can be understood as the proper time on the worldline that emerges in the degeneration of the worldsheet. The original contour of integration in $t$ around the neighborhood of the divisor is
\be
\int_{t_*}^\infty \d t\ t^{-\frac{3}{2}}\,\ex^{-2 t p_e^2}\,.
\ee
Since $t$ is the proper time on the worldsheet, we should Wick rotate it to Lorentzian time, which gives the integral
\be
\int_{t_*}^{t_*+i\infty} \d t\ t^{-\frac{3}{2}}\,\ex^{-2 t p_e^2}\,,
\ee
i.e.\ a contour that turns vertically into the complex plane. This contour is oscillatory, but does converge absolutely for purely imaginary choices of $p_e$ and defines the physical amplitude. It is not the same as the integral from $t_*$ to $t_*-i \infty$, which would correspond to the opposite sign of the $i \varepsilon$-prescription.

Since $t$ is rotated counterclockwise into the complex plane, it follows that $p_e$ is rotated clockwise into the complex plane.
For physical momenta, we should thus choose the branch choice of $\sqrt{p_e^2}$ obtained by rotating $p_e$ back, i.e.\ 
\be
\sqrt{-\omega_e^2}=2\sqrt{p_e^2}:=2\ex^{-\frac{\pi i}{2}} \sqrt{\ex^{\pi i} p_e^2}=-2p_e\,  \sgn(\Im p_e)=-i \omega_e \sgn(\Re \omega_e)\,,\label{eq:branch choice}
\ee
where we expressed the result in terms of $\omega$ using \eqref{eq:p omega relation}.
Thus, not surprisingly, the branch is determined by whether the intermediate particle has positive or negative energy.

\paragraph{Polynomial boundedness.} Another surprising property that is manifest in \eqref{eq:c=1 momentum space intersection numbers} is that each sheet of the amplitude is a polynomial in the external momenta. In particular, every sheet of the amplitude is a polynomial of degree at most $2 \dim_\CC \bM_{g,n}=6g-6+2n$ in the external momenta.
This is very similar to what happens in the minimal string \cite{Belavin:2006ex} or the Virasoro minimal string \cite{Collier:2023cyw}, but highly surprising from the worldsheet perspective, where the momentum dependence is much more involved at the level of the integrand. The amplitudes of the $\mathbb{C}$LS also grow only polynomially, but have a number of poles. $\mathbb{C}$LS amplitudes do not have sequential poles (i.e.\ the residues do not have further poles in other parameters). Thus also $c=1$ amplitudes are polynomially bounded and don't have poles, which implies that they are in fact polynomial on each sheet.

We shall see below that the degree of the polynomial is in fact lower and is only $4g-3+n$ (although we only know how to prove this for $g=0$). This indicates that \eqref{eq:Feynman rules integer shifts} has a large amount of cancellation built in.

\paragraph{Self-loop vanishing.} Consider a graph with a self-loop in the expansion \eqref{eq:c=1 momentum space intersection numbers}. Then we have a loop momentum $q$ associated to the self-loop. None of the other edge momenta depends on $q$. We have
\be
\int_{\RR/\ZZ} \d q\, B_{2d+2}(\{q\})=0\,,
\ee
and thus the contribution vanishes. This integral can be evaluated easily by appealing to \eqref{eq:Bernoulli polynomial identity}. Thus we can omit graphs with self-loops in \eqref{eq:c=1 momentum space intersection numbers}.

\paragraph{Dilaton equation.} $c=1$ string theory amplitudes satisfy a dilaton equation. It is inherited from its counterpart in $\CC$LS. It originates from the fact that the zero-momentum vertex operator is the identity operator in the free boson theory and the marginal operator in the Liouville theory. Thus inserting it shifts the string coupling and therefore modifies the amplitude in a universal way. See \cite[section 3.2]{Collier:2024kwt} for its worldsheet derivation, which is completely analogous in the case of the $c=1$ string.\footnote{More generally, one may consider arbitrary insertions of marginal operators in Liouville CFT. 
In the context of the $c=1$ string, this strategy led to the so-called resonance computation of tree-level scattering amplitudes in \cite{DiFrancesco:1991ocm,DiFrancesco:1991daf}, where the Liouville correlator at special (resonant) kinematics reduces to that of a free linear dilaton theory. This predates the discovery of the Liouville DOZZ formula.}
We have
\be
\hat{\mathsf{S}}_{g,n+1}(\omega_1,\dots,\omega_n,\omega_{n+1}=0)=\bigg[2g-2+n-\frac{1}{2}\sum_{i=1}^n \sqrt{-\omega_i^2}\bigg]\hat{\mathsf{S}}_{g,n}(\omega_1,\dots,\omega_n)\,. \label{eq:dilaton equation}
\ee
This property can also be proven directly from the intersection number expression \eqref{eq:c=1 momentum space intersection numbers} by relating it to topological recursion.

\paragraph{Sphere two-point function.} As usual, extra care has to be taken when computing the worldsheet two-point function. It can be obtained by assuming that the dilaton equation \eqref{eq:dilaton equation} extends to the two-point function, which gives
\be
\hat{\mathsf{S}}_{0,2}(\omega_1,\omega_2)=-\frac{1}{\sqrt{-\omega_1^2}}\,. \label{eq:worldsheet two-point function}
\ee
For a direct worldsheet calculation, one can use the prescription put forward in \cite{Maldacena:2001km, Erbin:2019uiz}. The worldsheet correlator contains two delta functions, one from Liouville theory and one from the free boson theory. The prescription tells us that the effect of dividing by the M\"obius little group on the worldsheet is to remove one delta function in the worldsheet conformal weight. This leads to \eqref{eq:worldsheet two-point function}, up to an overall constant normalization. The steps are exactly parallel to \cite[section 2.5]{Collier:2024kwt}.

\paragraph{Full $S$-matrix from the worldsheet.} The worldsheet amplitudes $\mathsf{S}_{g,n}$ are the perturbative contributions to the connected part of the S-matrix. We get the full S-matrix by also including disconnected contributions, i.e.
\begin{subequations}
\begin{align}
\mathbb{S}^{1 \to 1}&=\sum_{g=0}^\infty g_\text{s}^{2g}\begin{tikzpicture}[baseline={([yshift=-.5ex]current bounding box.center)},scale=.6]
    \node[shape=circle,draw=black, very thick, fill=gray, inner sep=.5pt] (A) at (0,0) {$\mathsf{S}_{g,2}$};
    \draw[very thick] (A) to (1.5,0) node[right] {2};
    \draw[very thick] (A) to (-1.5,0) node[left] {1};
\end{tikzpicture} \,, \\
\mathbb{S}^{1 \to 2}&=\sum_{g=0}^\infty g_\text{s}^{2g+1}\begin{tikzpicture}[baseline={([yshift=-.5ex]current bounding box.center)},scale=.6]
    \node[shape=circle,draw=black, very thick, fill=gray, inner sep=.5pt] (A) at (0,0) {$\mathsf{S}_{g,3}$};
    \draw[very thick] (A) to (1.5,1) node[right] {2};
    \draw[very thick] (A) to (1.5,-1) node[right] {3};
    \draw[very thick] (A) to (-1.5,0) node[left] {1};
\end{tikzpicture}\,, \\
\mathbb{S}^{1 \to 3}&=\sum_{g=0}^\infty g_\text{s}^{2g+2}\begin{tikzpicture}[baseline={([yshift=-.5ex]current bounding box.center)},scale=.6]
    \node[shape=circle,draw=black, very thick, fill=gray, inner sep=.5pt] (A) at (0,0) {$\mathsf{S}_{g,4}$};
    \draw[very thick] (A) to (1.5,1) node[right] {2};
    \draw[very thick] (A) to (1.5,0) node[right] {3};
    \draw[very thick] (A) to (1.5,-1) node[right] {4};
    \draw[very thick] (A) to (-1.5,0) node[left] {1};
\end{tikzpicture}\,, \\
\mathbb{S}^{2 \to 2}&=\sum_{g=0}^\infty g_\text{s}^{2g+2}\!\begin{tikzpicture}[baseline={([yshift=-.5ex]current bounding box.center)},scale=.6]
    \node[shape=circle,draw=black, very thick, fill=gray, inner sep=.5pt] (A) at (0,0) {$\mathsf{S}_{g,4}$};
    \draw[very thick] (A) to (1.5,1) node[right] {3};
    \draw[very thick] (A) to (1.5,-1) node[right] {4};
    \draw[very thick] (A) to (-1.5,1) node[left] {1};
    \draw[very thick] (A) to (-1.5,-1) node[left] {2};
\end{tikzpicture}+ \!\!\sum_{g_1,g_2=0}^\infty \!\!g_\text{s}^{2g_1+2g_2} \!\!\left[\hspace{-.1cm}\begin{tikzpicture}[baseline={([yshift=-.5ex]current bounding box.center)},scale=.6]
    \node[shape=circle,draw=black, very thick, fill=gray, inner sep=.5pt] (A) at (0,1) {$\mathsf{S}_{g_1,2}$};
    \node[shape=circle,draw=black, very thick, fill=gray, inner sep=.5pt] (B) at (0,-1) {$\mathsf{S}_{g_2,2}$};
    \draw[very thick] (A) to (1.5,1) node[right] {3};
    \draw[very thick] (A) to (-1.5,1) node[left] {1};
    \draw[very thick] (B) to (1.5,-1) node[right] {4};
    \draw[very thick] (B) to (-1.5,-1) node[left] {2};
\end{tikzpicture}\hspace{-.2cm}
+\hspace{-.2cm}\begin{tikzpicture}[baseline={([yshift=-.5ex]current bounding box.center)},scale=.6]
    \node[shape=circle,draw=black, very thick, fill=gray, inner sep=.5pt] (A) at (0,1) {$\mathsf{S}_{g_1,2}$};
    \node[shape=circle,draw=black, very thick, fill=gray, inner sep=.5pt] (B) at (0,-1) {$\mathsf{S}_{g_2,2}$};
    \draw[very thick] (A) to (2,-1) node[right] {4};
    \draw[very thick] (A) to (-1.5,1) node[left] {1};
    \fill[white] (1,0) circle (.15);
    \draw[very thick] (B) to (2,1) node[right] {3};
    \draw[very thick] (B) to (-1.5,-1) node[left] {2};
\end{tikzpicture}\hspace{-.1cm}\right]\, ,
\end{align} \label{eq:full S matrix from the worldsheet}%
\end{subequations}
and so forth. We also denote $\begin{tikzpicture}[baseline={([yshift=-.5ex]current bounding box.center)},scale=.6]
    \node[shape=circle,draw=black, very thick, fill=gray, inner sep=.3pt] (A) at (0,0) {\small $\mathsf{S}_{0,2}$};
    \draw[very thick] (A) to (1.5,0);
    \draw[very thick] (A) to (-1.5,0);
\end{tikzpicture}=\begin{tikzpicture}[baseline={([yshift=-.5ex]current bounding box.center)},scale=.6]
    \draw[very thick] (-1.5,0) to (1.5,0);
\end{tikzpicture}$, given by \eqref{eq:worldsheet two-point function}. We are not including $0 \to n$ or $n \to 0$ amplitudes as connected contributions, since they are entirely supported on vanishing momentum. They in particular do not contribute to unitarity cuts.\footnote{In fact, in the normalization of the amplitude where the two-point function is normalized as $\mathbb{S}^{1 \to 1}(\omega_1,\omega_2)=\delta(\omega_1+\omega_2)+\mathcal{O}(g_\text{s}^2)$, they vanish because that normalization differs from the worldsheet normalization by a leg factor of $\sqrt{\omega}$, which vanishes at zero energy.}
For the $n \to n$ amplitude, the leading order is given by the identity for indistinguishable particles, i.e.\
\be
\mathbb{S}^{n \to n}=\underbrace{\begin{tikzpicture}[baseline={([yshift=-.5ex]current bounding box.center)},scale=.6]
    \draw[very thick] (-1.5,1) to (1.5,1);
    \draw[very thick] (-1.5,.5) to (1.5,.5);
    \draw[very thick] (-1.5,0) to (1.5,0);
    \node at (0,-.35) {$\vdots$};
    \draw[very thick] (-1.5,-1) to (1.5,-1);
\end{tikzpicture}\ +\ \begin{tikzpicture}[baseline={([yshift=-.5ex]current bounding box.center)},scale=.6]
    \draw[very thick] (-1.5,1) to (1.5,.5);
    \fill[white] (0,.75) circle (.25);
    \draw[very thick] (-1.5,.5) to (1.5,1);
    \draw[very thick] (-1.5,0) to (1.5,0);
    \node at (0,-.35) {$\vdots$};
    \draw[very thick] (-1.5,-1) to (1.5,-1);
\end{tikzpicture}\ +\ \begin{tikzpicture}[baseline={([yshift=-.5ex]current bounding box.center)},scale=.6]
    \draw[very thick] (-1.5,1) to (1.5,0);
    \fill[white] (-.5,.6667) circle (.25);
    \fill[white] (.5,.3333) circle (.25);
    \draw[very thick] (-1.5,.5) to (1.5,1);
    \draw[very thick] (-1.5,0) to (1.5,.5);
    \node at (0,-.35) {$\vdots$};
    \draw[very thick] (-1.5,-1) to (1.5,-1);
\end{tikzpicture}\ + \ldots }_{\id}\, +\,  \mathcal{O}(g_\text{s}^2)\,. \label{eq:S identity contribution worldsheet}
\ee
It is conventional to also define the $\TT$-matrix, related to the full $\mathbb{S}$-matrix by $\mathbb{S}=\id+i \TT$. Thus, to compute $i\TT$ from the worldsheet, we simply have to remove the free part from the expansion. We will denote perturbative contributions by $\TT^{m \to n}_k$, defined as follows
\be
\TT^{m \to n}=\sum_{\begin{subarray}{c}  k=1\\ k+m+n \in 2 \ZZ \end{subarray}}^\infty g_\text{s}^{k} \TT^{m \to n}_k\,.
\ee

\paragraph{Geometric interpretation.} It is interesting to note that the Feynman rules in the form \eqref{eq:Feynman rules integer shifts} only involve the quantum volumes of the Virasoro minimal string. Let us recall from \cite{Collier:2023cyw} that it has a very natural origin. \eqref{eq:quantum volume definition} can be also written as
\be 
\mathsf{V}_{g,n}^{(b)}(\boldsymbol{p})=\int_{\bM_{g,n}} \text{td}(\mathcal{M}_{g,n})\,\mathrm{e}^{\omega(\boldsymbol{p})} \label{eq:quantum volumes index theorem}
\ee
with $\omega=\frac{c}{24}\kappa_1+\sum_{j=1}^n (h_j-\frac{c}{24}) \psi_j$ and the standard parametrizations $h_j=\frac{c-1}{24}-p_j^2$ and $c=1+6(b+b^{-1})^2$. Here, $\text{td}(\mathcal{M}_{g,n})$ is the Todd class of the tangent bundle of moduli space (with boundary class contributions removed). \eqref{eq:quantum volumes index theorem} has the form of an index theorem and therefore, the quantum volumes solve a counting problem. It can be motivated from canonical quantization of chiral 3d gravity.

This indicates that \eqref{eq:Feynman rules integer shifts} can also be understood as an index theorem, but on a moduli space that differs on the boundary divisors. Indeed, the main term in \eqref{eq:Feynman rules integer shifts} with $\Gamma$ the trivial graph takes exactly the same form as the quantum volume. Other graphs are entirely supported on the boundary of moduli space. This captures the different behaviour of the worldsheet correlators near degeneration limits of moduli space.\footnote{Something analogous happens in the minimal string case, where the relevant compactifications are weighted compactifications considered by Hassett \cite{Hassett}, see also \cite{Eberhardt:2023rzz, Artemev:2024rck}.} 
We discuss a natural moduli space associated to the $c=1$ string in section~\ref{subsec:r spin curves}.

\subsection{Examples}
Let us compute a few examples of $c=1$ amplitudes. Even though the formula \eqref{eq:Feynman rules integer shifts} is conceptually much cleaner, it is in practice more efficient to implement \eqref{eq:c=1 momentum space intersection numbers}. We also compare our results with the literature.

\paragraph{Tree-level four-point function.} At tree-level, the intersection number expressions are straightforward to evaluate. We have for example at four points,
\be
\hat{\mathsf{s}}_{0,4}(\boldsymbol{p})=\frac{1}{2}-\sum_{i=1}^4 p_i^2+\frac{1}{2}\sum_{1 \le i<j \le 4} B_2(\{p_i+p_j\})\,, \label{eq:g=0, n=4 discrete result}
\ee
where the first two terms come from the trivial graph (contact diagram) and the other terms come from exchange diagrams.
Using that $B_2(x)=\frac{1}{6}-x+x^2$ and restricting to the first Brillouin zone gives (using momentum conservation)
\be
\hat{\mathsf{S}}_{0,4}(\boldsymbol{\omega})=1+\frac{i}{4} \sum_{1 \le i<j \le 4} (\omega_i+\omega_j) \sgn\big(\Re(\omega_i+\omega_j)\big)\,,
\ee
where we used the prescription \eqref{eq:branch choice} to express it in terms of physical kinematics. This reproduces the known formulas of the tree-level four-point function amplitude \cite{Polchinski:1991uq,Balthazar:2017mxh}. Our prescription for selecting the branch in \eqref{eq:branch choice} originates from a Lorentzian spacetime signature. It thus differs slightly from the prescription used e.g.\ in \cite{Balthazar:2017mxh}, which is appropriate for a Euclidean target space signature. For instance, the $3\to 1$ tree-level amplitude reads (with $\omega_4\leq0$)
\begin{equation}
\hat{\mathsf{S}}_{0,4}^{(3\to1)}(\boldsymbol{\omega})=1+i(\omega_1+\omega_2+\omega_3) \,,
\end{equation}
whereas the $2\to2$ tree-level amplitude is given by (with $\omega_3,\omega_4\leq0$)
\begin{equation}
\hat{\mathsf{S}}_{0,4}^{(2\to2)}(\boldsymbol{\omega})=1+i \max \{ \omega_1,\omega_2,-\omega_3,-\omega_4 \} \,.
\end{equation}
In these cases, the $\CC$LS correlator was checked numerically in \cite{Collier:2024kwt}, which in view of the relation between $\CC$LS correlators and $c=1$ amplitudes reproduces the numerical check of $c=1$ amplitudes performed in \cite{Balthazar:2017mxh}.

\paragraph{Tree-level five-point function.} We will write the following amplitudes in terms of the energies $\omega_i$. They can be specialized to physical kinematics using the prescription \eqref{eq:branch choice}. For the five-point function, one gets
\begin{multline}
\mathsf{S}_{0,5}(\boldsymbol{\omega})=2+\frac{1}{4}\sum_{i=1}^5 \omega_i^2-\frac{1}{2}\sum_{1 \le i<j \le 5} \sqrt{\smash[b]{-(\omega_i+\omega_j)^2}}\\
+\frac{1}{8}\sum_{\begin{subarray}{c} 1 \le i<j \le 5 \\ 1 \le k<\ell \le 5 \\ \{i,j\} \cap \{k,\ell\}=\emptyset \end{subarray}} \sqrt{\smash[b]{-(\omega_i+\omega_j)^2}} \sqrt{\smash[b]{-(\omega_k+\omega_\ell)^2}}\,.
\end{multline}
The different discontinuous looking pieces come from the different possible graphs one can write down for a five point amplitude. The square roots can be removed by applying the prescription \eqref{eq:branch choice}.
\paragraph{Torus one-point amplitude.}  There are two stable graphs: the trivial one, and the self-loop. The latter doesn't contribute, as discussed in section~\ref{subsec:remarks}. Thus, we simply have
\be \label{eq:S11}
\mathsf{S}_{1,1}(\omega_1)=\int_{\bM_{1,1}} \ex^{\frac{1}{2}\kappa_1}=\frac{1}{48}\,.
\ee
\paragraph{Torus two-point amplitude.} Things become more interesting for the torus two-point amplitude. There are three stable graphs without self-loop, displayed in Table~\ref{tab:stable graphs g=1 and n=2}.
\begin{table}
\begin{center}
\begin{tabular}{c|ccc}
$\Gamma$  & \parbox[t][1cm][t]{3.2cm}{\begin{tikzpicture}[scale=.6]
    \node[shape=circle,draw=black, very thick] (A) at (0,0) {$1$};
    \draw[very thick] (A) to (-1.5,0) node[left] {1};
    \draw[very thick] (A) to (1.5,0) node[right] {2};
\end{tikzpicture}} &\parbox[t][1cm][t]{3.2cm}{\begin{tikzpicture}[scale=.6]
    \node[shape=circle,draw=black, very thick] (A) at (0,0) {$0$};
    \node[shape=circle,draw=black, very thick] (B) at (0,1.5) {$1$};
    \draw[very thick] (A) to (-1.5,0) node[left] {1};
    \draw[very thick] (A) to (1.5,0) node[right] {2};
    \draw[very thick] (A) to (B);
\end{tikzpicture}}  & \parbox[t][1cm][t]{3.2cm}{\begin{tikzpicture}[scale=.6]
    \node[shape=circle,draw=black, very thick] (A) at (0,0) {$0$};
    \node[shape=circle,draw=black, very thick] (B) at (-1.5,0) {$0$};
    \draw[very thick] (A) to (1,0) node[right] {2};
    \draw[very thick] (B) to (-2.5,0) node[left] {1};
    \draw[very thick, bend left=30] (A) to (B);
    \draw[very thick, bend right=30] (A) to (B);
\end{tikzpicture}} \\
\hline
$|\text{Aut}(\Gamma)|$ & 1 & 1 & 2
\end{tabular}
\end{center}
\caption{Stable graphs for $g=1$ and $n=2$.} \label{tab:stable graphs g=1 and n=2}
\end{table}
They all have physically a different interpretation. The first one corresponds to a 1-loop renormalization of the propagator, the second to a tadpole diagram and the third to a genuine 1-loop amplitude. The contributions from the first two diagrams are regular, while the loop integral in the last diagram introduces a non-analytic contribution. More specifically, the first diagram contributes
\be
\int_{\bM_{1,2}} \ex^{-\sum_{k \ge 1} \frac{B_k \kappa_k}{k\, k!}-p_1^2(\psi_1+\psi_2)}=\frac{1}{72}(6p_1^4-6p_1^2+1)\,. \label{eq:g=1, n=2 first diagram}
\ee
The second diagram contributes
\begin{align}
 \int_{\bM_{1,1}}\big(1+\tfrac{1}{2}\kappa_1\big)\big(B_{2}(0)-\tfrac{1}{2}B_4(0) \psi_\bullet\big)=\frac{1}{240}\,. \label{eq:g=1, n=2 second diagram}
\end{align}
Finally, the last diagram involving the loop integration contributes\footnote{This integral can be most easily computed by applying \eqref{eq:Bernoulli polynomial identity} to both terms in the integrands, which produces a double sum. The integral projects it to a single sum, which is again of the same form as the LHS of \eqref{eq:Bernoulli polynomial identity} and thus can be rewritten in terms of a Bernoulli polynomial.}
\be
\frac{1}{2} \int_{\RR/\ZZ} \d q \, B_2(\{q\})B_2(\{q+p_1\})=-\frac{1}{12}B_4(\{p_1\}) \,. \label{eq:g=1, n=2 third diagram}
\ee
In total, we get after restricting to the first Brillouin zone and changing to the parametrization in terms of the energy
\be
\mathsf{S}_{1,2}(\boldsymbol{\omega})=\frac{1}{48}\big(1+2\omega_1^2+(-\omega_1^2)^{\frac{3}{2}}\big)\overset{\omega_1 \geqslant 0}{=} \frac{1}{48}(1+i \omega_1)(1-i \omega_1+\omega_1^2)\,. \label{eq:S12}
\ee
Notice again that the degree is one lower than what would naively be expected from the polynomial boundedness discussed in section~\ref{subsec:remarks}.
\paragraph{Torus three-point amplitude.} Computing the torus three-point function with general choices of $\omega_i$ is cumbersome. Let us thus assume that $\omega_1,\, \omega_2 \ge 0$ and $\omega_3 \le 0$ (i.e.\ we are computing the $2 \to 1$ amplitude). There are seven graphs without self-loops that contribute to the amplitude. They are
\begin{multline}
\begin{tikzpicture}[baseline={([yshift=-.5ex]current bounding box.center)},scale=.6]
    \node[shape=circle,draw=black, very thick] (A) at (0,0) {$1$};
    \draw[very thick] (A) to (1,0) node[right] {1};
    \draw[very thick] (A) to (-1,-1) node[below] {3};
    \draw[very thick] (A) to (-1,1) node[above] {2};
\end{tikzpicture} \, , \
\begin{tikzpicture}[baseline={([yshift=-.5ex]current bounding box.center)},scale=.6]
    \node[shape=circle,draw=black, very thick] (A) at (0,0) {$1$};
    \node[shape=circle,draw=black, very thick] (B) at (-1.5,0) {$0$};
    \draw[very thick] (A) to (1,0) node[right] {1};
    \draw[very thick] (B) to (-2.5,-1) node[below] {3};
    \draw[very thick] (B) to (-2.5,1) node[above] {2};
    \draw[very thick] (A) to (B);
\end{tikzpicture}+\text{perm}\, , \
\begin{tikzpicture}[baseline={([yshift=-.5ex]current bounding box.center)},scale=.6]
    \node[shape=circle,draw=black, very thick] (A) at (0,0) {$1$};
    \node[shape=circle,draw=black, very thick] (B) at (-1.5,0) {$0$};
    \draw[very thick] (B) to (-1.5,-1) node[below] {3};
    \draw[very thick] (B) to (-1.5,1) node[above] {1};
    \draw[very thick] (B) to (-2.5,0) node[left] {2};
    \draw[very thick] (A) to (B);
\end{tikzpicture} \, , \
\begin{tikzpicture}[baseline={([yshift=-.5ex]current bounding box.center)},scale=.6]
    \node[shape=circle,draw=black, very thick] (A) at (0,0) {$0$};
    \node[shape=circle,draw=black, very thick] (B) at (-1.5,0) {$0$};
    \draw[very thick] (A) to (1,0) node[right] {1};
    \draw[very thick] (B) to (-2.5,-1) node[below] {3};
    \draw[very thick] (B) to (-2.5,1) node[above] {2};
    \draw[very thick, bend left=30] (A) to (B);
    \draw[very thick, bend right=30] (A) to (B);
\end{tikzpicture}+\text{perm}\, , \\
\begin{tikzpicture}[baseline={([yshift=-.5ex]current bounding box.center)},scale=.6]
    \node[shape=circle,draw=black, very thick] (A) at (0,0) {$0$};
    \node[shape=circle,draw=black, very thick] (C) at (1.5,0) {$1$};
    \node[shape=circle,draw=black, very thick] (B) at (-1.5,0) {$0$};
    \draw[very thick] (A) to (0,1) node[above] {1};
    \draw[very thick] (B) to (-1.5,1) node[above] {2};
    \draw[very thick] (B) to (-2.5,0) node[left] {3};
    \draw[very thick] (A) to (B);
    \draw[very thick] (A) to (C);
\end{tikzpicture}+\text{perm}\, ,
\ \begin{tikzpicture}[baseline={([yshift=-.5ex]current bounding box.center)},scale=.6]
    \node[shape=circle,draw=black, very thick] (C) at (-3,0) {$0$};
    \node[shape=circle,draw=black, very thick] (B) at (-1.5,0) {$0$};
    \node[shape=circle,draw=black, very thick] (A) at (0,0) {$0$};
    \draw[very thick] (A) to (1,0) node[right] {1};
    \draw[very thick] (C) to (-4,-1) node[below] {3};
    \draw[very thick] (C) to (-4,1) node[above] {2};
    \draw[very thick, bend left=30] (A) to (B);
    \draw[very thick, bend right=30] (A) to (B);
    \draw[very thick] (B) to (C);
\end{tikzpicture}+\text{perm}\, , \ \begin{tikzpicture}[baseline={([yshift=-.5ex]current bounding box.center)},scale=.6]
    \node[shape=circle,draw=black, very thick] (A) at (0,0) {$0$};
    \node[shape=circle,draw=black, very thick] (B) at (-1.5,1) {$0$};
    \node[shape=circle,draw=black, very thick] (C) at (-1.5,-1) {$0$};
    \draw[very thick] (A) to (1,0) node[above] {1};
    \draw[very thick] (B) to (-2.5,1) node[above] {2};
    \draw[very thick] (C) to (-2.5,-1) node[below] {3};
    \draw[very thick] (A) to (B);
    \draw[very thick] (A) to (C);
    \draw[very thick] (B) to (C);
\end{tikzpicture}\,, \label{eq:Feynman diagrams g=1, n=3}
\end{multline}
where `perm' denotes the diagrams where the external legs are permuted.
Most of these diagrams are fairly straightforward to evaluate. In practice, to avoid the case distinctions caused by the fractional part function, it is simplest to specialize the external momenta to specific transcendental values (such as odd Riemann zeta values), evaluate the integrals, and then restore the general dependence on the momenta. One obtains the compact result
\be
\hat{\mathsf{S}}_{1,3}(\boldsymbol{\omega})=-\frac{1}{48} (\omega_1+\omega_2-i)(\omega_1+\omega_2-2 i)(\omega_1^2-i \omega_1+\omega_2^2-i \omega_2+1)\  . \label{eq:S13}
\ee
Notice that the dilaton equation \eqref{eq:dilaton equation} reduces \eqref{eq:S13} to \eqref{eq:S12}. Moreover the polynomial degree is only four --- two lower than naively expected. Finally, we observe as a general feature of higher genus amplitudes that they have zeros, i.e.\ always nicely factorize. These zeros can be derived from the MQM description \eqref{eq:MPR n to 1}, but are obscured in the matrix integral (section~\ref{sec:matrix integral}) and worldsheet description.

\paragraph{Higher-points, higher loops.} It becomes increasingly cumbersome to list the diagrams by hand. In practice, we use the program \texttt{admcycles} \cite{admcycles} to generate a list of all stable graphs together with their automorphism groups, which we then import into \texttt{Mathematica} and use our implementation of the quantum volumes of the Virasoro minimal string \cite{Collier:2023cyw} to evaluate the remaining intersection numbers and integrals. We have computed the following amplitudes from summing over diagrams. For simplicity, we restrict to the $n-1 \to 1$ amplitude (or $1 \to n-1$ amplitude by applying \eqref{eq:flipping all momenta}). For $\omega_1,\dots,\omega_{n-1} \ge 0$, we find
\begin{subequations}
\begin{align}
\hat{\mathsf{S}}_{0,n}(\boldsymbol{\omega})&=(1+ie_1)_{n-3}\,,\ n \le 8\,, \label{eq:S0n result}\\
\hat{\mathsf{S}}_{1,n}(\boldsymbol{\omega})&=\frac{(1+ie_1)_{n-1}}{48}\,\big(1- i e_1-2 e_2+e_1^2\big)\, , \  n \le 6\,, \label{eq:S1n result}\\
\hat{\mathsf{S}}_{2,n}(\boldsymbol{\omega})&=\frac{(1+ie_1)_{n+1}}{23040}\,\big(7-12 i e_1-20 e_2+5 e_1^2+16 e_2^2+8 e_4+20 i e_2 e_1-8 e_3 e_1\nonumber\\
&\qquad-10 i e_1^3-12 e_2 e_1^2+3 e_1^4\big)\,,\ n \le 5\,, \label{eq:S2n result}\\
\hat{\mathsf{S}}_{3,n}(\boldsymbol{\omega})&=\frac{(1+ie_1)_{n+3}}{23224320}\big(93-205 i e_1-294 e_2-144 e_2^3+336 e_2^2-42 e_2 e_1^2-42 e_1^4\nonumber\\
&\qquad-217 i e_1^3+504 i e_2 e_1-336 i e_2^2 e_1+144 e_2^2 e_1^2-54 e_2 e_1^4+252 i e_2 e_1^3\nonumber\\
&\qquad-63 i e_1^5+9 e_1^6\big)\,,\ n \le 3\,, \label{eq:S3n result}\\
\hat{\mathsf{S}}_{4,n}(\boldsymbol{\omega})&=\frac{(1+ie_1)_{n+5}}{22295347200}\big(1143-2912 i e_1-854 e_1^2-2980 i e_1^3-1883 e_1^4-812 i e_1^5\nonumber\\
&\qquad-450 e_1^6-180 i e_1^7+15 e_1^8\big)\, ,\ n \le 2\,,  \label{eq:S4n result}\\
\hat{\mathsf{S}}_{5,0}(\boldsymbol{\omega})&=\frac{511}{155713536}\,.\label{eq:S50 result}
\end{align}
\label{eq:Sgn direct computation}%
\end{subequations}
Here, $(a)_n=a(a+1) \cdots (a+n-1)$ is the rising Pochhammer symbol and $e_i$ is the $i$-th elementary symmetric polynomial in $\omega_1,\dots,\omega_{n-1}$. Of course, $e_i=0$ for $i>n-1$. The number of diagrams grow very quickly; for example at $(g,n)=(2,5)$, we have 38,034 diagrams without self-loop (counting permutations of external legs as different diagrams).

We should note that the polynomial degree in $p_i$ of \eqref{eq:Sgn direct computation} is only $4g-3+n$ (as opposed to the naive upper bound of $6g-6+2n$ that we discussed in section~\ref{subsec:remarks}). Thus, there are many very non-trivial cancellations between the intersection numbers that lead to a much lower bound than expected. We also note a very simple pattern of zeros in the amplitudes. It is partially explained from the dilaton equation \eqref{eq:dilaton equation}.

\subsection{Perturbative unitarity}
\label{subsec:perturbative unitarity}

In this section, we give a direct proof that the (non-discrete) $c=1$ amplitudes defined by \eqref{eq:c=1 momentum space intersection numbers} and then analytically continued via the prescription explained in section~\ref{subsec:remarks} respect spacetime unitarity to any order in perturbation theory. This was already proven in \cite{Moore:1991zv} within $c=1$ MQM. At the non-perturbative level, there are subtle issues with unitarity, see \cite{Balthazar:2019rnh, Sen:2020ruy}. We will not discuss them here.

\paragraph{Cutting rules and spacetime unitarity.}
Spacetime unitarity is the statement $\mathbb{S}^\dag \mathbb{S}=\id$ where $\mathbb{S}$ is the full non-perturbative S-matrix with arbitrarily many states and possibly disconnected processes. All its initial and final states are physical multi-particle states. We can expand $\mathbb{S}$ perturbatively order by order in the string coupling and extract a cutting rule for the genus-$g$ contribution. Let us write as usual $\mathbb{S}=\id+i \TT$. Then the statement of unitarity becomes the optical theorem,
\be
2\Im \TT=\frac{1}{i}(\TT-\TT^\dag)=\TT^\dag \TT\,. \label{eq:unitarity equation}
\ee
The optical theorem is equivalent to the Cutkosky cutting rules in perturbation theory, which express the imaginary part of the scattering amplitude in terms of unitarity cuts, expressed by the right hand side of the equation. For Feynman diagrams of ordinary QFTs, various direct proofs of cutting rules and hence of perturbative unitarity exist in the literature, most notably by 't~Hooft and Veltman using position space Feynman diagrams \cite{tHooft:1972qbu, tHooft:1973wag} and later using time-ordered perturbation theory \cite{Hannesdottir:2022xki}. None of these proofs seem to directly carry over to the $c=1$ string.

Instead, we will work with the so-called holomorphic cutting rules, as advertised in \cite{Hannesdottir:2022bmo}. They are also equivalent to the Cutkosky cutting rules. To derive those, one expresses $\TT^\dag$ on the right-hand side of \eqref{eq:unitarity equation} in terms of $\TT$ as $\TT^\dag=\TT(\id+i \TT)^{-1}$ and thus
\be
2\Im \TT=\TT^2(1+i\TT)^{-1}=-\sum_{c=1}^\infty (-i \TT)^{c+1}\,. \label{eq:holomorphic unitarity equation}
\ee
Here, $c$ running over all positive integers corresponds to the number of unitarity cuts performed on an amplitude. In this form, the cutting rules will be simpler to derive from \eqref{eq:c=1 momentum space intersection numbers}, since they do not involve complex conjugation.

\paragraph{Discontinuity from intersection numbers.}
Let us analyze the discontinuity from the intersection number expression \eqref{eq:c=1 momentum space intersection numbers} and the corresponding full S-matrix \eqref{eq:full S matrix from the worldsheet}. This will require us to look at successively more complicated situations.
By inspection of \eqref{eq:c=1 momentum space intersection numbers}, we see that the vertex factors involve only quadratic polynomials in the momenta $p_i$. Therefore, they do not lead to discontinuities. The discontinuities arise fully from the edge factors, which feature the fractional part function $\{x\}$, which is discontinuous. This is already consistent with the cutting rules, which involve cutting the diagram along the edges.

\paragraph{Cutting a single edge.} Let's begin with the situation that doesn't involve any loop integrals. Consider a single internal edge, involving the Bernoulli polynomial $B_{2d+2}(\{p_e\})$. When restricting this to the first Brillouin zone, we can replace this by $B_{2d+2}(\sqrt{p_e^2})$, as discussed around \eqref{eq:c=1 continuous version intersection numbers genus 0}. The only term in $B_{2d+2}(\sqrt{p_e^2})$ involving an odd power of the argument is $-(d+1)(p_e^2)^{d+\frac{1}{2}}$. Therefore, for the purpose of computing the discontinuity, we can replace the corresponding edge in \eqref{eq:c=1 momentum space intersection numbers} by
\be
-\sum_{d=0}^\infty \frac{(p_e^2)^{d+\frac{1}{2}}(-\psi_\bullet-\psi_\circ)^d}{d!}=-\sqrt{p_e^2}\, \ex^{-p_e^2(\psi_\bullet+\psi_\circ)}\,. \label{eq:discontinuous part propagator}
\ee
This can also be seen in \eqref{eq:edge factor rewriting asymptotic series}, where only the term $\ell=0$ contributes to the discontinuity at $p_e=0$.
The exponentials $\ex^{-p_e^2 \psi_\bullet}$ and $\ex^{-p_e^2 \psi_\circ}$ combine with the other exponentials in \eqref{eq:c=1 momentum space intersection numbers} of the external states to give the full S-matrix on the left and the right of the cut.
For tree-level amplitudes, there is at most one edge with momentum $p_e$. Cutting along it gives the full discontinuity around $p_e=0$. It is defined as the difference of \eqref{eq:discontinuous part propagator} with standard sign choice as in \eqref{eq:branch choice} and opposite sign choice. Thus the factor $-\sqrt{p_e^2}$ becomes in terms of the energy
\be
2 \times \frac{1}{2} \sqrt{-\omega_e^2}=-i \omega_e\,. \label{eq:discontinuity factor i}
\ee
Here, we assumed that $\omega_e>0$, since we are cutting in a causal way.
Let us now consider the amplitude in terms of the physical energies $\omega_i$ and with the delta-function inserted, see \eqref{eq:S delta function strip off normalization}. Thus, the discontinuity that we just derived for a single diagram with topology $\Gamma$ and a single cut $c$ is
\begin{align}
\mathop{\text{Disc}}_c \mathsf{S}_{\Gamma}(\omega_1,\dots,\omega_n)&=-\mathcal{N} i \omega_e \, \hat{\mathsf{S}}_{\Gamma_\text{L}}( \boldsymbol{\omega}_\text{L})\hat{\mathsf{S}}_{\Gamma_\text{R}}(\boldsymbol{\omega}_\text{R}) \delta\Big(\sum\nolimits_i \omega_i\Big)\\
&=-\frac{i}{\mathcal{N}} \int_0^\infty \d \omega_e \, \omega_e \, \mathsf{S}_{\Gamma_\text{L}}(\boldsymbol{\omega}_\text{L})\mathsf{S}_{\Gamma_\text{R}}(\boldsymbol{\omega}_\text{R})\,,
\end{align}
where $\boldsymbol{\omega}_\text{L}$ denotes the energies of the on-shell states on the left, including the cut edge with energy $-\omega_e$, and $\boldsymbol{\omega}_\text{R}$ denotes the energies of the on-shell states on the right, including the cut edge with energy $\omega_e$.
The automorphism factor of the diagram is just the product of the left and the right part in this case and correctly factorizes.

\paragraph{Cutting loops.} We next analyze the case where we cut along a loop integral. To get oriented, it is useful to look at a specific integral first, for which we choose the triangle diagram, that is the last diagram in \eqref{eq:Feynman diagrams g=1, n=3}. It takes the form
\be
\int_0^1 \d q\ B_2(\{q\})B_2(\{q-p_1\})B_2(\{q-p_1-p_2\})\,.
\ee
The integrand has a discontinuity when $q=0$, $q=p_1$ or $q=p_1+p_2=-p_3$. The integral will have a discontinuity when any two of these points collide, i.e.\ when $p_1=0$, $p_2=0$ or $p_3=0$. Only the discontinuous part of the integrand can contribute to the discontinuous part of the integral. So for the purpose of computing the discontinuity around $p_1=0$, we can replace $B_2(\{q\}) \to - \sqrt{q^2}$ and $B_2(\{q-p_1\}) \to -\sqrt{(q-p_1)^2}$ (after restricting to the first Brillouin zone). Furthermore, we can restrict the integral over $q$ to the chamber bounded by $q=0$ and $q=p_1$, since the complement will not contribute to the discontinuity. Thus, the discontinuous part of the integral around $p_1$ can be written as
\be
\int_{\min(0,p_1)}^{\max(0,p_1)} \d q\, \sqrt{q^2} \sqrt{(q-p_1)^2} B_2(\{q-p_1-p_2\})\,.
\ee
Let us translate again to on-shell kinematics in terms of the energies with the prescription \eqref{eq:branch choice}. This leads to
\begin{align}
\mathop{\text{Disc}}_c \mathsf{S}_{\Gamma}(\boldsymbol{\omega})&=-\frac{\mathcal{N}i}{4}\int_0^{\omega_1} \d \omega \, \omega(\omega_1-\omega) \, \hat{\mathsf{S}}_{\Gamma_\text{L}}( \boldsymbol{\omega}_\text{L}) \hat{\mathsf{S}}_{\Gamma_\text{R}}(\boldsymbol{\omega}_\text{R}) \delta\Big(\sum\nolimits_i \omega_i\Big) \\
&=-\frac{i}{4\mathcal{N}} \int_0^\infty \d \omega\, \d \omega'\, \omega \omega'\, \mathsf{S}_{\Gamma_\text{L}}(\boldsymbol{\omega}_\text{L}) \mathsf{S}_{\Gamma_\text{R}}(\boldsymbol{\omega}_\text{R})\,.
\end{align}
This generalizes to more complicated diagrams. We can produce discontinuous pieces in the answer by cutting along propagators of a loop integral. Cutting amounts to replacing the propagators with their discontinuous part. The $\psi$-classes associated to the propagators factorize again into exponentials as in \eqref{eq:discontinuous part propagator}, which reconstructs the full scattering amplitude left and right of the cut.

We thus have in general for the discontinuity of a unitarity cut $c$,
\be
\mathop{\text{Disc}}_c \mathsf{S}_{\Gamma}(\boldsymbol{\omega})=-\frac{4i}{\mathcal{N}}\,  \frac{|\text{Aut}(\Gamma_\text{L})||\text{Aut}(\Gamma_\text{R})|}{|\text{Aut}(\Gamma)|}\int_0^\infty \prod_{i=1}^{n_c} \frac{\d \omega_i\, \omega_i}{4}\mathsf{S}_{\Gamma_\text{L}}(\boldsymbol{\omega}_\text{L})\mathsf{S}_{\Gamma_\text{R}}(\boldsymbol{\omega}_\text{R})\,.
\ee
Here $n_c$ is the number of edges participating in the unitarity cut $c$. We will in the following set
\be
\mathcal{N}=-4i\,, \label{eq:N value}
\ee
so that the prefactor becomes unity.
Let us also denote a unitarity cut as
\be
\begin{tikzpicture}[baseline={([yshift=-.5ex]current bounding box.center)},scale=.6]
    \node[shape=circle,draw=black, very thick, fill=gray, inner sep=7pt] (A) at (0,0) {};
    \node[shape=circle,draw=black, very thick, fill=gray, inner sep=7pt] (B) at (-2.2,0) {};
    \draw[very thick] (A) to (1.5,1);
    \draw[very thick] (A) to (1.5,-1);
    \node at (1,.1) {$\vdots$};
    \node at (-3.2,.1) {$\vdots$};
    \draw[very thick] (B) to (-3.7,1);
    \draw[very thick] (B) to (-3.7,-1);
    \draw[very thick, bend left=45] (A) to (B);
    \draw[very thick, bend right=20] (A) to (B);
    \draw[very thick, bend right=45] (A) to (B);
    \node at (-1.3,-.05) {$\vdots$};
    \node at (-.9,-.05) {$\vdots$};
    \draw[very thick, red, dashed] (-1.1,-1) to (-1.1,1);
\end{tikzpicture}\,.
\ee
Here, a gray vertex is meant to denote any subdiagram. Furthermore, we read the diagram from left to right, i.e.\ the external legs on the left are incoming and the external legs on the right are outgoing. We define the unitarity cut as the integral over the on-shell phase space with the measure
\be
\frac{1}{n_c!}\prod_{i=1}^{n_c} \Big(\frac{1}{4}\omega_i\, \d \omega_i\Big)\,, \label{eq:on-shell measure}
\ee
where $\omega_1,\dots,\omega_{n_c}$ are the intermediate energies, which are real and positive for on-shell kinematics. The factor $\frac{1}{n_c!}$ accounts for the fact that we are treating the internal particles as indistinguishable.

We can also check that this normalization is compatible with the two-point function that we derived in \eqref{eq:worldsheet two-point function}, which is
\be
\mathsf{S}_{0,2}(\omega_1,\omega_2)=\frac{\mathcal{N}}{i \omega_1} \delta(\omega_1+\omega_2)=\frac{4}{\omega_2}\, \delta(\omega_1+\omega_2)\,,
\ee
which is precisely the inverse of the measure \eqref{eq:on-shell measure}.

\paragraph{An example.} To derive the full cutting rules in the form \eqref{eq:holomorphic unitarity equation}, we have to look more closely at the combinatorics of diagrams. Let us first illustrate this at an example, for which we are again going to use the one-loop three-point function, whose Feynman diagrams are listed in \eqref{eq:Feynman diagrams g=1, n=3}. We are interested in deriving the full discontinuous part of the diagram. Suppose that $\omega_1,\, \omega_2 \ge 0$ are incoming and $\omega_3 \le 0$ is outgoing. The first and third diagram do not give a discontinuous contribution. The second, fourth and fifth diagram together with all its perturbation have a unique possible unitarity cut. Thus, so far we have the following contribution to $ \mathop{\text{Disc}} i \TT^{2 \to 1}_3$,
\begin{multline}
\begin{tikzpicture}[baseline={([yshift=-.5ex]current bounding box.center)},scale=.6]
    \node[shape=circle,draw=black, very thick] (A) at (.3,0) {$1$};
    \node[shape=circle,draw=black, very thick] (B) at (-1.5,0) {$0$};
    \draw[very thick] (A) to (1.8,0) node[right] {3};
    \draw[very thick] (B) to (-2.5,-1) node[left] {2};
    \draw[very thick] (B) to (-2.5,1) node[left] {1};
    \draw[very thick] (A) to (B);
    \draw[very thick, red, dashed] (-.6,-1) to (-.6,1);
\end{tikzpicture}\hspace{-.1cm}+\hspace{-.2cm}\begin{tikzpicture}[baseline={([yshift=-.5ex]current bounding box.center)},scale=.6]
    \node[shape=circle,draw=black, very thick] (A) at (.3,0) {$0$};
    \node[shape=circle,draw=black, very thick] (B) at (-1.5,0) {$0$};
    \draw[very thick] (A) to (1.8,0) node[right] {3};
    \draw[very thick] (B) to (-2.5,-1) node[left] {2};
    \draw[very thick] (B) to (-2.5,1) node[left] {1};
    \draw[very thick, bend left=30] (A) to (B);
    \draw[very thick, bend right=30] (A) to (B);
    \draw[very thick, red, dashed] (-.6,-1) to (-.6,1);
\end{tikzpicture}\hspace{-.1cm}+\hspace{-.2cm}
\begin{tikzpicture}[baseline={([yshift=-.5ex]current bounding box.center)},scale=.6]
    \node[shape=circle,draw=black, very thick] (A) at (.3,0) {$0$};
    \node[shape=circle,draw=black, very thick] (C) at (.3,1.5) {$1$};
    \node[shape=circle,draw=black, very thick] (B) at (-1.5,0) {$0$};
    \draw[very thick] (A) to (1.8,0) node[right] {3};
    \draw[very thick] (B) to (-2.5,1) node[left] {1};
    \draw[very thick] (B) to (-2.5,-1) node[left] {2};
    \draw[very thick] (A) to (B);
    \draw[very thick] (A) to (C);
    \draw[very thick, red, dashed] (-.6,-1) to (-.6,1);
\end{tikzpicture}\\
+\Bigg[\begin{tikzpicture}[baseline={([yshift=-.5ex]current bounding box.center)},scale=.6]
    \node[shape=circle,draw=black, very thick] (A) at (.3,0) {$0$};
    \node[shape=circle,draw=black, very thick] (B) at (-1.5,.52) {$1$};
    \draw[very thick] (A) to (1.8,0) node[right] {3};
    \draw[very thick] (A) to (-2.5,-.8) node[left] {2};
    \draw[very thick] (B) to (-2.5,.8) node[left] {1};
    \draw[very thick] (A) to (B);
    \draw[very thick, red, dashed] (-.6,0) to (-.6,.6);
\end{tikzpicture}\hspace{-.1cm}+\hspace{-.2cm}\begin{tikzpicture}[baseline={([yshift=-.5ex]current bounding box.center)},scale=.6]
    \node[shape=circle,draw=black, very thick] (A) at (.3,0) {$0$};
    \node[shape=circle,draw=black, very thick] (B) at (-1.5,-.52) {$0$};
    \draw[very thick] (A) to (1.8,0) node[right] {3};
    \draw[very thick] (A) to (-2.5,.8) node[left] {1};
    \draw[very thick] (B) to (-2.5,-.8) node[left] {2};
    \draw[very thick, bend left=20] (A) to (B);
    \draw[very thick, bend right=20] (A) to (B);
    \draw[very thick, red, dashed] (-.6,-.8) to (-.6,.2);
\end{tikzpicture}\hspace{-.1cm}+\hspace{-.2cm}\begin{tikzpicture}[baseline={([yshift=-.5ex]current bounding box.center)},scale=.6]
    \node[shape=circle,draw=black, very thick] (A) at (.3,0) {$0$};
    \node[shape=circle,draw=black, very thick] (B) at (-1.5,.52) {$0$};
    \node[shape=circle,draw=black, very thick] (C) at (-1.5,2) {$1$};
    \draw[very thick] (B) to (C);
    \draw[very thick] (A) to (1.8,0) node[right] {3};
    \draw[very thick] (A) to (-2.5,-.8) node[left] {2};
    \draw[very thick] (B) to (-2.5,.8) node[left] {1};
    \draw[very thick] (A) to (B);
    \draw[very thick, red, dashed] (-.6,0) to (-.6,.6);
\end{tikzpicture}\hspace{-.1cm}+(1 \leftrightarrow 2)\Bigg]\,.\label{eq:g=1, n=3 simple diagrams}
\end{multline}
In the cases where we cut a loop, the factor $\frac{1}{n_c!}$ always arises from the automorphism factor of the diagram.

Let us discuss the penultimate diagram in \eqref{eq:Feynman diagrams g=1, n=3}. It can be cut in two ways to obtain a discontinuous contribution. This means that it also contains a contribution of the form $\sqrt{-\omega_3^2}\sqrt{-\omega_3^2} \times \cdots$, obtained by taking the discontinuous piece of both edges. This piece is obviously not discontinuous and thus has to be subtracted from the unitarity cuts. This diagram then leads to the discontinuity
\begin{align}
&\begin{tikzpicture}[baseline={([yshift=-.5ex]current bounding box.center)},scale=.6]
    \node[shape=circle,draw=black, very thick] (C) at (-3,0) {$0$};
    \node[shape=circle,draw=black, very thick] (B) at (-1.2,0) {$0$};
    \node[shape=circle,draw=black, very thick] (A) at (.6,0) {$0$};
    \draw[very thick] (A) to (1.8,0) node[right] {3};
    \draw[very thick] (C) to (-4,-1) node[left] {2};
    \draw[very thick] (C) to (-4,1) node[left] {1};
    \draw[very thick, bend left=30] (A) to (B);
    \draw[very thick, bend right=30] (A) to (B);
    \draw[very thick] (B) to (C);
    \draw[very thick, red, dashed] (-.3,-1) to (-.3,1);
\end{tikzpicture}
\hspace{-.1cm}+\hspace{-.3cm}
\begin{tikzpicture}[baseline={([yshift=-.5ex]current bounding box.center)},scale=.6]
    \node[shape=circle,draw=black, very thick] (C) at (-3,0) {$0$};
    \node[shape=circle,draw=black, very thick] (B) at (-1.2,0) {$0$};
    \node[shape=circle,draw=black, very thick] (A) at (.6,0) {$0$};
    \draw[very thick] (A) to (1.8,0) node[right] {3};
    \draw[very thick] (C) to (-4,-1) node[left] {2};
    \draw[very thick] (C) to (-4,1) node[left] {1};
    \draw[very thick, bend left=30] (A) to (B);
    \draw[very thick, bend right=30] (A) to (B);
    \draw[very thick] (B) to (C);
    \draw[very thick, red, dashed] (-2.1,-1) to (-2.1,1);
\end{tikzpicture}
\hspace{-.1cm}-\hspace{-.2cm}
\begin{tikzpicture}[baseline={([yshift=-.5ex]current bounding box.center)},scale=.6]
    \node[shape=circle,draw=black, very thick] (C) at (-3,0) {$0$};
    \node[shape=circle,draw=black, very thick] (B) at (-1.2,0) {$0$};
    \node[shape=circle,draw=black, very thick] (A) at (.6,0) {$0$};
    \draw[very thick] (A) to (1.8,0) node[right] {3};
    \draw[very thick] (C) to (-4,-1) node[left] {2};
    \draw[very thick] (C) to (-4,1) node[left] {1};
    \draw[very thick, bend left=30] (A) to (B);
    \draw[very thick, bend right=30] (A) to (B);
    \draw[very thick] (B) to (C);
    \draw[very thick, red, dashed] (-2.1,-1) to (-2.1,1);
    \draw[very thick, red, dashed] (-.3,-1) to (-.3,1);
\end{tikzpicture}\nonumber\\
&\qquad+\Bigg[\begin{tikzpicture}[baseline={([yshift=-.5ex]current bounding box.center)},scale=.6]
    \node[shape=circle,draw=black, very thick] (A) at (.3,0) {$0$};
    \node[shape=circle,draw=black, very thick] (B) at (-1.5,.6) {$0$};
    \node[shape=circle,draw=black, very thick] (C) at (-3.2,.6) {$0$};
    \draw[very thick, bend left=25] (B) to (C);
    \draw[very thick, bend right=25] (B) to (C);
    \draw[very thick] (A) to (1.5,0) node[right] {3};
    \draw[very thick] (A) to (-4.1,-.6) node[left] {2};
    \draw[very thick] (C) to (-4.1,.6) node[left] {1};
    \draw[very thick] (A) to (B);
    \draw[very thick, red, dashed] (-.6,0) to (-.6,.8);
\end{tikzpicture}\hspace{-.1cm}+\hspace{-.2cm}\begin{tikzpicture}[baseline={([yshift=-.5ex]current bounding box.center)},scale=.6]
    \node[shape=circle,draw=black, very thick] (A) at (.3,0) {$0$};
    \node[shape=circle,draw=black, very thick] (B) at (-1.5,.6) {$0$};
    \node[shape=circle,draw=black, very thick] (C) at (-3.2,.6) {$0$};
    \draw[very thick, bend left=25] (B) to (C);
    \draw[very thick, bend right=25] (B) to (C);
    \draw[very thick] (A) to (1.5,0) node[right] {3};
    \draw[very thick] (A) to (-4.1,-.6) node[left] {2};
    \draw[very thick] (C) to (-4.1,.6) node[left] {1};
    \draw[very thick] (A) to (B);
    \draw[very thick, red, dashed] (-2.35,0) to (-2.35,1.2);
\end{tikzpicture}\hspace{-.1cm}\nonumber\\
&\qquad\qquad\qquad\qquad\qquad-\hspace{-.2cm}\begin{tikzpicture}[baseline={([yshift=-.5ex]current bounding box.center)},scale=.6]
    \node[shape=circle,draw=black, very thick] (A) at (.3,0) {$0$};
    \node[shape=circle,draw=black, very thick] (B) at (-1.5,.6) {$0$};
    \node[shape=circle,draw=black, very thick] (C) at (-3.2,.6) {$0$};
    \draw[very thick, bend left=25] (B) to (C);
    \draw[very thick, bend right=25] (B) to (C);
    \draw[very thick] (A) to (1.5,0) node[right] {3};
    \draw[very thick] (A) to (-4.1,-.6) node[left] {2};
    \draw[very thick] (C) to (-4.1,.6) node[left] {1};
    \draw[very thick] (A) to (B);
    \draw[very thick, red, dashed] (-.6,0) to (-.6,.8);
    \draw[very thick, red, dashed] (-2.35,0) to (-2.35,1.2);
\end{tikzpicture}\hspace{-.1cm}+(1 \leftrightarrow 2) \Bigg]\,. \label{eq:g=1, n=3 penultimate diagram}
\end{align}
The prefactor of the subtracted diagram is
\be
-2 \times 2\times (-i)^2 \times\mathcal{N}^{-2} \times \Big(\frac{i}{2}\Big)\times \Big(\frac{i}{2}\Big)^3 \times 4^3=-1\,.
\ee
The first factor of 2 comes from the fact that we have to subtract the contribution twice, once for both single cuts. The second factor of 2 comes because we are subtracting the discontinuity, which is given by multiplying the edge factor \eqref{eq:discontinuous part propagator} by a factor of 2. The factor of $(-i)^2$ comes from the expression for the discontinuity, see \eqref{eq:discontinuity factor i}. $\mathcal{N}^{-2}$ accounts for the relative normalization of the diagrams. $(\frac{i}{2}) \times (\frac{i}{2})^3$ are the factors we obtain from setting $p=\frac{i}{2} \omega$ in the two cuts, respectively. Finally, a factor of $4^3$ is eaten up by the measure \eqref{eq:on-shell measure}. The automorphism factor of the diagram precisely compensates the factor $\frac{1}{n_c!}$ in \eqref{eq:on-shell measure}. We then set again $\mathcal{N}=-4i$.

Finally, we need to analyze the triangle diagram, which features so-called anomalous thresholds.
We can always cut the two propagators joining the upper and lower vertices to the right vertex.
Following the same logic as for the previous diagram, the full discontinuity is obtained by performing all possible cuts and subtracting overcounting.
\begin{multline}
2\begin{tikzpicture}[baseline={([yshift=-.5ex]current bounding box.center)},scale=.6]
    \node[shape=circle,draw=black, very thick] (A) at (0,0) {$0$};
    \node[shape=circle,draw=black, very thick] (B) at (-2,1) {$0$};
    \node[shape=circle,draw=black, very thick] (C) at (-2,-1) {$0$};
    \draw[very thick] (A) to (1.5,0) node[right] {3};
    \draw[very thick] (B) to (-3.5,1) node[left] {1};
    \draw[very thick] (C) to (-3.5,-1) node[left] {2};
    \draw[very thick] (A) to (B);
    \draw[very thick] (A) to (C);
    \draw[very thick] (B) to (C);
    \draw[very thick, red, dashed] (-1,-1.25) to (-1,1.25);
\end{tikzpicture}\\
+\Bigg[2\begin{tikzpicture}[baseline={([yshift=-.5ex]current bounding box.center)},scale=.6]
    \node[shape=circle,draw=black, very thick] (A) at (-1.8,0) {$0$};
    \node[shape=circle,draw=black, very thick] (B) at (1.8,0) {$0$};
    \node[shape=circle,draw=black, very thick] (C) at (0,1.5) {$0$};
    \draw[very thick] (A) to (B);
    \draw[very thick] (A) to (C);
    \draw[very thick] (B) to (C);
    \draw[very thick] (A) to (-3,0) node[left] {$2$};
    \draw[very thick] (C) to (-3,1.5) node[left] {$1$};
    \draw[very thick] (B) to (3,0) node[right] {$3$};
    \draw[very thick, red, dashed] (-.9,-.5) to (-.9,1.2);
\end{tikzpicture}\hspace{-.1cm}-4\begin{tikzpicture}[baseline={([yshift=-.5ex]current bounding box.center)},scale=.6]
    \node[shape=circle,draw=black, very thick] (A) at (-1.8,0) {$0$};
    \node[shape=circle,draw=black, very thick] (B) at (1.8,0) {$0$};
    \node[shape=circle,draw=black, very thick] (C) at (0,1.5) {$0$};
    \draw[very thick] (A) to (B);
    \draw[very thick] (A) to (C);
    \draw[very thick] (B) to (C);
    \draw[very thick] (A) to (-3,0) node[left] {$2$};
    \draw[very thick] (C) to (-3,1.5) node[left] {$1$};
    \draw[very thick] (B) to (3,0) node[right] {$3$};
    \draw[very thick, red, dashed] (-.9,-.5) to (-.9,1.2);
    \draw[very thick, red, dashed] (.9,-.5) to (.9,1.2);
\end{tikzpicture}+(1 \leftrightarrow 2)\Bigg] .  \label{eq:g=1, n=3 triangle diagram}
\end{multline}
In this case, the factors $\frac{1}{n_c!}$ did not arise from the automorphism factors of the diagram which led to non-trivial multiplicities.

Summing over \eqref{eq:g=1, n=3 simple diagrams}, \eqref{eq:g=1, n=3 penultimate diagram} and \eqref{eq:g=1, n=3 triangle diagram} leads to the following formula for the discontinuity of $i\TT_3^{2 \to 1}$,
\begin{align}
    &\begin{tikzpicture}[baseline={([yshift=-.5ex]current bounding box.center)},scale=.6]
    \node[shape=circle,draw=black, very thick, fill=gray, inner sep=.5pt] (A) at (.3,0) {$\mathsf{S}_{1,2}$};
    \node[shape=circle,draw=black, very thick, fill=gray, inner sep=.5pt] (B) at (-1.5,0) {$\mathsf{S}_{0,3}$};
    \draw[very thick] (A) to (1.8,0) node[right] {3};
    \draw[very thick] (B) to (-2.5,-1) node[left] {2};
    \draw[very thick] (B) to (-2.5,1) node[left] {1};
    \draw[very thick] (A) to (B);
    \draw[very thick, red, dashed] (-.6,-1) to (-.6,1);
\end{tikzpicture}\hspace{-.1cm}+\hspace{-.2cm}\begin{tikzpicture}[baseline={([yshift=-.5ex]current bounding box.center)},scale=.6]
    \node[shape=circle,draw=black, very thick, fill=gray, inner sep=.5pt] (A) at (.3,0) {$\mathsf{S}_{0,3}$};
    \node[shape=circle,draw=black, very thick, fill=gray, inner sep=.5pt] (B) at (-1.5,0) {$\mathsf{S}_{0,4}$};
    \draw[very thick] (A) to (1.8,0) node[right] {3};
    \draw[very thick] (B) to (-2.5,-1) node[left] {2};
    \draw[very thick] (B) to (-2.5,1) node[left] {1};
    \draw[very thick, bend left=30] (A) to (B);
    \draw[very thick, bend right=30] (A) to (B);
    \draw[very thick, red, dashed] (-.6,-1) to (-.6,1);
\end{tikzpicture}\hspace{-.1cm}-\hspace{-.2cm}\begin{tikzpicture}[baseline={([yshift=-.5ex]current bounding box.center)},scale=.6]
    \node[shape=circle,draw=black, very thick, fill=gray, inner sep=.5pt] (C) at (-3,0) {$\mathsf{S}_{0,3}$};
    \node[shape=circle,draw=black, very thick, fill=gray, inner sep=.5pt] (B) at (-1.2,0) {$\mathsf{S}_{0,3}$};
    \node[shape=circle,draw=black, very thick, fill=gray, inner sep=.5pt] (A) at (.6,0) {$\mathsf{S}_{0,3}$};
    \draw[very thick] (A) to (1.8,0) node[right] {3};
    \draw[very thick] (C) to (-4,-1) node[left] {2};
    \draw[very thick] (C) to (-4,1) node[left] {1};
    \draw[very thick, bend left=30] (A) to (B);
    \draw[very thick, bend right=30] (A) to (B);
    \draw[very thick] (B) to (C);
    \draw[very thick, red, dashed] (-2.1,-1) to (-2.1,1);
    \draw[very thick, red, dashed] (-.3,-1) to (-.3,1);
\end{tikzpicture}\hspace{-.1cm}\nonumber\\
&\qquad+\Bigg[2\begin{tikzpicture}[baseline={([yshift=-.5ex]current bounding box.center)},scale=.6]
    \node[shape=circle,draw=black, very thick, fill=gray, inner sep=.5pt] (A) at (.3,0) {$\mathsf{S}_{0,3}$};
    \node[shape=circle,draw=black, very thick, fill=gray, inner sep=.5pt] (B) at (-1.5,.52) {$\mathsf{S}_{1,2}$};
    \draw[very thick] (A) to (1.8,0) node[right] {3};
    \draw[very thick] (A) to (-2.5,-.8) node[left] {2};
    \draw[very thick] (B) to (-2.5,.8) node[left] {1};
    \draw[very thick] (A) to (B);
    \draw[very thick, red, dashed] (-.6,-1) to (-.6,1);
\end{tikzpicture}\hspace{-.1cm}+3\begin{tikzpicture}[baseline={([yshift=-.5ex]current bounding box.center)},scale=.6]
    \node[shape=circle,draw=black, very thick, fill=gray, inner sep=.5pt] (A) at (.3,0) {$\mathsf{S}_{0,4}$};
    \node[shape=circle,draw=black, very thick, fill=gray, inner sep=.5pt] (B) at (-1.5,.52) {$\mathsf{S}_{0,3}$};
    \draw[very thick] (A) to (1.8,0) node[right] {3};
    \draw[very thick] (A) to (-2.5,-.8) node[left] {2};
    \draw[very thick] (B) to (-2.5,.8) node[left] {1};
    \draw[very thick, bend left=20] (A) to (B);
    \draw[very thick, bend right=20] (A) to (B);
    \draw[very thick, red, dashed] (-.6,-1) to (-.6,1);
\end{tikzpicture}-6\begin{tikzpicture}[baseline={([yshift=-.5ex]current bounding box.center)},scale=.6]
    \node[shape=circle,draw=black, very thick, fill=gray, inner sep=.5pt] (A) at (.3,0) {$\mathsf{S}_{0,3}$};
    \node[shape=circle,draw=black, very thick, fill=gray, inner sep=.5pt] (B) at (-1.5,.6) {$\mathsf{S}_{0,3}$};
    \node[shape=circle,draw=black, very thick, fill=gray, inner sep=.5pt] (C) at (-3.2,.6) {$\mathsf{S}_{0,3}$};
    \draw[very thick, bend left=25] (B) to (C);
    \draw[very thick, bend right=25] (B) to (C);
    \draw[very thick] (A) to (1.5,0) node[right] {3};
    \draw[very thick] (A) to (-4.1,-.6) node[left] {2};
    \draw[very thick] (C) to (-4.1,.6) node[left] {1};
    \draw[very thick] (A) to (B);
    \draw[very thick, red, dashed] (-.6,-1.2) to (-.6,1.2);
    \draw[very thick, red, dashed] (-2.35,-1.2) to (-2.35,1.2);
\end{tikzpicture}\nonumber\\
&\qquad\qquad\qquad\qquad-12\begin{tikzpicture}[baseline={([yshift=-.5ex]current bounding box.center)},scale=.6]
    \node[shape=circle,draw=black, very thick, fill=gray, inner sep=.5pt] (A) at (-1.8,0) {$\mathsf{S}_{0,3}$};
    \node[shape=circle,draw=black, very thick, fill=gray, inner sep=.5pt] (B) at (1.8,0) {$\mathsf{S}_{0,3}$};
    \node[shape=circle,draw=black, very thick, fill=gray, inner sep=.5pt] (C) at (0,1.5) {$\mathsf{S}_{0,3}$};
    \draw[very thick] (A) to (B);
    \draw[very thick] (A) to (C);
    \draw[very thick] (B) to (C);
    \draw[very thick] (A) to (-3,0) node[left] {$2$};
    \draw[very thick] (C) to (-3,1.5) node[left] {$1$};
    \draw[very thick] (B) to (3,0) node[right] {$3$};
    \draw[very thick, red, dashed] (-.9,-.5) to (-.9,2);
    \draw[very thick, red, dashed] (.9,-.5) to (.9,2);
\end{tikzpicture}+(1 \leftrightarrow 2)\Bigg]\,. \label{eq:g=1 n=3 resummation in Sgn}
\end{align}
Here we extended the cuts to also cut through the external legs. This changes of course the multiplicities to correctly account for the factor $\frac{1}{n_c!}$. Otherwise the cutting has no effect since the two-point function is precisely the inverse of the integration measure \eqref{eq:on-shell measure}. We can further recognize this formula as building up the full disconnected scattering matrices. For example the part left of the unitarity cut in the second, third and fourth diagram make up $i \TT^{1 \to 1}$ at order $g_\text{s}^2$. We can write the full discontinuity as
\begin{align}
    \mathop{\text{Disc}} i\TT^{2 \to 1}_3&= \begin{tikzpicture}[baseline={([yshift=-.5ex]current bounding box.center)},scale=.5]
    \node[shape=circle,draw=black, very thick, fill=gray, inner sep=.5pt] (A) at (1.5,0) {$i\TT^{1 \to 1}_2$};
    \node[shape=circle,draw=black, very thick, fill=gray, inner sep=.5pt] (B) at (-1.5,0) {$i\TT^{2 \to 1}_1$};
    \draw[very thick] (A) to (3,0) node[right] {3};
    \draw[very thick] (B) to (-2.5,-1) node[left] {2};
    \draw[very thick] (B) to (-2.5,1) node[left] {1};
    \draw[very thick] (A) to (B);
    \draw[very thick, red, dashed] (0,-1) to (0,1);
\end{tikzpicture}\hspace{-.1cm}+\hspace{-.2cm}\begin{tikzpicture}[baseline={([yshift=-.5ex]current bounding box.center)},scale=.5]
    \node[shape=circle,draw=black, very thick, fill=gray, inner sep=.5pt] (B) at (-1.5,0) {$i\TT^{2 \to 2}_2$};
    \node[shape=circle,draw=black, very thick, fill=gray, inner sep=.5pt] (A) at (1.5,0) {$i\TT^{2 \to 1}_1$};
    \draw[very thick] (A) to (3,0) node[right] {3};
    \draw[very thick] (B) to (-2.5,-1) node[left] {2};
    \draw[very thick] (B) to (-2.5,1) node[left] {1};
    \draw[very thick, red, dashed] (0,-1) to (0,1);
    \draw[very thick, bend left=30] (A) to (B);
    \draw[very thick, bend right=30] (A) to (B);
\end{tikzpicture}\hspace{-.1cm}+\hspace{-.2cm}\begin{tikzpicture}[baseline={([yshift=-.5ex]current bounding box.center)},scale=.5]
    \node[shape=circle,draw=black, very thick, fill=gray, inner sep=.5pt] (B) at (-1.5,0) {$i\TT^{2 \to 3}_1$};
    \node[shape=circle,draw=black, very thick, fill=gray, inner sep=.5pt] (A) at (1.5,0) {$i\TT^{3 \to 1}_2$};
    \draw[very thick] (A) to (3,0) node[right] {3};
    \draw[very thick] (B) to (-2.5,-1) node[left] {2};
    \draw[very thick] (B) to (-2.5,1) node[left] {1};
    \draw[very thick, bend left=30] (A) to (B);
    \draw[very thick, bend right=30] (A) to (B);
    \draw[very thick] (A) to (B);
    \draw[very thick, red, dashed] (0,-1) to (0,1);
\end{tikzpicture}\nonumber\\
&\qquad- \hspace{-.2cm}\begin{tikzpicture}[baseline={([yshift=-.5ex]current bounding box.center)},scale=.5]
    \node[shape=circle,draw=black, very thick, fill=gray, inner sep=.5pt] (A) at (-1.5,0) {$i\TT^{2 \to 1}_1$};
    \node[shape=circle,draw=black, very thick, fill=gray, inner sep=.5pt] (B) at (1.5,0) {$i\TT^{1 \to 2}_1$};
    \node[shape=circle,draw=black, very thick, fill=gray, inner sep=.5pt] (C) at (4.5,0) {$i\TT^{2 \to 1}_1$};
    \draw[very thick] (C) to (6,0) node[right] {3};
    \draw[very thick] (A) to (-2.5,-1) node[left] {2};
    \draw[very thick] (A) to (-2.5,1) node[left] {1};
    \draw[very thick] (A) to (B);
    \draw[very thick, bend left=30] (B) to (C);
    \draw[very thick, bend right=30] (B) to (C);
    \draw[very thick, red, dashed] (0,-1) to (0,1);
    \draw[very thick, red, dashed] (3,-1) to (3,1);
\end{tikzpicture}\hspace{-.1cm} - \hspace{-.2cm}\begin{tikzpicture}[baseline={([yshift=-.5ex]current bounding box.center)},scale=.5]
    \node[shape=circle,draw=black, very thick, fill=gray, inner sep=.5pt] (A) at (-1.5,0) {$i\TT^{2 \to 3}_1$};
    \node[shape=circle,draw=black, very thick, fill=gray, inner sep=.5pt] (B) at (1.5,0) {$i\TT^{3 \to 2}_1$};
    \node[shape=circle,draw=black, very thick, fill=gray, inner sep=.5pt] (C) at (4.5,0) {$i\TT^{2 \to 1}_1$};
    \draw[very thick] (C) to (6,0) node[right] {3};
    \draw[very thick] (A) to (-2.5,-1) node[left] {2};
    \draw[very thick] (A) to (-2.5,1) node[left] {1};
    \draw[very thick] (A) to (B);
    \draw[very thick, bend left=30] (A) to (B);
    \draw[very thick, bend right=30] (A) to (B);
    \draw[very thick, bend left=30] (B) to (C);
    \draw[very thick, bend right=30] (B) to (C);
    \draw[very thick, red, dashed] (0,-1) to (0,1);
    \draw[very thick, red, dashed] (3,-1) to (3,1);
\end{tikzpicture}\!\!\\
&=\big((-i \TT)^2\big)^{2 \to 1}_3+\big((-i \TT)^3\big)^{2 \to 1}_3\,.
\end{align}
This is precisely the sum that appears in \eqref{eq:holomorphic unitarity equation} on the RHS and demonstrates unitarity of the $2 \to 1$ process in perturbation theory to this order. In this case, we can at most have two holomorphic cuts.

\paragraph{General case.} The pattern that we described in detail above for the case $(g,n)=(1,3)$ extends to the general case by elementary combinatorics.

We want to extract the full discontinuity of a diagram by an inclusion-exclusion argument over all possible unitarity cuts. A product of $c$ edge discontinuities contributes a factor $(\sqrt{-\omega_e^2})^c$, which is discontinuous for odd $c$ and analytic for even $c$.

As above, summing over all cuts also includes the product of pairs of edge discontinuities, which is not discontinuous and has to be subtracted as in \eqref{eq:g=1, n=3 penultimate diagram}. Consider now a triple of cuts, which leads to an odd power $(\sqrt{-\omega_e^2})^{3}$ and thus should be included in the discontinuity. So far, this discontinuous piece appears $2 \times 3-2^2 \times 3=-6$ times in the sum, since we add it three times from the single cut and subtract it 3 times by subtracting the pair of cuts. The factor of $2$ and $2^2$ arises because we are taking the discontinuity, which is twice \eqref{eq:discontinuous part propagator}. Thus, we should add the triple cuts back, so that the contribution appears $2 \times 3-2^2 \times 3+2^3=2$ times in the final equation for the discontinuity, as desired. In general, we should take the alternating sum over cuts with an odd number of cuts appearing with a plus sign and an even number of cuts appearing with a minus sign. Consider now a system of $c$ cuts. There are $\binom{c}{k}$ ways to choose a subset of $k$ cuts from those $c$. Thus, a product of the discontinuous pieces of $c$ cuts appears then
\be
\sum_{k=1}^c \binom{c}{k} (-1)^{k+1}2^k=1-(-1)^c
\ee
times, i.e.\ we precisely pick up the odd number of edge discontinuities.
Thus the discontinuity of a single diagram is given by summing over all cuts (by default we always extend the cut to also cut through external lines as in \eqref{eq:g=1 n=3 resummation in Sgn}, which only affects the automorphism factors)
\be
\mathop{\text{Disc}} \mathsf{S}_\Gamma(\boldsymbol{\omega})=\!\!\sum_{c_1,\dots,c_{k-1}} \!\!\! (-1)^{k} \frac{\prod_j n_{c_j}! \prod_i |\text{Aut}(\Gamma_i)|}{|\text{Aut}(\Gamma)|} \begin{tikzpicture}[baseline={([yshift=-.5ex]current bounding box.center)},scale=.5]
    \node[shape=circle,draw=black, very thick, fill=gray, inner sep=1pt] (A) at (-2.2,0) {$\Gamma_1$};
    \node[shape=circle,draw=black, very thick, fill=gray, inner sep=1pt] (B) at (0,0) {$\Gamma_2$};
    \node[shape=circle,draw=black, very thick, fill=gray, inner sep=1pt] (C) at (2.2,0) {$\Gamma_3$};
    \node[shape=circle,draw=black, very thick, fill=gray, inner sep=1pt] (D) at (4.4,0) {$\Gamma_k$};
    \draw[very thick] (A) to (-3.5,1);
    \draw[very thick] (A) to (-3.5,-1);
    \node at (5.4,.2) {$\vdots$};
    \node at (-3.2,.2) {$\vdots$};
    \draw[very thick, bend left=45] (A) to (B);
    \draw[very thick, bend left=20] (A) to (B);
    \draw[very thick, bend right=45] (A) to (B);
    \node at (-1.3,-.05) {$\vdots$};
    \node at (-.9,-.05) {$\vdots$};
    \draw[very thick, bend left=45] (B) to (C);
    \draw[very thick, bend left=20] (B) to (C);
    \draw[very thick, bend right=45] (B) to (C);
    \node at (.9,-.05) {$\vdots$};
    \node at (1.3,-.05) {$\vdots$};
    \draw[very thick, bend left=45] (C) to (D);
    \draw[very thick, bend left=20] (C) to (D);
    \draw[very thick, bend right=45] (C) to (D);
    \node at (3.1,-.05) {$\vdots$};
    \node at (3.5,-.05) {$\vdots$};
    \draw[very thick] (D) to (5.7,1);
    \draw[very thick] (D) to (5.7,-1);
    \fill[white] (1.1,-1) rectangle (3.3,1);
    \draw[very thick, red, dashed] (-1.1,-1) to (-1.1,1);
    \draw[very thick, red, dashed] (1.1,-1) to (1.1,1);
    \draw[very thick, red, dashed] (3.3,-1) to (3.3,1);
    \node at (2.2,0) {$\cdots$};
\end{tikzpicture}\, .
\ee
Here $k-1 \ge 1$ is the number of cuts applied on the diagram.
The automorphism groups $\text{Aut}(\Gamma_i)$ are subgroups of $\text{Aut}(\Gamma)$. The combinatorial prefactor counts the permutations of the intermediate legs that lead to different individual diagrams $\Gamma_i$ as in \eqref{eq:full S matrix from the worldsheet}.
We now sum over all graphs $\Gamma \in \mathcal{G}_{g,n}$ for fixed $(g,n)$. For each diagram, the sum over all cuts produces all ways of factoring $\Gamma$ into ordered sequences of subgraphs $\Gamma_1,\dots,\Gamma_k$ with $k \ge 2$. The combinatorial prefactor ensures that after summing over $\Gamma$, each subgraph $\Gamma_i$ runs independently over all connected graphs compatible with the external and intermediate leg assignments. The constraint that all cuts are distinct means that none of the $\Gamma_i$ can represent the free propagation \eqref{eq:S identity contribution worldsheet}; equivalently, each factor contributes to the $\TT$-matrix rather than the full $\mathbb{S}$-matrix. This is because the sum over graphs $\Gamma$ already runs over all topologies, so that restricting to a given set of cuts simply partitions each topology into all compatible factorizations; the completeness of the graph sum on each side of the cut then ensures that each factor independently reconstructs the full $\TT$-matrix at the appropriate order.

Summing over $(g,n)$ and including disconnected contributions then assembles the full $\TT$-matrix at each order in $g_\text{s}$. The alternating sign from the inclusion-exclusion gives
\be
\mathop{\text{Disc}} i\TT=\sum_{k\ge 2} (-1)^k (i \TT)^k=\sum_{c=1}^\infty (-i \TT)^{c+1}\,,
\ee
valid order-by-order in perturbation theory. Since $\mathop{\text{Disc}} \TT=2i \Im \TT$, this reproduces \eqref{eq:holomorphic unitarity equation} and gives a direct proof of unitarity of our intersection number expression \eqref{eq:c=1 momentum space intersection numbers}.

\subsection{The tree level amplitude rewritten} \label{subsec:tree level rewriting}
The formula \eqref{eq:c=1 momentum space intersection numbers} as a sum over stable graphs is intuitive to work with and makes the duality with the matrix integral transparent. However, it doesn't reflect the starting point of the worldsheet integrals \eqref{eq:c=1 S-matrix worldsheet definition} which only include an integral over the full moduli space. It also obscures some of the analytic and geometric features.
It is thus useful to rewrite the formula in terms of a single intersection number on $\bM_{g,n}$ that doesn't involve a sum over stable graphs. It makes the geometric interpretation of the formula much more transparent. This is relatively straightforward for $g=0$, where one does not have to deal with loop momenta. For higher genera, such a rewriting will require us to pass to a more general moduli space that we will discuss in section~\ref{subsec:r spin curves}.

\paragraph{Boundary classes.} We can write the sum over stable graphs in terms of a single intersection number that also involves pushforwards of cohomology classes from the boundary of moduli space. To explain this, let us recall some basic facts about the topology of $\bM_{g,n}$. The Deligne-Mumford compactification $\bM_{g,n}$ describes stable nodal surfaces. In particular, $\bM_{g,n}$ contains boundary divisors describing surfaces with exactly one node, corresponding to stable graphs with exactly one edge. There are two kinds: separating divisors where the surface disconnects when we remove the nodal point and the non-separating divisor where the surface stays connected upon removal of the node (but with genus one less). The former is labelled by the genus $h$ of one component and the index set $I \subset \{1,\dots,n\}$ of the labels of the external legs that end up on one side. We denote the corresponding divisor by $\mathscr{D}_{h,I}=\mathscr{D}_{g-h,I^c} \cong \bM_{h,|I|+1} \times \bM_{g-h,|I^c|+1}$.\footnote{For the case $g \in 2 \NN$ and $n=0$, the divisor $\mathscr{D}_{g/2,\emptyset}$ has an extra $\ZZ_2$ automorphism and is isomorphic to $(\bM_{g/2,1} \times \bM_{g/2,1})/\ZZ_2$ with the $\ZZ_2$ exchanging both copies.} The isomorphism follows from the fact that we can still vary the moduli of both parts of the surface. The latter exists only for $g \ge 1$ and is denoted by $\mathscr{D}_\text{irr} \cong \bM_{g-1,n+2}/\ZZ_2$. The $\ZZ_2$ exchanges the two nodal points. For these boundary divisors, we have the inclusion maps $\xi_{h,I}:\mathscr{D}_{h,I} \longrightarrow \bM_{g,n}$ and $\xi_\text{irr}:\mathscr{D}_\text{irr} \longrightarrow \bM_{g,n}$. We can use these maps in particular to push forward the psi-classes $\psi_\bullet$ and $\psi_\circ$ of the two nodes of the nodal surface. These pushforwards are denoted by $(\xi_{h,I})_*(\alpha)$ or $(\xi_\text{irr})_*(\alpha)$.

In physical terms, such boundary classes should be thought of as being localized entirely on the relevant divisor, i.e.\ in physical terms they represent contact terms. In fact, in terms of differential forms, this pushforward means that we are wedging the form $\alpha$ with a delta-function localized on the relevant divisor. In particular all analytic features of the amplitude that come from the boundary of moduli space originate in such terms.
\paragraph{Rewriting as exponential.} Let us start with the genus 0 expression that doesn't involve loop integrals. On $\bM_{0,n}$, we have the following identity, where $\alpha_p \in \CC[[\psi_\bullet,\psi_\circ]]$ is an arbitrary combination of $\psi_\bullet$ and $\psi_\circ$ classes (with $\alpha_p=\alpha_{-p}$),
\be
\sum_{\Gamma \in \mathcal{G}_{0,n}} \frac{1}{|\text{Aut}(\Gamma)|} (\xi_\Gamma)_* \prod_{e \in \mathcal{E}_\Gamma} \frac{1-\ex^{-(\psi_\bullet+\psi_\circ)\alpha_{p_e}}}{\psi_\bullet+\psi_\circ}=\exp\bigg(\frac{1}{2}\sideset{}{'}\sum_{I \subset [n]} (\xi_I)_*(\alpha_{p_I})\bigg) \label{eq:tree level sum over graphs identity}
\ee
On the LHS, we take the sum over all stable graphs with an arbitrary edge factor described by the class $\alpha_{p_e}$, depending on the edge momentum $p_e$. Here, $\xi_\Gamma: \bM_\Gamma \longrightarrow \bM_{g,n}$ is the generalization of the inclusion of boundary divisors to any boundary stratum described by the stable graph $\Gamma$. We can rewrite this in terms of just a single exponential and the pushforward of boundary classes from the boundary divisors, which for genus 0 are just labelled by subsets $I \subset [n]=\{1,\dots,n\}$. We wrote $\xi_I \equiv \xi_{0,I}=\xi_{0,I^c}$. The factor of $\frac{1}{2}$ removes the double counting from summing over $I$ and $I^c$ separately. Stability of the surfaces described by the boundary divisor imply also that $2 \le |I| \le n-2$, which is denoted by the prime next to the sum. This identity arises from repeated application of the intersection excess formula \cite{Fulton:intersection}
\be
(\xi_I)^*(\xi_I)_*(\alpha)=\alpha \, c_1(\mathscr{N})=(-\psi_\bullet-\psi_\circ)\alpha\,,
\ee
where $\mathscr{N}$ is the normal bundle of the boundary divisor, whose first Chern class is given by $-\psi_\bullet-\psi_\circ$.
We refer to appendix~\ref{app:algebraic geometry} for some further details. This implies that we can rewrite \eqref{eq:c=1 continuous version intersection numbers genus 0} into a single exponential as follows,
\begin{multline}
    \hat{\mathsf{s}}_{0,n}(\boldsymbol{p})=\int_{\bM_{0,n}}\exp\bigg(-\sum_{k=1}^\infty \frac{B_k \kappa_k}{k\, k!} -\sum_{i=1}^n p_i^2 \psi_i\\+\frac{1}{2} \sideset{}{'}\sum_{I \subset [n]} (\xi_I)_*\mathcal{F}(\{p_I\},-\psi_\bullet-\psi_\circ)\bigg)\,. \label{eq:genus 0 rewriting discrete}
\end{multline}
Here,
\be
\mathcal{F}(p,x)=
\frac{1}{x}\log \sum_{d=0}^\infty \frac{B_{2d}(p)}{d!} x^d =\frac{1}{x} \log \bigg(\frac{\partial_p}{\ex^{\partial_p}-1}\, \ex^{x p^2} \bigg)\in \CC[[x]]\,. \label{eq:mathcal F definition}
\ee
We used the following identity, valid as a power series in $x$,
\be
\sum_{d=0}^\infty \frac{B_{2d}(p)x^d}{d!}=\frac{\partial_p}{\ex^{\partial_p}-1}\,\ex^{x p^2}\,. \label{eq:Bernoulli polynomial edge factor identity}
\ee
We can prove this as follows. All of the following equalities are understood as formal power series in $x$, i.e.\ in the ring $\mathbb{Q}[p][[x]]$.
Recall the standard expansion of Bernoulli polynomials in terms of Bernoulli numbers
\be
B_d(p)=\sum_{j=0}^d \binom{d}{j} B_j\, p^{d-j}=\sum_{j=0}^d \frac{B_j\,\partial_p^j}{j!} p^d=\sum_{j=0}^\infty \frac{B_j\,\partial_p^j}{j!} p^d=\frac{\partial_p}{\ex^{\partial_p}-1} p^d\,,
\ee
where we used the generating function of Bernoulli numbers in the last step. Thus we conclude
\be
\sum_{d=0}^\infty \frac{B_{2d}(p)x^d}{d!}
 =\frac{\partial_p}{\ex^{\partial_p}-1}\sum_{d=0}^\infty \frac{(x p^2)^d}{d!}
 =\frac{\partial_p}{\ex^{\partial_p}-1}\,\ex^{x p^2}\,,
\ee
as claimed in \eqref{eq:Bernoulli polynomial edge factor identity}.
From \eqref{eq:genus 0 rewriting discrete}, it is straightforward to pass to the physical (non-discretized) amplitude with the result
\begin{multline}
    \hat{\mathsf{S}}_{0,n}(\boldsymbol{\omega})=\int_{\bM_{0,n}}\exp\bigg(-\sum_{k=1}^\infty \frac{B_k \kappa_k}{k\, k!} +\frac{1}{4}\sum_{i=1}^n \omega_i^2 \psi_i\\+\frac{1}{2} \sideset{}{'}\sum_{I \subset [n]} (\xi_I)_*\mathcal{F}(-\tfrac{i}{2} \omega_I\, \sgn(\Re \omega_I),-\psi_\bullet-\psi_\circ)\bigg)\,. \label{eq:genus 0 rewriting first form}
\end{multline}
One can simplify this a bit further thanks to the following identity valid in $\mathrm{H}^2(\bM_{0,n})$,
\be
\sum_{i=1}^n \omega_i^2 \psi_i-\frac{1}{2} \sideset{}{'} \sum_{I \subset [n]} \omega_I^2 \delta_{I}=0\,,
\ee
with $\delta_{I}=(\xi_{I})_*(1)$ the Poincar\'e dual of the boundary divisor. This combination is tantalizingly similar to the class that appears in a similar limit of the minimal string \cite{Eberhardt:2023rzz}. It vanishes because of momentum conservation and the relation\footnote{This identity is simple to derive as follows. $\psi_i=c_1(\LL_i)$, where $\LL_i$ is the line bundle over $\bM_{g,n}$ whose fiber is the cotangent space at the $i$-th marked point. We can easily write down a meromorphic section of $\LL_i$ for genus 0. Pick two marked points $z_j$ and $z_k$ with $i,j,k$ all different. Then $\alpha=\big(\frac{1}{z-z_j}-\frac{1}{z-z_k}\big)\d z\big|_{z_i}$ is a meromorphic section. The first Chern class is given by the zero divisor minus the pole divisor of the section. Clearly, the section doesn't have zeros or poles on the bulk of moduli space. So we should investigate what happens on the boundary divisors $\mathscr{D}_{I}$, where the surface splits into two parts. If $j$ and $k$ end up on different components of the nodal surface (i.e.\ exactly one of $j$ or $k$ is in $I$), then $\alpha$ is non-zero. For $j$ and $k$ both in $I$, $\big(\frac{1}{z-z_j}-\frac{1}{z-z_k}\big)\d z$ vanishes identically on the component of the surface not containing $j$ and $k$. Therefore, the section $\alpha$ vanishes on the divisors where $i \not \in I$ and $j,k \in I$, or $i \in I$ and $j,k \not\in I$. Since $I$ and $I^c$ parametrize the same divisor, we just have to keep one of the cases, which leads to the claimed formula.}
\be
\psi_i=\sideset{}{'}\sum_{I,\, j,k \not \in I,\, i \in I} \delta_I
\ee
on $\mathrm{H}^2(\bM_{0,n})$. Thus we can write the alternative formula
\be
    \hat{\mathsf{S}}_{0,n}(\boldsymbol{\omega})=\int_{\bM_{0,n}}\exp\bigg(-\sum_{k=1}^\infty \frac{B_k \kappa_k}{k\, k!}+\frac{1}{2} \sideset{}{'}\sum_{I \subset [n]} (\xi_I)_*\mathcal{G}(-\tfrac{i}{2} \omega_I\, \sgn(\Re \omega_I),-\psi_\bullet-\psi_\circ)\bigg) \label{eq:genus 0 rewriting}
\ee
with 
\be
\mathcal{G}(p,x)=
-p^2+\mathcal{F}(p,x) =\frac{1}{x} \log \bigg(\ex^{-xp^2}\,\frac{\partial_p}{\ex^{\partial_p}-1}\, \ex^{x p^2} \bigg)\in \CC[[x]]\,.
\ee
$\mathcal{G}(p,x)$ has the property that the coefficient of $x^d$ is a polynomial of degree $d+1$ in $p$. This follows from the observation that taking $d$ derivatives of $\ex^{x p^2}$ in $p$ will bring down a polynomial prefactor of order $d$ in both $x$ and $p$, while the Gaussian parts cancel.\footnote{In fact, that polynomial is a Hermite polynomial as follows from Rodrigues' formula.} 
Since $(\xi_I)_*$ raises the cohomological degree by 2, this formula makes it manifest that $\mathsf{S}_{0,n}(\boldsymbol{\omega})$ is a piecewise polynomial of degree $n-3$ (as opposed to the naive $2n-6$ that we observed above).

In particular, for the $n-1 \to 1$ scattering where $\omega_1,\dots,\omega_{n-1} \ge 0$, we have $\omega_I \ge 0$ for $n \not \in I$ and $\omega_I \le 0$ for $n \in I$. Thus there are no case distinction and the result is polynomial. It also has to be symmetric in the $\omega_1,\dots,\omega_{n-1}$. With the help of the dilaton equation \eqref{eq:dilaton equation}, we can uniquely recursively determine it to be \eqref{eq:S0n result}.

\subsection{An intersection number on the moduli space of \texorpdfstring{$r$}{r}-spin curves} \label{subsec:r spin curves}
We now discuss a similar rewriting of the Feynman rules at any genus in terms of a single intersection number. This will require us to pass to a different moduli space that also incorporates the loop momenta in \eqref{eq:c=1 momentum space intersection numbers}. Since the loop momenta are continuous variables, this suggests that the appropriate moduli space is the universal Jacobian. This is a rather subtle object and we will instead use a much better studied moduli space: the moduli space of $r$-spin curves $\bM^r_{g,n}$. This moduli space will furthermore discretize the momentum integrals and we will take a limit $r \to \infty$ in the end. 
\paragraph{Moduli space of $r$-spin curves.} This moduli space parametrizes jointly curves and particular line bundles. More precisely, it parametrizes objects $(\Sigma,z_1,\dots,z_n,L)$, where $(\Sigma,z_1,\dots,z_n)$ is a stable curve and $L$ is a line bundle with $L^{\otimes r} \cong \mathcal{O}$, i.e.\ an $r$-th root of the trivial line bundle.\footnote{This moduli space is further generalized in the literature, where one also considers $r$-th roots of line bundles other than the trivial one, but we will not need this generalization.} In other words, we are giving the additional data of a flat $\ZZ_r$-bundle on the Riemann surface. For $r \to \infty$, this intuitively becomes a $\mathrm{U}(1)$-bundle and should recover the universal Jacobian. It is however much more controlled to work at finite $r$. Physically, the finite-$r$ amplitudes roughly speaking both discretize and compactify the Euclidean time direction in target space, which gets turned into something akin to a spin chain. 

The compactification of this moduli space works very analogously to the case of the ordinary moduli space of curves and was constructed in \cite{Jarvis:2000, Chiodo:2008}. Over a smooth surface, there are $r^{2g}$ possible choices of such $r$-th roots. One then modifies the orbifold structure of $\bM^r_{g,n}$ appropriately and includes a $\ZZ_r$ stabilizer group of the generic $r$-spin curve that multiplies the fiber by overall roots of unity, as well as an additional $\ZZ_r$ stabilizer for every node. With this orbifold structure, the forgetful map
\be 
\pi: \bM^r_{g,n} \longrightarrow \bM_{g,n}
\ee
is then an $r^{2g-1}$-fold covering in the orbifold sense. 

\paragraph{Discretizing the momentum integrals.} Let $a_i \in \ZZ$ and identify the momenta by $p_i=\frac{a_i}{r}$. We can discretize the momentum integrals in \eqref{eq:c=1 momentum space intersection numbers} and define the regulated amplitude as $\mathsf{s}_{g,n}^r(\frac{1}{r}\boldsymbol{a})=\int_{\bM_{g,n}} \widetilde{\Omega}_{g,n}^r(\boldsymbol{a})$, with 
\begin{multline} 
\widetilde{\Omega}^r_{g,n}(\boldsymbol{a})=\sum_{\Gamma \in \mathcal{G}_{g,n}} \sum_{w \in \mathcal{W}_{\Gamma,r}} \frac{1}{|\Aut(\Gamma)|} \frac{1}{r^{h_1(\Gamma)}} (\xi_\Gamma)_* \bigg[ \prod_{v \in \mathcal{V}_\Gamma}\ex^{-\sum_k\frac{B_{k}\kappa_{k}}{k \, k!}}\prod_{i=1}^n \ex^{-\frac{a_i^2 \psi_i}{r^2}} \\
    \times\!\prod_{e=(\bullet,\circ) \in \mathcal{E}_\Gamma} \sum_{d=0}^\infty \frac{B_{2d+2}(\frac{a_e}{r})(-\psi_\bullet-\psi_\circ)^d}{(d+1)!}\bigg]\,.  \label{eq:c=1 discretized Feynman rules}
\end{multline}
Here, we used the convention employed in the mathematical literature. $w \in \mathcal{W}_{\Gamma,r}$ runs over so-called weightings of a stable graph. It associates to every half-edge an element in $\{0,\dots,r-1\} \cong \ZZ/r\ZZ$. For an external leg, the weighting agrees with the external $a_i \bmod r$. For any vertex, the weighting labels have to be conserved mod $r$ and for every edge, the two half-legs have to satisfy $a+a' \in r \ZZ$. In other words, a weighting is precisely a discretized version of momentum assignments. The existence of weightings of course requires the external momenta to be conserved mod $r$ and hence overall momentum conservation is naturally realized in \eqref{eq:c=1 discretized Feynman rules}. The factor $r^{-h_1(\Gamma)}$ in \eqref{eq:c=1 discretized Feynman rules} is present, since $r^{-1}$ is the step width in the discretization of the $L=h_1(\Gamma)$ loop integrals in \eqref{eq:c=1 momentum space intersection numbers}.

\paragraph{Lifting to the moduli space of $r$-spin curves.} We claim that $\widetilde{\Omega}_{g,n}^r(\boldsymbol{a})$ is a pushforward of the form $\widetilde{\Omega}_{g,n}^r(\boldsymbol{a})=\pi_* \Omega^r_{g,n}$ for $\Omega^r_{g,n} \in \mathrm{H}^\bullet(\bM_{g,n}^r)$ a cohomology class on the moduli space of $r$-spin curves. This is fairly simple, since the covering $\pi$ over each stratum has degree $r^d$ with
\begin{align} 
d&=\sum_v (2g_v-1)\\
&=2\sum_v g_v-|\mathcal{V}_\Gamma| \\
&=2(g-h_1(\Gamma))-\big(|\mathcal{E}_\Gamma|+1-h_1(\Gamma)\big) \\
&=2g-1-h_1(\Gamma)-|\mathcal{E}_\Gamma|\ .
\end{align}
Here we used Euler's formula $h_1(\Gamma)=|\mathcal{E}_\Gamma|-|\mathcal{V}_\Gamma|+1$ for a connected graph. Therefore, we simply have to divide by the degree, which gives
\begin{multline} 
\Omega^r_{g,n}(\boldsymbol{a})=\frac{1}{r^{2g-1}}\sum_{\Gamma \in \mathcal{G}_{g,n}} \sum_{w \in \mathcal{W}_{\Gamma,r}} \frac{r^{|\mathcal{E}_\Gamma|}}{|\Aut(\Gamma)|}  (\xi_{\Gamma,w})_* \bigg[ \prod_{v \in \mathcal{V}_\Gamma}\ex^{-\sum_k\frac{B_{k}\kappa_{k}}{k \, k!}}\prod_{i=1}^n \ex^{-\frac{a_i^2 \psi_i}{r^2}} \\
    \times\!\prod_{e=(\bullet,\circ) \in \mathcal{E}_\Gamma} \sum_{d=0}^\infty \frac{B_{2d+2}(\frac{a_e}{r})(-\psi_\bullet-\psi_\circ)^d}{(d+1)!}\bigg]\,.  \label{eq:c=1 graph sum r spin moduli space}
\end{multline}
Here, the embedding $\xi_{\Gamma,w}$ depends on both the graph and the weighting $w$.

\paragraph{Exponentiation.} At this point, we can exponentiate the formula as in the degree 0 case. The only slight wrinkle is that the first Chern class of the normal bundle to the boundary divisors in $\bM_{g,n}^r$ is given by $\mathscr{N}=-\frac{\psi_\bullet+\psi_\circ}{r}$ because of the extra $\ZZ_r$ orbifold group that every node carries. This will lead to some additional factors of $r$. We claim that
\be 
\Omega^r_{g,n}(\boldsymbol{a})=\frac{1}{r^{2g-1}} \exp\bigg(-\sum_k \frac{B_k \kappa_k}{k\, k!}-\sum_{i=1}^n \frac{a_i^2 \psi_i}{r^2}+r \sum_{q=0}^{r-1} (\xi_q)_* \mathcal{F}(\tfrac{q}{r},-\psi_\bullet-\psi_\circ)\bigg)\ , \label{eq:c=1 integrand discretized and exponentiated}
\ee
where $\mathcal{F}$ was defined in \eqref{eq:mathcal F definition}. Here $\xi_q$ is the inclusion of the union of all boundary divisors where `momentum' $q$ is associated to the node. We sum over $q$ to account for every possible momentum flowing through the node. In particular, this implements the loop momentum integral. To go back from the exponentiated class \eqref{eq:c=1 integrand discretized and exponentiated} to the graph sum \eqref{eq:c=1 graph sum r spin moduli space}, one applies the same logic as in Appendix~\ref{app:algebraic geometry}, but on the moduli space of $r$-spin curves. The last term in the exponent has a factor of $r$ to reproduce the factor $r^{|\mathcal{E}_\Gamma|}$ in \eqref{eq:c=1 graph sum r spin moduli space}: every time it is pulled back (i.e.\ a new node and therefore edge in the stable graph is generated), we get an additional factor of $r$. If the same term is pulled back $k$ times, one applies the excess intersection formula \eqref{eq:excess intersection} as in \eqref{eq:pushforward product}. Because the first Chern class of the normal bundle has an extra $\frac{1}{r}$, we get the final power $r^{-(k-1)+k}=r$, which again matches with $r^{|\mathcal{E}_\Gamma|}$ for the single node.

Therefore, we can write the compact formula
\be 
\hat{\mathsf{s}}_{g,n}(\boldsymbol{p})=\lim_{r \to \infty} \frac{1}{r^{2g-1}}\int_{\bM^r_{g,n}} \exp\bigg(-\sum_k \frac{B_k \kappa_k}{k\, k!}-\sum_{i=1}^n \frac{a_i^2 \psi_i}{r^2}+r \sum_{q=0}^{r-1} (\xi_q)_* \mathcal{F}(\tfrac{q}{r},-\psi_\bullet-\psi_\circ)\bigg)\ .
\ee
\paragraph{The $r \to \infty$ limit.} The previous formulas suggest that we should consider in some sense the moduli space described by the $r \to \infty$ limit of $\bM_{g,n}^r$. It describes stable surfaces, together with choices of \emph{flat} $\mathrm{U}(1)$ bundles. For a smooth surface, this space agrees with the universal Jacobian $\mathcal{J}_{g,n}^0$. A generic such curve by definition has a $\RR/\ZZ$ stabilizer. 
Whenever a puncture is pinched, we lose the corresponding cycle and the real dimension of the moduli space drops by 1, but the node introduces another $\RR/\ZZ$ stabilizer. 
We leave the development of this perspective for the future.

\subsection{Factorization of \texorpdfstring{$c=1$}{c=1} strings}
\label{subsec:higher genus rewriting}
At genus 0, the sum over stable graphs could be rewritten as a single intersection number \eqref{eq:genus 0 rewriting} by exponentiating the separating boundary classes, both for the discrete and the non-discrete amplitudes. At higher genus, achieving a similar exponentiation required us to pass to a more general moduli space in order to incorporate the loop momenta. 

We now develop an alternative characterization of the integrand on moduli space by analyzing its factorization properties. This captures the recursive structure of the amplitudes through factorization axioms that encode how the integrand degenerates at the boundary of moduli space. This is also the structure underlying the topological recursion of the dual matrix integral described in section~\ref{sec:matrix integral} via \cite{Dunin-Barkowski:2012kbi}. The $c=1$ string integrand however does not constitute a cohomological field theory (CohFT) \cite{Givental:2001,Teleman:2012}. The $c=1$ string has the infinite-dimensional state space $V=\mathrm{L}^2(\RR/\ZZ)$, which already takes it outside the scope of that classification.

\paragraph{Pullbacks.} The integrands can be viewed as a collection of cohomology classes $\Omega_{g,n} \in \mathrm{H}^\bullet(\bM_{g,n},\CC) \otimes (V^*)^{\otimes n}$, subject to symmetry in the external arguments and compatibility with gluing. Here, $V=\mathrm{L}^2(\RR/\ZZ)$ and
\begin{multline}
\Omega_{g,n}(\boldsymbol{p})=\sum_{\Gamma \in \mathcal{G}_{g,n}}\frac{\delta_\ZZ(\sum_i p_i)}{|\text{Aut}(\Gamma)|}\int_{\RR/\ZZ} \d^L \boldsymbol{q}\,  (\xi_\Gamma)_*\bigg[ \prod_{v \in \mathcal{V}_\Gamma}\ex^{-\sum_k\frac{B_{k}\kappa_{k}}{k \, k!}}\prod_{i=1}^n \ex^{-p_i^2 \psi_i} \\
    \times\!\prod_{e=(\bullet,\circ) \in \mathcal{E}_\Gamma} \sum_{d=0}^\infty \frac{B_{2d+2}(\{p_e\})(-\psi_\bullet-\psi_\circ)^d}{(d+1)!}\bigg]\ . \label{def:Omegagn}
\end{multline}
This is just the integrand of \eqref{eq:c=1 momentum space intersection numbers}, together with the Dirac comb delta function. The pushforward $\xi_\Gamma:\bM_\Gamma \longrightarrow \bM_{g,n}$ includes the boundary stratum labelled by the graph into the full moduli space, so that $\mathsf{s}_{g,n}(\boldsymbol{p})=\int_{\bM_{g,n}} \Omega_{g,n}(\boldsymbol{p})$.\footnote{With this definition $\Omega_{g,n}$ does not belong to $\mathrm{H}^\bullet(\bM_{g,n},\CC) \otimes (V^*)^{\otimes n}$ since it is not periodic in the external momenta due to the $\psi$-class contributions in \eqref{def:Omegagn}. We can instead consider the renormalized integrand $\ex^{\sum_{i=1}^n p_i^2 \psi_i} \Omega_{g,n}(\boldsymbol{p}) \in \mathrm{H}^\bullet(\bM_{g,n},\CC) \otimes (V^*)^{\otimes n}$. This merely corresponds to a renormalization of the external operators. We thus continue to work with $\Omega_{g,n}$ in order to not introduce a new object.} The compatibility conditions take the following form. For every stable separating divisor $\mathscr{D}_{h,I} \subset \bM_{g,n}$ with gluing map $\xi_{h,I}:\bM_{h,|I|+1} \times \bM_{g-h,|I^c|+1} \longrightarrow \bM_{g,n}$,
\begin{align}
    \xi_{h,I}^*\,\Omega_{g,n}(\boldsymbol{p}) &= \int_{\RR/\ZZ}\d q\  \mathcal{P}(q,-\psi_\bullet-\psi_\circ) \ \Omega_{h,|I|+1}(\boldsymbol{p}_I,q)\otimes \Omega_{g-h,|I^c|+1}(\boldsymbol{p}_{I^c},-q)\,. \label{eq:separating factorization}
\end{align}
For the non-separating divisor $\mathscr{D}_\mathrm{irr} \subset \bM_{g,n}$ with gluing map $\xi_\mathrm{irr}:\bM_{g-1,n+2} \longrightarrow \bM_{g,n}$,
\begin{align}
    \xi_\mathrm{irr}^*\,\Omega_{g,n}(\boldsymbol{p}) &= \int_{\RR/\ZZ}\d q\  \mathcal{P}(q,-\psi_\bullet-\psi_\circ) \ \Omega_{g-1,n+2}(\boldsymbol{p},q,-q)\,. \label{eq:non-separating factorization}
\end{align}
Here the propagator is built from the edge factor \eqref{eq:c=1 momentum space intersection numbers} dressed with the external-leg exponentials:
\be
\mathcal{P}(q,-\psi_\bullet-\psi_\circ)=\ex^{q^2(\psi_\bullet+\psi_\circ)}\sum_{d=0}^\infty \frac{B_{2d}(\{q\}) (-\psi_\bullet-\psi_\circ)^d}{d!}\,. \label{eq:propagator}
\ee
These properties express the factorization of string amplitudes on the boundary of moduli space (see the boundary class discussion in subsection~\ref{subsec:tree level rewriting}). The integral over $q \in \RR/\ZZ$ sums over the internal momentum flowing through the node, with momentum conservation imposed at each vertex.

\paragraph{Derivation of the factorization properties.} We now derive the factorization properties \eqref{eq:separating factorization} and \eqref{eq:non-separating factorization} from the stable graph sum \eqref{eq:c=1 momentum space intersection numbers}. We present the argument for separating factorization in detail; the non-separating case is analogous and we indicate the minor differences at the end.

Fix a separating divisor $\mathscr{D}_{h,I} \subset \bM_{g,n}$ with gluing map $\xi_{h,I}$. We split the graph sum according to whether $\Gamma$ contains a separating edge $e$ of type $(h,I)$.
Every graph with such an edge decomposes uniquely as $\Gamma=\Gamma_\text{L} \cup_e \Gamma_\text{R}$ with $\Gamma_\text{L} \in \mathcal{G}_{h,|I|+1}$ and $\Gamma_\text{R} \in \mathcal{G}_{g-h,|I^c|+1}$, and the automorphism factors satisfy $|\Aut(\Gamma)|=|\Aut(\Gamma_\text{L})|\,|\Aut(\Gamma_\text{R})|$.\footnote{For $(h,I)=(g-h,I^c)$ (which requires $n=0$ and $g=2h$), there is an extra $\ZZ_2$ exchanging the two sides, absorbed by $|\Aut(\mathscr{D}_{h,\emptyset})|=2$.} The integrand factors between the two subgraphs, and the separating edge contributes the factor, see \eqref{eq:c=1 momentum space intersection numbers}
\be
\sum_{d=1}^\infty \frac{B_{2d}(\{p_e\})(-\psi_\bullet-\psi_\circ)^{d-1}}{d!}=\frac{\ex^{-q^2(\psi_\bullet+\psi_\circ)}\mathcal{P}(q,-\psi_\bullet-\psi_\circ)-1}{-\psi_\bullet-\psi_\circ}\,.
\ee
The gluing map factors as $\xi_\Gamma=\xi_{h,I} \circ (\xi_{\Gamma_\text{L}}\times \xi_{\Gamma_\text{R}})$. Summing over all such pairs $(\Gamma_\text{L},\Gamma_\text{R})$ reconstructs the graph sums for $\Omega_{h,|I|+1}$ and $\Omega_{g-h,|I^c|+1}$, giving for this contribution
\begin{multline}
\Omega_{g,n}^{(1)}= (\xi_{h,I})_*\bigg[\int_0^1 \d q\  \frac{\mathcal{P}(q,-\psi_\bullet-\psi_\circ)-\ex^{q^2(\psi_\bullet+\psi_\circ)}}{-\psi_\bullet-\psi_\circ}\\ \times \Omega_{h,|I|+1}(\boldsymbol{p}_I,q)\otimes \Omega_{g-h,|I^c|+1}(\boldsymbol{p}_{I^c},-q)\bigg]\,. \label{eq:Omega edge}
\end{multline}
We compensated the $\psi$-classes on the two factors by a corresponding factor $\ex^{(\psi_\bullet+\psi_\circ)q^2}$.
Applying $\xi_{h,I}^*$ and using the self-intersection formula $\xi_{h,I}^*(\xi_{h,I})_*\beta = (-\psi_\bullet-\psi_\circ)\,\beta$ (see the excess intersection formula in subsection~\ref{subsec:tree level rewriting}), we obtain
\begin{align}
\xi_{h,I}^*\,\Omega_{g,n}^{(1)} &= \int_0^1 \d q\ \big(\mathcal{P}(q,-\psi_\bullet-\psi_\circ)-\ex^{q^2(\psi_\bullet+\psi_\circ)}\big)\, \Omega_{h,|I|+1}(\boldsymbol{p}_I,q)\otimes \Omega_{g-h,|I^c|+1}(\boldsymbol{p}_{I^c},-q)\,. \label{eq:pullback with edge}
\end{align}
For the remaining graphs (those without a separating edge of type $(h,I)$), the pullback $\xi_{h,I}^*(\xi_\Gamma)_*$ produces transverse intersections (no excess factor) and the vertex factor restricts as
\be
\xi_{h,I}^*\bigg[\ex^{-\sum_k \frac{B_k \kappa_k}{k\,k!}-\sum_i p_i^2 \psi_i}\bigg] = \ex^{-\sum_k \frac{B_k}{k\, k!}(\kappa_{\text{L},k}+\kappa_{\text{R},k})-\sum_{i \in I} p_i^2 \psi_{\text{L},i}-\sum_{i \in I^c}p_i^2 \psi_{\text{R},i}}\,. \label{eq:trivial graph pullback}
\ee
The result is simply the product of the contributions to $\Omega_{h,|I|+1}$ and $\Omega_{g-h,|I^c|+1}$ but without the nodal leg factors $\ex^{-q^2(\psi_\bullet+\psi_\circ)}$, which can be compensated by multiplying by $\ex^{q^2(\psi_\bullet+\psi_\circ)}$.

When adding the two contributions, the first piece contributes $\mathcal{P}(q,-\psi_\bullet-\psi_\circ)-\ex^{q^2(\psi_\bullet+\psi_\circ)}$ times $\Omega_{h,|I|+1}\otimes \Omega_{g-h,|I^c|+1}$ from \eqref{eq:pullback with edge}, while the second piece contributes $+\ex^{q^2(\psi_\bullet+\psi_\circ)}$ times the same product. The two exponential terms cancel, leaving the full pullback
\begin{multline}
\xi_{h,I}^*\,\Omega_{g,n}(\boldsymbol{p}) = \int_0^1 \d q\ \mathcal{P}(q,-\psi_\bullet-\psi_\circ)\,  \Omega_{h,|I|+1}(\boldsymbol{p}_I,q)\otimes \Omega_{g-h,|I^c|+1}(\boldsymbol{p}_{I^c},-q)\,. \label{eq:full sep pullback}
\end{multline}
For the non-separating degeneration, the argument is analogous and one obtains \eqref{eq:non-separating factorization}. We have also verified the properties \eqref{eq:separating factorization} and \eqref{eq:non-separating factorization} directly with the help of \texttt{admcycles} \cite{admcycles} and various numerical choices of external momenta for all stable cases $(g,n)$ with $2g-2+n \le 5$. 
\paragraph{Consistency with the exponential formula.} We can verify \eqref{eq:full sep pullback} also directly at genus 0 from the expression \eqref{eq:genus 0 rewriting first form}. Write
\be
\Omega_{0,n}(\boldsymbol{p})=\exp\bigg(-\sum_{k=1}^\infty \frac{B_k \kappa_k}{k\, k!} -\sum_{i=1}^n p_i^2 \psi_i+\frac{1}{2} \sideset{}{'}\sum_{I \subset [n]} (\xi_I)_*\mathcal{F}(\{p_I\},-\psi_\bullet-\psi_\circ)\bigg)\,. \label{eq:Omega exp}
\ee
Since $\xi_I^*$ is a ring homomorphism, it suffices to pull back each term in the exponent. The $\kappa$- and $\psi$-classes restrict to the appropriate component; the boundary classes are computed by the excess intersection formula as in subsection~\ref{subsec:tree level rewriting}. The self-intersection $\{J,J^c\}=\{I,I^c\}$ contributes $(-\psi_\bullet-\psi_\circ)\,\mathcal{F}(\{p_I\},-\psi_\bullet-\psi_\circ)$ to the exponent. Exponentiating and using $\mathcal{P}(p,x)=\ex^{-p^2x}\, \exp\big(x\mathcal{F}(\{p\},x)\big)$ (cf.\ \eqref{eq:Bernoulli polynomial edge factor identity}) gives
\be
\xi_I^*\,\Omega_{0,n}(\boldsymbol{p}) = \mathcal{P}(p_I,-\psi_\bullet-\psi_\circ)\, \Omega_{0,|I|+1}(\boldsymbol{p}_I,-p_I)\otimes \Omega_{0,|I^c|+1}(\boldsymbol{p}_{I^c},p_I)\,,\label{eq:sep pullback genus 0 exp}
\ee
which matches \eqref{eq:full sep pullback} at genus 0 (where the momentum integral localizes to $q=p_I$). In the same way, we could also derive the higher-genus factorization properties in the moduli space of $r$-spin curves. Pushing the result back down to the ordinary moduli space of curves and taking the large $r$ limit recovers \eqref{eq:sep pullback genus 0 exp}.

\paragraph{Comparison with the Virasoro Minimal string.}
The Virasoro minimal string (VMS) defines a CohFT. Explicitly, we have for the VMS with $b=1$ \cite{Collier:2023cyw}
\be
\Omega^\text{VMS}_{g,n}(\boldsymbol{p})=\ex^{-\sum_k \frac{B_k \kappa_k}{k\, k!}-\sum_i p_i^2 \psi_i}\ ,
\ee
which satisfies
\be
\xi_{h,I}^* \Omega^\text{VMS}_{g,n}(\boldsymbol{p})=\Omega^\text{VMS}_{h,|I|+1}(\boldsymbol{p}_I,0)\Omega^\text{VMS}_{g-h,|I^c|+1}(\boldsymbol{p}_{I^c},0)\ . \label{eq:VMS factorization}
\ee
and similarly for the non-separating divisor. Thus, the factorization does not require a $\psi$-dependent propagator as in \eqref{eq:propagator}, which is part of the definition of a CohFT.

\paragraph{Obstruction for a CohFT.} One might have expected that also $c=1$ strings define a CohFT. In fact, in certain cases, one can remove the $\psi$-class dependence and recover the CohFT structure. For example, we can remove the prefactor in \eqref{eq:propagator} by multiplication by $\e^{-q^2 \psi_\bullet}$, which in particular factorizes appropriately. However, the infinite sum cannot be removed and $c=1$ strings does not define a CohFT. This can be seen as follows. The propagator is of the form $F(\psi_\bullet+\psi_\circ)$ for an explicit function $F$. We view this as a multiplication operator in $V$ and treat $\psi_\bullet$ and $\psi_\circ$ as formal power series. The question whether the $\psi$-dependence can be removed is the question whether we can write the propagator as $F(\psi_\bullet+\psi_\circ)=R(\psi_\bullet)^\mathsf{T} \eta^{-1} R(\psi_\circ)$ for some operators $R$ and $\eta$. $R$ describes the change of basis to the CohFT basis. $\eta$ is the two-point function normalization in the CohFT basis. This is very constraining. Since is should hold as a formal power series in $\psi_\bullet$ and $\psi_\circ$, we can first put $\psi_\bullet=\psi_\circ=0$. Since $F(0)=1$, multiplication by it is the identity operator. Therefore, $\eta=R(0) R(0)^\mathsf{T}$. We can redefine $R(0)^{-1}R(\psi)$ as $R(\psi)$, which leads to the simpler problem of writing $F(\psi_\bullet+\psi_\circ)=R(\psi_\bullet)^\mathsf{T}R(\psi_\circ)$ with $R(0)=\id$. Putting $\psi_\bullet=0$ tell us that $R(\psi)=F(\psi)$ is the same multiplication operator. Thus everything reduces to the question whether $F(\psi_\bullet+\psi_\circ)=F(\psi_\bullet)F(\psi_\circ)$, which is easily be seen to be false. In fact, the only solutions to this equation are exponentials, $F(\psi)=\e^{f(q) \psi}$ for $f \in V$.

\section{The \texorpdfstring{$c=1$}{c=1} matrix integral} \label{sec:matrix integral}

\subsection{Minimal strings and matrix integrals} \label{subsec:minimal strings matrix integrals}
A central paradigm that emerged in the early 2000s for minimal string theory, and that has recently resurfaced in the study of JT gravity, the VMS, and the $\CC$LS, is their duality with double-scaled Hermitian matrix integrals.
In particular, two-matrix integrals of the special class
\begin{align}\label{eq:2matrixintegral}
\mathcal{Z}
=
\int_{\mathbb{R}^{2N^2}} [\d M_1][\d M_2]\,
\ex^{-N\tr\left(V_1(M_1)+V_2(M_2)-M_1M_2\right)} \, ,
\end{align}
provide a broad class of matrix models that capture exact descriptions of their worldsheet string theory duals.
Such models arise in the matrix-integral description of the $(p,q)$ A-series minimal strings, where one takes a double-scaling limit by sending $N\to\infty$ while zooming into a distinguished region of the eigenvalue distribution.\footnote{In contrast, the D-series $(p,q)$ minimal string is conjectured to be dual to a 4-matrix integral that is not exactly solvable \cite{DiFrancesco:1990jd,Rodriguez:2025rte}.}
In this limit, observables admit a genus expansion that is recursively determined by the spectral curve through topological recursion.

For the A-series $(p,q)$ minimal strings, this construction leads to algebraic genus-zero spectral curves with finitely many singularities, engineered from polynomial matrix potentials and rational double scalings \cite{Seiberg:2003nm,Seiberg:2004at}. 
By contrast, in the complex Liouville string one is led to an irrational version of this story: the dual description is again given by a double-scaled two-matrix integral, but now with non-polynomial data, a non-algebraic spectral curve, and infinitely many branch points and nodal singularities.
This more general setting still admits topological recursion, and it is in this broader framework that the $c=1$ spectral curve should be viewed.

In the preceding section we showed that the $c=1$ string amplitudes can be obtained by extracting the residue of a particular class of poles in the $\CC$LS amplitudes.
Implementing this procedure directly at the level of the intersection-number formula for the $\CC$LS, we derived the momentum-space Feynman rules for the $c=1$ string.
Notably, this operation still leads to a spectral curve of the same general type that arises from double-scaled two-matrix models.
In other words, although the construction proceeds through a residue of the $\CC$LS amplitudes, the resulting $c=1$ theory itself naturally fits into the broader two-matrix integral framework described above.

In this section we describe the double-scaled two-matrix integral that captures the (discretized) S-matrix elements of the $c=1$ string. More precisely, we describe how these matrix elements satisfy the same type of topological recursion relation that arises from the loop equations of double-scaled Hermitian two-matrix integrals \cite{Eynard:2002kg,Chekhov:2006vd}.
In fact, the matrix integral most directly computes the same generalizations of the $c=1$ S-matrix elements that are computed by the Feynman rules \eqref{eq:Feynman rules integer shifts}, which we denoted by $\hat{\mathsf{s}}_{g,n}$.
In this formulation the total momentum is integer-quantized rather than strictly conserved, reflecting a discrete, rather than continuous, translation symmetry in target space.

The generalized amplitudes can be written as
\begin{align}
\mathsf{s}_{g,n}(p_1,\ldots,p_n) &= \sum_{m\in\mathbb{Z}}\ex^{2\pi i m(p_1+\ldots+p_n)}\hat{\mathsf{s}}_{g,n}(p_1,\ldots,p_n)\nonumber\\
&= \delta_{\mathbb{Z}}(p_1+\ldots+p_n)\hat{\mathsf{s}}_{g,n}(p_1,\ldots,p_n)\, .\label{eq:mathcalS definition}
\end{align}
In the second line we rewrote the sum over $m$ as a Dirac comb. The function $\hat{\mathsf{s}}_{g,n}$ is polynomially bounded and free of poles but exhibits discontinuities as required by perturbative unitarity, as discussed in section \ref{subsec:perturbative unitarity}. It is only the sector where total momentum is fully conserved, restricted to the first Brillouin zone, that corresponds to the S-matrix elements of the conventional $c=1$ string. Moreover recall that the physical momenta $\omega_j$ of the $c=1$ string are related to the Liouville momenta $p_j$ by $p_j = \frac{i\omega_j}{2}$, so this quantization condition corresponds to integer quantization of imaginary total physical momentum.

\paragraph{Matrix integral technology.} Let us briefly review a few basic facts about matrix integrals in the class \eqref{eq:2matrixintegral}. For a more comprehensive review, see \cite{Collier:2024lys}.
Observables in matrix integrals are correlation functions of operators
\begin{equation}
    \Big\langle \prod_{j=1}^n \mathcal{O}_j \Big\rangle = \frac{1}{\mathcal{Z}} \int [\d M_1] [\d M_2] \, \ex^{-N \text{tr} (V_1(M_1) + V_2(M_2) - M_1M_2)} \prod_{j=1}^n \mathcal{O}_j(M_1,M_2) \, .
\end{equation}
A convenient way of organizing observables of matrix integrals is in terms of resolvents. For instance, the resolvent associated with the matrix $M_1$ of the two-matrix integral \eqref{eq:2matrixintegral} is defined as
\begin{equation} \label{eq:resolvent def}
    R(x) =  \tr\left(\frac{1}{x-M_1}\right)  \,.
\end{equation}
For example, the expectation value of this resolvent serves as a generating function for single-trace observables, or moments, of the form $\langle \text{tr} M_1^n \rangle$.

More generally, one can consider multi-resolvent correlators with respect to one of the matrices,
\(
R(x_1,\ldots,x_n)\equiv \prod_{j=1}^n R(x_j),
\)
which serve as generating functions for multi-trace observables.
A crucial property of the connected parts of these multi-resolvents is that they admit a topological genus expansion in powers of $\frac{1}{N^2}$:
\begin{align}
    \langle R(x_1,\ldots,x_n) \rangle_{\text{c}} = \sum_{g=0}^\infty N^{2-2g-n} R_{g,n}(x_1,\ldots,x_n) \,.
\end{align}
After taking the double-scaling limit --- a calibrated limit in which $N\to\infty$ while zooming into a particular region of the eigenvalue distribution --- this expansion becomes the perturbative genus expansion of the dual string theory, with $\frac{1}{N}$ playing the role of the string coupling $g_{\text{s}}$.

While we could also consider the resolvent associated with the second matrix $M_2$, and consider correlators of products of these two kinds of resolvents, the topological recursion solution of the matrix integral \eqref{eq:2matrixintegral} discussed in this section concerns observables with respect to a single chosen matrix, say $M_1$. Thus, we will restrict to observables built out of only the resolvent \eqref{eq:resolvent def}.

A special feature of the class of two-matrix integrals \eqref{eq:2matrixintegral} is that they may be diagonalized and expressed in terms of the eigenvalues of the matrices.
At large $N$, the eigenvalue distribution of the matrix $M_1$ typically develops a continuous support, which can be described by a density function $\rho(x)$.
The resolvent $R(x)$, which has poles at the eigenvalues of $M_1$, after integrating over $M_1$ develops branch cuts in the complex $x$-plane, reflecting the support of the eigenvalue density. Thus, the resolvents $R(x_1,\ldots,x_n)$ become multi-valued functions of the complex variables $x_j$. The multi-sheeted cover of the complex $x$-plane defined by the branch cuts of the resolvents is the central geometric object (a Riemann surface $\Sigma$) in the solution of these matrix integrals, and is known as the spectral curve of the matrix integral.

\paragraph{Spectral curves and topological recursion.}
Observables in large-$N$ matrix integrals satisfy the so-called loop equations, which are the Schwinger-Dyson identities obtained from infinitesimal changes of variables in the matrix integral.
A precise approach to solving the specific class of two-matrix integrals \eqref{eq:2matrixintegral} is provided by topological recursion, which is a recursive procedure that solves for the multi-resolvents $R_{g,n}$, with the use of the loop equations, by harnessing the analytic structure of these functions in the complex variables $x_j$.
This approach provides a solution for the resolvents to all orders in the $\frac{1}{N}$ expansion, and thus provides a complete solution to all orders in perturbation theory in $\frac{1}{N}$ (or $g_\text{s}$ after double-scaling) for these matrix integrals.

In particular, the leading-order loop equation leads to an algebraic equation for the planar resolvent $R_{0,1}(x)$, giving an implicit defining equation for the spectral curve of the form
\begin{equation}
    P(x,y(x))=0 \,,
\end{equation}
where the variable $y(x)=V_1'(x)-R_{0,1}(x)$ is related to the potential and the planar resolvent of $M_1$.
For example, for polynomial potentials $V_1(x)$ and $V_2(y)$, the function $P(x,y)$ is a polynomial in $x$ and $y$.
This implicit definition of the spectral curve can be made more explicit by introducing a (uniformizing) parametrization of the curve in terms of a complex variable $z\in\Sigma$, and writing the spectral curve as a pair of functions $(\xx(z),\yy(z))$ satisfying the algebraic relation
\begin{equation}
    P(\xx(z),\yy(z))=0 \,.
\end{equation}
Another consequence of a different planar loop equation is that the two-point planar resolvent $R_{0,2}(x_1,x_2)$ takes a universal form,
\begin{equation}
    R_{0,2}(\xx(z),\xx(z')) \d\xx(z)\d\xx(z') = \frac{\d z \d z'}{(z-z')^2} - \frac{\d \xx(z) \d \xx(z')}{(\xx(z)-\xx(z'))^2}  \,.
\end{equation}
It is convenient to define the resolvent differentials,
\begin{equation}
    \omega_{g,n}(z_1,\ldots,z_n) = R_{g,n}(\xx(z_1),\ldots,\xx(z_n)) \d\xx(z_1)\cdots \d\xx(z_n) \,,
\end{equation}
which turn out to be well-defined meromorphic multi-differentials on the spectral curve $\Sigma$, with a slightly modified definition for the cases $(g,n)=(0,1)$ and $(0,2)$:
\begin{subequations}
\begin{align}
    \omega_{0,1}(z) &= \big(R_{0,1}(\xx(z)) - V_1'(\xx(z)) \big) \d \xx(z) = -\yy(z)\d\xx(z) \,, \\
    \omega_{0,2}(z_1,z_2) &= \Big( R_{0,2}(\xx(z_1),\xx(z_2)) + \frac{1}{(\xx(z_1)-\xx(z_2))^2} \Big)\d \xx(z_1) \d \xx(z_2) = \frac{\d z_1\, \d z_2}{(z_1-z_2)^2}\,.
\end{align}
\end{subequations}
The remarkable property of the class of matrix integrals \eqref{eq:2matrixintegral} is that the higher-genus $n$-point resolvent differentials $\omega_{g,n}$ can be recursively reconstructed solely from the knowledge of the spectral curve, without having to explicitly solve all of the loop equations at higher genus. The resulting recursion relation is known as \emph{topological recursion}.
Its explicit form will be shown below, but the main point is that it reconstructs the resolvents $\omega_{g,n}$ by taking residues at the branch points of the map $\xx(z)$, defined by the condition $\d \xx(z)=0$, where different sheets of the spectral curve meet.

\paragraph{Connection to intersection theory.}
The differentials $\omega_{g,n}$ produced by topological recursion are closely linked to intersection theory on the moduli space $\bM_{g,n}$. In fact, it is known that for a given spectral curve, $\omega_{g,n}$ can be written as a sum over stable graphs of intersection numbers, structurally similar to the intersection expressions discussed in section~\ref{subsec:c=1 amplitudes as intersection numbers}. The stable graphs are naturally `colored' with each color $m$ corresponding to a branch point $z_m^*$ with $\d \xx(z_m^*)=0$ in the topological recursion \cite{Eynard:2011ga}.

\subsection{The \texorpdfstring{$c=1$}{c=1} spectral curve}
The discretized $c=1$ string amplitudes described in the previous section admit a presentation in terms of spectral curve data required for topological recursion.
In particular, the seed resolvent differentials of the $c=1$ string spectral curve can be extracted from the intersection number expressions for the amplitudes, and take the following form
\begin{align}\label{eq:c1 spec 1}
\omega_{0,1}(z) = 2\sqrt{2}\sin(z)^2\, \d z \,,\quad \omega_{0,2}(z_1,z_2) = \frac{\d z_1 \, \d z_2}{(z_1-z_2)^2} \,.
\end{align}
The spectral curve may be parametrized by
\begin{align}\label{eq:c1 spec param}
\xx(z) = 2\sqrt{2}\cos(z) \,,  \quad \yy(z) = \sin(z) \,,
\end{align}
in terms of which $\omega_{0,1}(z) = -\yy(z)\d\xx(z)$.
Its geometric properties are somewhat similar to those of the $\CC$LS spectral curve discussed in \cite{Collier:2024lys}, but we do not know of a way to derive \eqref{eq:c1 spec 1} directly by degenerating the $\CC$LS spectral curve, even though the string amplitudes are related by degeneration as described in section~\ref{subsec:degenerating CLS amplitudes}.
The image of \eqref{eq:c1 spec param} satisfies the algebraic relation $\xx^2/8 + \yy^2 = 1$, so the underlying curve is a smooth genus-zero curve, isomorphic to $\mathbb{CP}^1$ (the Riemann sphere).
Since $\xx(z)$ and $\yy(z)$ are periodic, the coordinate $z\in\CC$ realizes this curve as an infinite-sheeted covering of the sphere.
Thus, the $c=1$ spectral curve has infinitely many sheets.
Its branch points, with $\d\xx(z^*_m) = 0$, are located at
\begin{align}
    z^*_m = \pi m \,, \quad m\in\mathbb{Z} \,.
\end{align}
The local Galois involution at each such branch point satisfying $\xx(z) = \xx(\sigma_m(z))$ and $\sigma_m(z^*_m) = z^*_m$ in the neighborhood of the branch point, can be taken to be $\sigma_m(z) = 2\pi m - z$.

\paragraph{Topological recursion.} The resolvents of the $c=1$ matrix integral are recursively determined from the initial data \eqref{eq:c1 spec 1} by topological recursion \cite{Eynard:2002kg,Chekhov:2006vd}
\begin{align}
\omega_{g,n}(z_1,\boldsymbol{z}) &= \sum_{m\in\mathbb{Z}}\Res_{z=z_m^*}K_m(z_1,z)\Big(\omega_{g-1,n+1}(z,\sigma_m(z),\boldsymbol{z})\nonumber\\
&\quad+ \sum_{h=0}^g\sum_{\substack{I\cup I^c = \{2,\ldots,n\}\\ (h,I) \ne (0,\emptyset)\\(g-h,I^c)\ne (0,\emptyset)}}\omega_{h,1+|I|}(z,\boldsymbol{z}_{I})\omega_{g-h,1+|I^c|}(\sigma_m(z),\boldsymbol{z}_{I^c})\Big)\, ,\label{eq:omega topological recursion}
\end{align}
where $\boldsymbol{z}=\{z_2,\ldots,z_n\}$, and the recursion kernel associated with the $m^{\text{th}}$ branch point is given by
\begin{align}
K_m(z_1,z) = \frac{1}{8\sqrt{2}\sin^2(z)}\left(\frac{1}{z_1-z}-\frac{1}{z_1-\sigma_m(z)}\right) \,.
\end{align}

\paragraph{Equivalence to the intersection number expressions.} It is straightforward to prove the equivalence of the intersection number expression to the topological recursion. For this, it is simplest to start from the position space Feynman rules \eqref{eq:c=1 position space intersection numbers}. As mentioned above, the theorem \cite[Theorem 4.1]{Eynard:2011ga} expresses $\omega_{g,n}$ in terms of intersection numbers on moduli space. It is then simple to check that \eqref{eq:c1 spec 1} indeed leads to \eqref{eq:c=1 position space intersection numbers}. The computation is entirely analogous to the $\CC$LS case explained in \cite[Appendix B]{Collier:2024lys} and thus we omit the details. We note that the `positions' $m$ in the position space Feynman rules are naturally associated with the branch points $z_m^*$, which is why we use the same label for both.

\paragraph{Examples.}
Let us work out the simplest nontrivial examples of resolvent differentials obtained from topological recursion.
The three-point, genus-zero resolvent is given by
\begin{align}\label{eq:omega03 b1}
\omega_{0,3}(z_1,z_2,z_3) &= \sum_{m\in\mathbb{Z}}\Res_{z=z^*_m}
K_m(z_1,z)\big(\omega_{0,2}(z, z_2)\omega_{0,2}(\sigma_m(z), z_3) + (z_2\leftrightarrow z_3)\big) \nonumber\\
&= -\frac{1}{2\sqrt{2}}\sum_{m\in\mathbb{Z}} \frac{1}{(z_1 - \pi m)^2 (z_2 -\pi m)^2 (z_3 -\pi m)^2} \,,
\end{align}
The overall minus signs in these results arise from $\d\sigma_m(z)=-\d z$.
The one-point, genus-one resolvent is
\begin{align}\label{eq:omega11 b1}
\omega_{1,1}(z_1) &= \sum_{m\in\mathbb{Z}}\Res_{z=z^*_m}
K_m(z_1,z) \omega_{0,2}(z, \sigma_m(z)) = -\sum_{m\in\mathbb{Z}} \frac{(z_1-\pi  m)^2+3}{48\sqrt{2} (z_1-\pi  m)^4} \,.
\end{align}
Similarly,
\begin{align}\label{eq:omega12 b1}
\omega_{1,2}(z_1,z_2)
&=\sum_{m\in\mathbb{Z}} \frac{45 (z_{1,m}^4 + z_{2,m}^4) + 18 z_{1,m}^2 z_{2,m}^2 (z_{1,m}^2 + z_{2,m}^2) + 4 z_{1,m}^4 z_{2,m}^4 + 27 z_{1,m}^2 z_{2,m}^2}{576 z_{1,m}^6 z_{2,m}^6} \nonumber\\
&\quad+ \sum_{\substack{m,m'\in\mathbb{Z} \\ m\neq m'}} \left[\frac{1}{16\pi^4\Delta^4 z_{1,m}^2 z_{2,m'}^2}
+ \frac{1}{192 \, z_{1,m}^2 z_{2,m}^2} \left( \frac{3}{\pi^4\Delta^4} + \frac{1}{\pi^2\Delta^2} \right) \right] \,,
\end{align}
where we have defined $z_{j,m}\equiv z_j- \pi m$ and $\Delta\equiv (m-m')$ for clarity.
For the tree-level four-point resolvent, we obtain
\begin{multline} \label{eq:omega04 b1}
\omega_{0,4}(\boldsymbol{z})
= \sum_{m\in\mathbb{Z}} \frac{1}{8 z_{1,m}^2 z_{2,m}^2 z_{3,m}^2 z_{4,m}^2} \left( 1 + \sum_{j=1}^4\frac{3}{z_{j,m}^2} \right) \\+ \sum_{\substack{m,m'\in\mathbb{Z} \\ m\neq m'}} \left(\frac{1}{8\pi^2\Delta^2 z_{1,m}^2 z_{2,m}^2 z_{3,m'}^2 z_{4,m'}^2} + \text{2 perms} \right) \,.
\end{multline}

\paragraph{Relation between observables.}
The dictionary that relates these resolvent differentials back to the $c=1$ string scattering amplitudes takes the same form as in the VMS or $\CC$LS cases:
\begin{align}\label{eq:dictionary c1 resolvents to amps}
\mathsf{s}_{g,n}(p_1,\ldots,p_n)
&= \sum_{m_j\in\mathbb{Z}} \prod_{j=1}^n \Res_{z_j = \pi m_j} \frac{i \e^{2i p_j z_j}}{\sqrt{2}p_j} \omega_{g,n}(z_1,\ldots,z_n) \\
&= \left( \int_{\mathbb{R}-i\epsilon} - \int_{\mathbb{R}+i\epsilon} \right) \prod_{j=1}^n \frac{\e^{2i p_j z_j}}{2\sqrt{2}\pi p_j} \omega_{g,n}(z_1,\ldots,z_n) \,.
\end{align}
We may interpret the sum over residues at the branch points weighted by the exponential factors $\ex^{2\pi i m_j p_j}$ as a sort of discrete Fourier transform that takes us from position to momentum space. In the second line, we have opened the residue contours into the difference of horizontal contours lying just below and just above the real axis.
Note that convergence of this horizontal integral representation requires the momenta $p_j$ to be real-valued.

The inverse relation is given by
\be\label{eq:S to omega}
    \omega_{g,n}(z_1,\ldots,z_n) = \int_0^\infty \prod_{j=1}^n 2\sqrt{2}p_j\d p_j\, {\e}^{\mp 2ip_jz_j}\,\mathsf{s}_{g,n}(\pm p_1,\ldots,\pm p_n)\, ,
\ee
where the signs in the exponent are correlated with the corresponding signs in the momentum arguments of the amplitude $\mathsf{s}_{g,n}(\boldsymbol{p})$.
The integral with the upper choice of signs converges for $z_j$ in the lower half-plane, while the integral with the lower choice of signs converges for $z_j$ in the upper half-plane.\footnote{As a simple toy model to verify this relationship, consider a meromorphic function of the form
\be
    r(z) = \sum_{m\in\mathbb{Z}} \sum_{k\geq1} \frac{c_{m,k}}{(z-\pi m)^k} \,,
\ee
which has the same structural form as the resolvents $\omega_{g,n}$. The corresponding ``string amplitude" obtained from the residue dictionary is
\begin{align}
    a(p) = \sum_{m\in\mathbb{Z}} \Res_{z=\pi m} \frac{i\e^{2ipz}}{\sqrt{2}p} r(z) = -\sqrt{2} \sum_{m\in\mathbb{Z}} \sum_{k\geq1} \frac{c_{m,k}(2ip)^{k-2}}{(k-1)!} {\e}^{2\pi i m p} \, .
\end{align}
We can then verify that the inverse transform
\be
    \int_0^\infty 2\sqrt{2} p \, \d p\, {\e}^{\mp 2i pz} a(\pm p) \,,
\ee
which converges for $z$ in the lower and upper half-plane respectively, recovers the original function $r(z)$.}
This slight difference in the inverse dictionary, reflected in the sign choice of the momenta, can be interpreted as distinguishing incoming from outgoing states in the amplitude, albeit for these Euclidean momenta $p_j$.

\paragraph{Examples.} Let us explicitly verify this dictionary in a few of the first nontrivial cases.
For the tree-level three-point scattering amplitudes, we apply \eqref{eq:dictionary c1 resolvents to amps} to the resolvent \eqref{eq:omega03 b1} and evaluate the resulting residues to obtain
\begin{align}
\mathsf{s}_{0,3}(p_1,p_2,p_3)
= \sum_{m\in\mathbb{Z}} {\e}^{2\pi i m (p_1+p_2+p_3)}
= \delta_\ZZ(p_1+p_2+p_3)\mathsf{V}_{0,3}^{(b)}(p_1,p_2,p_3) \,,
\end{align}
where in the last equality we expressed the result in terms of the Dirac comb and noticed that the coefficient of the Dirac comb can trivially be written in terms of the sphere three-point VMS quantum volume $\mathsf{V}_{0,3}^{(b=1)}=1$. Similarly, we can compute
\begin{align}
\mathsf{s}_{1,1}(p_1)
= \sum_{m\in\mathbb{Z}} {\e}^{2\pi i m p_1} \frac{1}{48} (1 - 2p_1^2)
= \delta_\ZZ(p_1) \mathsf{V}_{1,1}^{(b=1)}(p_1) \,.
\end{align}
We see that in this case the amplitude is given by the corresponding VMS quantum volume at $c=25$, weighted by a Dirac comb that sets the momentum to be an integer.
If we restrict to the first Brillouin zone, we obtain $\mathsf{s}_{1,1}(p_1) \to  \delta(p_1) \frac{1}{48}$ in agreement with \eqref{eq:S11}.

Next, let us compute $\mathsf{s}_{1,2}(p_1,p_2)$. Consider the first term in \eqref{eq:omega12 b1}. Evaluating the residues as in \eqref{eq:dictionary c1 resolvents to amps} to pass to the string S-matrix element, we obtain
\begin{multline}
\sum_{m_j\in\mathbb{Z}} \prod_{j=1}^2 \Res_{z_j = \pi m_j} \frac{i \e^{2i p_j z_j}}{\sqrt{2}p_j} \sum_{\substack{m,m'\in\mathbb{Z} \\ m\neq m'}} \frac{1}{16\pi^4\Delta^4 z_{1,m}^2 z_{2,m'}^2}
= \sum_{\substack{m,m'\in\mathbb{Z} \\ m\neq m'}} {\e}^{2\pi i(mp_1+m'p_2)} \frac{1}{8\pi^4\Delta^4} \\
= \sum_{m\in\mathbb{Z}} {\e}^{2\pi i m (p_1+p_2)} \sum_{\Delta\neq0} \frac{\e^{2\pi i p_2 \Delta}}{8\pi^4\Delta^4} = -\frac{1}{12} \sum_{m\in\mathbb{Z}} {\e}^{2\pi i m (p_1+p_2)} B_4(\{p_2\}) \,,
\end{multline}
where in the second line we have reorganized the double sum over $m$ and $m'$ as a sum over $m$ and another over their difference $\Delta$.
In the last step, we made use of the Bernoulli polynomial identity \eqref{eq:Bernoulli polynomial identity}.
The remaining terms in \eqref{eq:omega12 b1} are evaluated in the same way, and assembling all contributions we arrive at the final result:
\begin{align}
\mathsf{s}_{1,2}(p_1,p_2)
&= \sum_{m\in\mathbb{Z}} {\e}^{2\pi i m (p_1+p_2)} \left[ -\frac{1}{12}B_4(\{p_2\}) + \frac{1}{48}\big(B_2(0)-B_4(0)\big) \right. \nonumber\\
&\quad \left. + \frac{1}{144}\Big( 2 - 6 (p_1^2 + p_2^2) + 3 (p_1^2 + p_2^2)^2 \Big) \right] \nonumber\\
&=\delta_\ZZ(p_1+p_2)\left[\mathsf{V}^{(b=1)}_{1,2}(p_1,p_2) -\frac{1}{12}B_4(\{p_2\})  + \frac{1}{240} \right] ~.
\end{align}
We notice that in this case the coefficient of the Dirac comb is not given solely by the corresponding VMS quantum volume, though it does make an appearance. This recovers \eqref{eq:g=1, n=2 first diagram}--\eqref{eq:g=1, n=2 third diagram}.
Restricting to the first Brillouin zone gives \eqref{eq:S12}.
A similar calculation, applying \eqref{eq:dictionary c1 resolvents to amps} to \eqref{eq:omega04 b1} to extract the tree-level four-point amplitude, yields
\begin{align}
\mathsf{s}_{0,4}&(p_1,p_2,p_3,p_4)
= \sum_{m\in\mathbb{Z}} {\e}^{2\pi i m \sum_j p_j} \left[ \frac{1}{2} - \sum_{j=1}^4 p_j^2 + \big( B_2(\{p_3+p_4\}) + \text{2 perms}\big) \right] \nonumber\\
&= \delta_\ZZ({\textstyle\sum_j} p_j) \left[  \mathsf{V}^{(b=1)}_{0,4}(p_1,p_2,p_3,p_4) +\big( B_2(\{p_3+p_4\}) + \text{2 perms}\big)\right] \,.
\end{align}
This matches \eqref{eq:g=0, n=4 discrete result}.

\subsection{Recursion relation for \texorpdfstring{$c=1$}{c=1} S-matrix elements}\label{subsec:recursion relation}
Topological recursion \eqref{eq:omega topological recursion} completely determines the perturbative resolvents $\omega_{g,n}$ from the spectral curve \eqref{eq:c1 spec 1}. Starting from the relations \eqref{eq:dictionary c1 resolvents to amps} and \eqref{eq:S to omega} between the resolvents and the generalized S-matrix elements, we can translate this into a recursion relation for the generalized S-matrix elements themselves. We have
\begin{align}
p_1 \mathsf{s}_{g,n}(p_1,\boldsymbol{p}) &= \sum_{m_1\in\mathbb{Z}}\Res_{z_1=\pi m_1}\frac{i\e^{2ip_1 z_1}}{\sqrt{2}}\left(\sum_{m_2,\ldots,m_n\in\mathbb{Z}}\prod_{j=2}^n\Res_{z_j=\pi m_j}\frac{i\e^{2ip_jz_j}}{\sqrt{2}p_j}\right)\omega_{g,n}(z_1,\boldsymbol{z})\nonumber\\
&= \sum_{m_1\in\mathbb{Z}}\Res_{z_1=\pi m_1}\frac{i\e^{2ip_1 z_1}}{\sqrt{2}}\left(\sum_{m_2,\ldots,m_n\in\mathbb{Z}}\prod_{j=2}^n\Res_{z_j=\pi m_j}\frac{i\e^{2ip_jz_j}}{\sqrt{2}p_j}\right)\sum_{m\in\mathbb{Z}}\Res_{z=\pi m}K_m(z_1,z)\nonumber\\
&\quad\times\Big(\omega_{g-1,n+1}(z,\sigma_m(z),\boldsymbol{z})+ \sideset{}{'}\sum_{h,I}\omega_{h,1+|I|}(z,\boldsymbol{z}_{I})\omega_{g-h,1+|I^c|}(\sigma_m(z),\boldsymbol{z}_{I^c})\Big)\, .
\end{align}
We wrote the amplitudes in terms of the resolvents and then used topological recursion for the resolvents. To proceed, we exchange the sum over $m_1$ with that over $m$ and rewrite $z = \pi m + 2\pi u$ to get
\begin{align}
p_1 \mathsf{s}_{g,n}(p_1,\boldsymbol{p})
&= -\frac{\pi}{4}\sum_{m\in\mathbb{Z}}\ex^{2\pi i mp_1}\Res_{u=0}\frac{\sin(4\pi u p_1)}{\sin(2\pi u )^2}\left(\sum_{m_2,\ldots,m_n\in\mathbb{Z}}\prod_{j=2}^n\Res_{z_j=\pi m_j}\frac{i\e^{2ip_jz_j}}{\sqrt{2}p_j}\right)\nonumber\\
&\quad\times \Big(\omega_{g-1,n+1}(2\pi u, -2\pi u,\boldsymbol{z}-\pi m)\nonumber\\
&\quad\quad\!\!+ \sideset{}{'}\sum_{h,I}\omega_{h,1+|I|}( 2\pi u,\boldsymbol{z}_{I}-\pi m)\omega_{g-h,1+|I^c|}(-2\pi u ,\boldsymbol{z}_{I^c}-\pi m)\Big)\, .\label{eq:s recursion in terms of omegas}
\end{align}
In writing it this way we used the fact that the resolvents are invariant under the simultaneous shifts of their arguments by a multiple of $\pi$. We now reexpress the resolvents in terms of the corresponding generalized amplitudes via \eqref{eq:S to omega} to arrive at
\begin{align}
p_1 \mathsf{s}_{g,n}(p_1,\boldsymbol{p})
&= 2\pi \left(\sum_{m\in\mathbb{Z}}\ex^{2\pi i m (p_1+\ldots+p_n)}\right) \Res_{u=0}\bigg\{\frac{\sin(4\pi u p_1)}{\sin(2\pi u)^2}\bigg[\int_0^\infty q \,\d q\,q'\,\d q'\, \ex^{-4\pi i u(q+q')}\nonumber\\
&\quad\quad\times\left(\mathsf{s}_{g-1,n+1}(q,-q',\boldsymbol{p})+ \sideset{}{'}\sum_{h,I}\mathsf{s}_{h,1+|I|}(q,\boldsymbol{p}_{I})\mathsf{s}_{g-h,1+|I^c|}(-q',\boldsymbol{p}_{I^c})\right)\nonumber\\
&\quad - \sum_{j=2}^n\int_0^\infty q \,\d q\,  \ex^{-4\pi i u (q+p_j)}\mathsf{s}_{g,n-1}(q,\boldsymbol{p}\setminus p_j) \bigg]\bigg\} \, .\label{eq:s recursion without deltas stripped}
\end{align}
As a reminder, $\boldsymbol{p} = \{p_2,\ldots, p_n\}$, and the sum in the second line runs over all decompositions $I\cup I^c = \{p_2,\ldots,p_n\}$, excluding the cases $(h,I) = (0,\emptyset)$ and $(h,I^c) = (g,\emptyset)$. The prime on this sum indicates that only \emph{stable} amplitudes $\mathsf{s}_{g,n}$ appear in the summand; in particular, amplitudes such as $\mathsf{s}_{0,3}$ and those of higher genus contribute, whereas $\mathsf{s}_{0,2}$ does not. The latter is instead accounted for by the third term in the recursion.

Notice in particular that the sum over $m$ in \eqref{eq:s recursion in terms of omegas} conspired to reconstruct the Dirac comb that enforces conservation of total momentum up to an integer, which cleanly factorizes in \eqref{eq:s recursion without deltas stripped}.
This makes it trivial to write down a recursion relation for the delta-stripped amplitudes $\hat{\mathsf{s}}_{g,n}$ themselves: 
\begin{align}
&p_1 \hat{\mathsf{s}}_{g,n}(p_1,\boldsymbol{p})\nonumber\\
&\ = 2\pi\Res_{u=0}\bigg\{\frac{\sin(4\pi u p_1)}{\sin(2\pi u)^2}\bigg[\sum_{m\in\mathbb{Z}}\int_0^\infty q\,\d q\, q'\,\d q'\,\ex^{-4\pi i u(q+q')}\ex^{2\pi i m(q-q'-p_1)}\hat{\mathsf{s}}_{g-1,n+1}(q,-q',\boldsymbol{p})\nonumber\\
&\ \quad + \sideset{}{'}\sum_{h,I}\sum_{m,m'\in\mathbb{Z}}\int_0^\infty q\,\d q\, q' \, \d q'\, \ex^{-4\pi i u(q+q')}\ex^{2\pi i m(q+p_I)}\ex^{2\pi i m'(-q'+p_{I^c})}\nonumber\\
&\qquad\qquad\qquad\qquad\qquad\times \hat{\mathsf{s}}_{h,1+|I|}(q,\boldsymbol{p}_{I})\hat{\mathsf{s}}_{g-h,1+|I^c|}(-q',\boldsymbol{p}_{I^c})\nonumber\\
&\ \quad - \sum_{j=2}^n \sum_{m\in\mathbb{Z}}\int_0^\infty q\, \d q\, \ex^{-4\pi i u(q+p_j)}\ex^{2\pi i m(q-p_1-p_j)}\hat{\mathsf{s}}_{g,n-1}(q,\boldsymbol{p}\setminus p_j)\bigg]\bigg\}\, ,\label{eq:shat recursion integral form}
\end{align}
where we used the shorthand $p_{I} = \sum_{j\in I} p_j$, and similarly for $p_{I^c}$. 
The recursion involves infinite sums over integers $m,m'$. The terms in the recursion with all such integers set to zero are collectively equivalent to the recursion relation for the VMS quantum volumes at $c=25$ ($b=1$). The terms in the recursion with $m,m'\ne 0$ are responsible for generating the corrections to the VMS quantum volumes involving Bernoulli polynomials depending on the fractional parts of partial sums of the momenta.

We can alternatively use the momentum-conserving delta functions in \eqref{eq:mathcalS definition} to further simplify the presentation and rewrite the recursion as follows: 
\begin{tcolorbox}[
  before skip=6pt,
  after skip=6pt,
]
\begingroup
\setlength{\abovedisplayskip}{0pt}
\setlength{\abovedisplayshortskip}{0pt}
\setlength{\belowdisplayskip}{0pt}
\setlength{\belowdisplayshortskip}{0pt}
\begin{align}
p_1 \hat{\mathsf{s}}_{g,n}(p_1,\boldsymbol{p})&= \half \Res_{u=0} \frac{\sinh(u p_1)}{\sinh(\frac{u}{2})^2} \bigg[ \int_0^1 \d q \hspace{-.5cm}\sum_{\substack{r\in\{q\}+\mathbb{Z}_{\geq 0}\\ r'\in \{q-p_1\}+\ZZ_{\ge 0}}} \hspace{-.5cm} rr'\ex^{u(r+r')}\hat{\mathsf{s}}_{g-1,n+1}(r,-r',\boldsymbol{p}) \! \nonumber\\
&\quad + {\sum_{h,I}}' \sum_{\substack{r \in \{-p_{I}\}+\mathbb{Z}_{\geq 0} \\ r'\in \{p_{I^c}\}+\mathbb{Z}_{\geq 0}}} \hspace{-.5cm} r r'\ex^{u(r+r')}\hat{\mathsf{s}}_{h,1+|I|}(r,\boldsymbol{p}_{I})\hat{\mathsf{s}}_{g-h,1+|I^c|}(-r',\boldsymbol{p}_{I^c})\nonumber\\
&\quad -  \sum_{j=2}^n\sum_{r\in \{p_1+p_j\}+\mathbb{Z}_{\geq 0}} r\ex^{u(r+p_j)}\hat{\mathsf{s}}_{g,n-1}(r,\boldsymbol{p}\setminus p_j)\bigg]\, .\label{eq:Mirzarec c1}
\end{align}
\endgroup
\end{tcolorbox}
\noindent
Compared to~\eqref{eq:shat recursion integral form}, we have performed the substitution $u\to iu/(4\pi)$, which Wick-rotates the recursion kernel from $2\pi\sin(4\pi u p_1)/\sin(2\pi u)^2$ to $\half\sinh(u p_1)/\sinh(u/2)^2$ and replaces $\ex^{-4\pi i u X}\to \ex^{uX}$ in the exponentials.
The Dirac combs in \eqref{eq:mathcalS definition} have reduced most of the integrals to sums. The lower bounds on the sums are important, as they introduce non-analyticities to the resulting amplitudes. Crucially, the recursion relation \eqref{eq:Mirzarec c1} does not close on physical $c=1$ string amplitudes where the total momentum is strictly conserved; it involves sums over sub-amplitudes where momentum is only conserved up to an integer.

Let us illustrate the derivation of the first line in \eqref{eq:Mirzarec c1} corresponding to the non-separating degeneration. From \eqref{eq:s recursion without deltas stripped}, we strip off  $\delta_\ZZ(q-q'-p_1)$, where $\delta_\ZZ$ denotes the Dirac comb. We can write the term in the square brackets as
\begin{align}
&\int_0^\infty q\,\d q\, q' \,\d q'\, \ex^{-4\pi i u(q+q')}\mathsf{s}_{g-1,n+1}(q,-q',\boldsymbol{p})\nonumber\\
&\quad=\sum_{r,r'=0}^\infty\int_0^1\d q \, \d q'\, (q+r)(q'+r') \ex^{-4\pi i u(q+q'+r+r')}\hat{\mathsf{s}}_{g-1,n+1}(q+r,-q'-r',\boldsymbol{p}) \nonumber\\
&\quad\qquad\times\delta_\ZZ(q-q'-p_1) \\
&\quad=\int_0^1\d q \, \d q'\!\!\!\!\sum_{\substack{r\in\{q\}+\mathbb{Z}_{\geq 0}\\ r'\in \{q'\}+\ZZ_{\ge 0}}}\!\!\!rr'\, \ex^{-4\pi i u (r+r')}\hat{\mathsf{s}}_{g-1,n+1}(r,-r',\boldsymbol{p}) \delta_\ZZ(q-q'-p_1) \, ,
\end{align}
where in the last line we shifted the sums over $r,r'$ by $-q$ and $-q'$ respectively.
Lastly, the integral over $q'$ can be done and puts $\{q'\}=\{q-p_1\}$, which after the Wick rotation reproduces the first line of \eqref{eq:Mirzarec c1}.

\begin{figure}[ht]
    \centering
    \begin{tikzpicture}[baseline={([yshift=-.5ex]current bounding box.center)},scale=1]

        \draw[thick, bleudefrance] (1,1.32) to[out=0, in = 0,looseness=.8] (1,1/16);
        \draw[thick, bleudefrance, dashed] (1,1.32) to[out=180, in = 180, looseness=.8] (1,1/16);

        \draw[thick, bleudefrance] (1-2/8,-1/16) to[out=-90,in=-90] (.3,-1/16);
        \draw[thick, bleudefrance, dashed] (1-2/8,-1/16) to[out=90,in=90] (.3,-1/16);

        \draw[fill=bleudefrance,draw=bleudefrance,opacity=.4] ({0+1/4*cos(60)},{1+1/2*sin(60)}) to[out=-30, in = 180] (1,1.32) to[out=0, in = 0,looseness=.8] (1,1/16) to[out=0, in = 90] (1-2/8,-1/16) to[out = -90, in = -90] (.3,-1/16) to[out=90, in = 0] (0,1/2) to[out=0, in = -30,looseness=.75] ({{0+1/4*cos(60)}},{1+1/2*sin(60)});

        \draw[very thick] (0,1) ellipse (1/4 and 1/2);
        \draw[very thick] (0,-1) ellipse (1/4 and 1/2);
        \draw[very thick] (0,1/2) to[out=0, in = 0] (0,-1/2);
        \draw[very thick] ({0+1/4*cos(60)},{1+1/2*sin(60)}) to[out= -30, in = 180] (2,3/2) to [out=0, in=90] (3,0) to [out=-90, in= 0] (2,-3/2) to[out=180, in = 30] ({{0+1/4*cos(-60)}},{-1+1/2*sin(-60)});

        \node at (0,1) {$p_1$};
        \node at (0,-1) {$p_2$};

        \draw[thick] (5/8,0) to[out=-30, in = 210] (11/8,0);
        \draw[thick] (1-2/8,-1/16) to[out=60, in = 120] (1+2/8,-1/16);

        \draw[thick] (14/8+1/8,0) to [out=-30, in=210] (20/8+1/8,0);
        \draw[thick] (17/8-2/8+1/8,-1/16) to[out=60, in = 120] (17/8+2/8+1/8,-1/16);

        \draw[ultra thick, candyapplered, ->] (1.0,0.7) to (1.6,0.7);
        \draw[ultra thick, candyapplered, ->] (0.5,-0.45) to (0.5,0.15);

        \node[scale = 1] at (1.75,-3/4) {$\hat{\mathsf{s}}_{g-1,n+1}$};
        \node[scale = 1] at (1.5,-2) {$+$};
        \node[scale = 0.8] at (1.5,-2.5) {$\text{(arrow reversal)}$};

        \begin{scope}[shift={(5,0)}]
            \draw[thick, bleudefrance] (1.7,1.48) to[out=0,in=0,looseness=.15] (1.7,-1.48);
            \draw[thick, bleudefrance, dashed] (1.7,1.48) to[out=180,in=180,looseness=.15] (1.7,-1.48);

            \draw[thick, bleudefrance] (1/4+.05,0) to[out=0, in =180] (1,1/4) to[out=0, in = 90] (3/2,0) to[out= -90, in =60,looseness=.4] (3/2,-1.45);
            \draw[thick, bleudefrance, dashed] (3/2,-1.45) to [out=150, in =-90,looseness=.4] (3/2-.05,0) to[out=90, in = 0] (1,1/4-0.05) to[out=180, in = 0] (1/4+.05,-.05);

            \draw[fill=bleudefrance, draw=bleudefrance, opacity=.4] ({0+1/4*cos(60)},{1+1/2*sin(60)}) to[out= -30, in=180] (3/4,1.28) to[out=0, in = 180,looseness=.9] (1.7,1.48) to[out=0,in=0,looseness=.15] (1.7,-1.48) to[out=180, in = -10] (3/2,-1.45) to[out=60, in = -90,looseness=.4] (3/2,0) to[out = 90, in = 0] (1,1/4) to[out=180, in = 0] (1/4+.05,0) to[out=90, in = 0] (0,1/2) to[out= 0, in = -30,looseness=.75] ({{0+1/4*cos(60)}},{1+1/2*sin(60)});

            \draw[fill=bleudefrance, draw=bleudefrance, opacity=.25] (3/2,-1.45) to[out=150, in =-90,looseness=.4] (3/2-.05,0) to[out=90, in = 0] (1,1/4-0.05) to[out=180, in = 0] (1/4+.05,-.05) to (1/4+.05,0) to[out=0, in =180] (1,1/4) to[out=0, in = 90] (3/2,0) to[out= -90, in =60,looseness=.4] (3/2,-1.45);

            \draw[very thick] (0,1) ellipse (1/4 and 1/2);
            \draw[very thick] (0,-1) ellipse (1/4 and 1/2);
            \draw[very thick] (0,1/2) to[out=0, in = 0] (0,-1/2);
            \draw[very thick] ({0+1/4*cos(60)},{1+1/2*sin(60)}) to[out= -30, in = 180] (2,3/2) to [out=0, in=90] (3,0) to [out=-90, in= 0] (2,-3/2) to[out=180, in = 30] ({{0+1/4*cos(-60)}},{-1+1/2*sin(-60)});

            \node at (0,1) {$p_1$};
            \node at (0,-1) {$p_2$};

            \draw[thick] (5/8,0) to[out=-30, in = 210] (11/8,0);
            \draw[thick] (1-2/8,-1/16) to[out=60, in = 120] (1+2/8,-1/16);

            \draw[thick] (14/8+1/8,0) to [out=-30, in=210] (20/8+1/8,0);
            \draw[thick] (17/8-2/8+1/8,-1/16) to[out=60, in = 120] (17/8+2/8+1/8,-1/16);

            \draw[ultra thick, candyapplered, ->] (1.65,0.35) to (2.25,0.35);
            \draw[ultra thick, candyapplered, ->] (0.5,-0.3) to (0.5,0.3);

            \node[scale = 1] at (.8,-1.85) {$\hat{\mathsf{s}}_{h,1+|I|}$};
            \node[scale = 1] at (2.75,-1.85) {$\hat{\mathsf{s}}_{g-h,1+|I^c|}$};

            \node[scale = 1] at (1.5,-2.5) {$+$};
            \node[scale = 0.8] at (1.5,-3) {$\text{(arrow reversal)}$};

        \end{scope}

        \begin{scope}[shift={(10,0)}]
            \draw[thick,bleudefrance] (1/2, 1.3) to[out=0, in =0,looseness=.2] (1/2,-1.3);
            \draw[thick,bleudefrance,dashed] (1/2, 1.3) to[out=180, in =180,looseness=.2] (1/2,-1.3);
            \draw[draw=bleudefrance, fill=bleudefrance, opacity=.4] ({0+1/4*cos(60)},{1+1/2*sin(60)}) to[out=-30, in = 180] (1/2,1.3) to[out=0, in = 0,looseness=.2] (1/2,-1.3) to[out=180, in = 30] ({{0+1/4*cos(-60)}},{-1+1/2*sin(-60)}) to[out=30, in =0, looseness=.75] (0,-1/2) to[out=0, in=0] (0,1/2) to[out=0, in = -30,looseness=.75] ({{0+1/4*cos(60)}},{1+1/2*sin(60)});

            \draw[very thick] (0,1) ellipse (1/4 and 1/2);
            \draw[very thick] (0,-1) ellipse (1/4 and 1/2);
            \draw[very thick] (0,1/2) to[out=0, in = 0] (0,-1/2);
            \draw[very thick] ({0+1/4*cos(60)},{1+1/2*sin(60)}) to[out= -30, in = 180] (2,3/2) to [out=0, in=90] (3,0) to [out=-90, in= 0] (2,-3/2) to[out=180, in = 30] ({{0+1/4*cos(-60)}},{-1+1/2*sin(-60)});

            \node at (0,1) {$p_1$};
            \node at (0,-1) {$p_2$};

            \draw[thick] (5/8,0) to[out=-30, in = 210] (11/8,0);
            \draw[thick] (1-2/8,-1/16) to[out=60, in = 120] (1+2/8,-1/16);

            \draw[thick] (14/8+1/8,0) to [out=-30, in=210] (20/8+1/8,0);
            \draw[thick] (17/8-2/8+1/8,-1/16) to[out=60, in = 120] (17/8+2/8+1/8,-1/16);

            \draw[ultra thick, candyapplered, ->] (0.4,0.35) to (1.0,0.35);

            \node[scale = 1] at (1.75,-3/4) {$\hat{\mathsf{s}}_{g,n-1}$};

            \node[scale = 1] at (1.5,-2) {$+$};
            \node[scale = 0.8] at (1.5,-2.5) {$\text{(arrow reversal)}$};

        \end{scope}

\end{tikzpicture}
\caption{The three distinct ways of embedding a pair of pants with a distinguished external cuff (labeled by $p_1$ above) into a surface. These correspond to the three classes of terms in the Mirzakhani-type recursion~\eqref{eq:Mirzarec c1} for the $c=1$ string amplitudes $\hat{\mathsf{s}}_{g,n}(\boldsymbol{p})$. In the $c=1$ string case there is a notion of ingoing versus outgoing momenta flowing through the diagram, although here the recursion is formulated for Euclidean momenta $\boldsymbol{p}$ with momentum conservation enforced modulo integers.}
\label{fig:Mirzarec c1}
\end{figure}
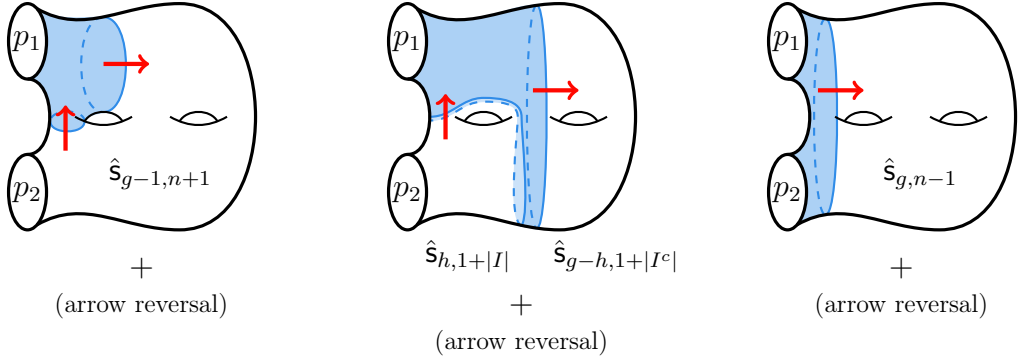

The recursion relations \eqref{eq:s recursion without deltas stripped}, \eqref{eq:shat recursion integral form}, \eqref{eq:Mirzarec c1}  are three-term relations of the same structural form as Mirzakhani’s recursive formula for the Weil-Petersson volumes of moduli space \cite{Mirzakhani:2006fta}, as depicted in figure \ref{fig:Mirzarec c1}. Here, however, the recursion computes a moduli space volume weighted by the string worldsheet path integral. The resulting family of weighted volumes encodes the physical S-matrix of the $c=1$ string in target spacetime.

\paragraph{Simplification.} The Mirzakhani-type recursion \eqref{eq:Mirzarec c1} can be further simplified.
It is useful to examine the general structure of the amplitudes $\hat{\mathsf{s}}$: as seen in the examples above, they consist of two classes of terms, polynomial contributions and terms involving Bernoulli polynomials whose arguments depend on the fractional parts of the momenta.
In what follows, we will consider terms of each type, performing the integer sums and evaluating the residue in the recursion \eqref{eq:Mirzarec c1}.

To begin, consider a contribution from the third term in the recursion (denoted by the superscript $(3)$)
\be
p_1\hat{\mathsf{s}}_{g,n}(p_1,\boldsymbol{p})^{(3)}
        = -\half\Res_{u=0}\frac{\sinh(u p_1)\ex^{u p_j}}{\sinh(\frac{u}{2})^2}\sum_{r \in \{p_1+p_j\}+\ZZ_{\ge 0}} r\ex^{u r}\hat{\mathsf{s}}_{g,n-1}(r,\boldsymbol{p}\setminus p_j)\, .
\ee
First, note that $\hat{\mathsf{s}}_{g,n-1}(r,\boldsymbol{p}\setminus p_j)$ becomes polynomial in its arguments when evaluated on points whose momenta differ by integers. In this situation the Bernoulli-type terms in the amplitude, which depend only on the fractional parts of the momenta, can be treated as constants. Since the amplitude is a finite polynomial in $r$ with coefficients that are independent of $r$, it suffices by linearity to evaluate the contribution for a single monomial $r^k$, with $k$ even and non-negative. One can perform the sum over $r$ for such a term by using
\be
    \sum_{r \in \{p_1+p_j\}+\ZZ_{\ge 0}} r^{k+1}\ex^{u r} =\partial_u^{k+1}\frac{\ex^{u \{p_1+p_j\}}}{1-\ex^{u}} \,. \label{eq:extract residue ex}
\ee
In order to compute the residue, we next consider the Laurent expansion of the RHS of \eqref{eq:extract residue ex} around $u=0$, using the identity
\be
\partial_u^a \frac{\ex^{u\alpha}}{1-\ex^u}=(-1)^{a+1}a!\,u^{-(a+1)} - \sum_{m=a+1}^\infty\frac{B_m(\alpha)}{m(m-a-1)!}u^{m-(a+1)}\, .
\ee
Only terms with non-positive powers of $u$ can contribute to the residue in \eqref{eq:extract residue ex}. Only the first two terms contribute and the last line of \eqref{eq:Mirzarec c1} becomes
\begin{align}
    p_1\hat{\mathsf{s}}_{g,n}(p_1,\boldsymbol{p})^{(3)}&=-\frac{1}{2}\Res_{u=0}\frac{\sinh(u p_1)\ex^{u p_j}}{\sinh(\frac{u}{2})^2}\bigg(\frac{(k+1)!}{u^{k+2}}-\frac{B_{k+2}(\{p_1+p_j\})}{k+2} \bigg) \nonumber\\
    &=\frac{B^{(2)}_{k+3}(p_j-p_1+1)-B^{(2)}_{k+3}(p_j+p_1+1)}{(k+2)(k+3)}+\frac{2p_1}{k+2} B_{k+2} (\{p_1+p_j\}) \,,
\end{align}
where we expressed the residue in terms of the second order Bernoulli polynomial\footnote{\texttt{NorlundB[n,2,x]} in \texttt{Mathematica}.}, defined as
\be
\frac{t^2 \, \ex^{x t}}{(\ex^t-1)^2}=\sum_{n=0}^\infty \frac{B_n^{(2)}(x)\, t^n}{n!}\,.
\ee
To streamline the notation in what follows it is convenient to introduce the following shorthands
\begin{align}
f_s(p) &\equiv \frac{B^{(2)}_s(1+p)-B^{(2)}_s(1-p)}{s!} \,\nonumber,\\
E_s(p) & \equiv \sum_{m\ne0}\frac{e^{2\pi i m p}}{(2\pi i m)^s} = - \frac{B_s(\{p\})}{s!}\, .\label{eq:modified bernoulli}
\end{align}
The polynomial $f_s(p)$ is only non-zero for odd $s$ due to the reflection property of the second-order Bernoulli polynomials, $B^{(2)}_s(x) = (-1)^s B_s^{(2)}(2-x)$. The function $E_s(p)$ depends only on the fractional part of $p$ and is even (odd) under reflection for even (odd) $s$. Since it only depends on the fractional part of the momenta it obeys the property $E_s(p_{a}) = (-1)^s E_s(p_{b})$ for any combination of momenta adding up to an integer, $p_{a}+p_{b} \in\mathbb{Z}$.

As noted earlier, the amplitudes $\hat{\mathsf{s}}_{g,n}$ may be regarded as polynomials in the momenta $\boldsymbol{p}$, with coefficients that depend on the fractional parts $\{p_I\}$ for all subsets $I$.
When evaluating the second and third terms in the recursion \eqref{eq:Mirzarec c1}, these fractional parts may be treated as constants.
Therefore, the contribution computed above can be written as
\begin{align}
p_1\hat{\mathsf{s}}_{g,n}(p_1,\boldsymbol{p})^{(3)}&=-\sum_{j=2}^n \sum_{k \ge 0} (k+1)! \bigg( \frac{f_{k+3}(p_1+p_j) + f_{k+3}(p_1-p_j)}{2} \nonumber\\
&\qquad+ 2p_1 E_{k+2}(p_1+p_j) \bigg) \,[p^k]\,\hat{\mathsf{s}}_{g,n-1}(p,\boldsymbol{p} \setminus p_j)\,.\label{eq:term3 simplified}
\end{align}
where $[p^k]$ extracts the coefficient of $p^k$ in the polynomial $\hat{\mathsf{s}}_{g,n-1}$, treating the fractional parts $\{p_I\}$ as constants.

This provides an efficient way to implement the third term of the recursion formula.
A notable feature of the result is that the contribution naturally splits into two cleanly separated parts: one involving fractional parts of the momenta (in $E_{k+2}(p_1+p_j)$), and one that does not.
The first contribution in \eqref{eq:term3 simplified} reproduces precisely the VMS recursion kernel, while the second accounts for the boundary contribution in the stable-graph expansion.

One can apply similar tricks to simplify the other contributions. Consider a contribution from the second term in the recursion \eqref{eq:Mirzarec c1} whose sub-amplitudes $\hat{\mathsf{s}}_{h,1+|I|}(r,\boldsymbol{p}_I)\hat{\mathsf{s}}_{g-h,1+|I^c|}(-r',\boldsymbol{p}_{I^c})$ contribute a monomial $r^k r'^{\ell}$ (again $k$ and $\ell$ are even and non-negative). Then the integrand contains $r^{k+1}r'^{\ell+1}$ and we have
\begin{align}
    &\half \Res_{u=0}\frac{\sinh(up_1)}{\sinh(\frac{u}{2})^2}\sum_{\substack{r\in\{-p_{I}\}+\mathbb{Z}_{\geq 0}\\ r'\in \{p_{I^c}\}+\mathbb{Z}_{\geq 0}}}r^{k+1}r'^{\ell+1} \ex^{u(r+r')} \nonumber\\
    &= \half \Res_{u=0}\frac{\sinh(up_1)}{\sinh(\frac{u}{2})^2}\left(\partial_u^{k+1}\frac{\e^{u\{-p_{I}\}}}{1-\e^u}\right)\left(\partial_u^{\ell+1}\frac{\e^{u\{p_{I^c}\}}}{1-\e^u}\right)\nonumber\\
    &=\half \Res_{u=0}\frac{\sinh(up_1)}{\sinh(\frac{u}{2})^2}\left((k+1)!u^{-(k+2)}-\sum_{m=k+2}^\infty \frac{1}{m(m-k-2)!} B_m(\{-p_{I}\})u^{m-(k+2)}\right)\nonumber\\
    &\quad\times\left((\ell+1)!u^{-(\ell+2)} - \sum_{m'=\ell+2}^\infty \frac{1}{m'(m'-\ell-2)!}B_{m'}(\{p_{I^c}\})u^{m'-(\ell+2)}\right)
    \, .
\end{align}
To proceed we now extract terms that contribute to the residue.
All together the residue is straightforwardly computed to be
\begin{align}
&\half \Res_{u=0}\frac{\sinh(up_1)}{\sinh(\frac{u}{2})^2}\sum_{\substack{r\in\{-p_{I}\}+\mathbb{Z}_{\geq 0}\\ r'\in \{p_{I^c}\}+\mathbb{Z}_{\geq 0}}}r^{k+1}r'^{\ell+1} \ex^{u(r+r')}\\
&= (k+1)!(\ell+1)!\bigg[f_{k+\ell+5}(p_1) + 2p_1 E_{k+2}(p_{I})E_{\ell+2}(p_{I^c})\nonumber\\
&\quad + \sum_{s=1}^{\ell+3}\tbinom{k+\ell+4-s}{\ell+3-s}f_s(p_1)E_{k+\ell+5-s}(p_{I})+ \sum_{s=1}^{k+3}\tbinom{k+\ell+4-s}{k+3-s}f_s(p_1)E_{k+\ell+5-s}(p_{I^c})\bigg]\, .
\end{align}
We see that the recursion generates terms that cleanly separate into those that are purely polynomial in $p_1$ (coming from the VMS recursion kernel), those that depend only on fractional parts of partial sums of the external momenta, and combinations of the two.
Thus we learn of the following contributions to the full amplitude from the second term in the recursion
\begin{align}\label{eq:term2 simplified}
    &p_1\hat{\mathsf{s}}_{g,n}(p_1,\boldsymbol{p})^{(2)}\nonumber\\
    &\qquad=\sideset{}{'}\sum_{h,I}\sum_{k,\ell\geq 0}(k+1)!(\ell+1)!\bigg(f_{k+\ell+5}(p_1) + 2p_1 E_{k+2}(p_{I})E_{\ell+2}(p_{I^c}) \nonumber\\
    &\qquad\quad + \sum_{s=1}^{\ell+3}\tbinom{k+\ell+4-s}{\ell+3-s}f_s(p_1)E_{k+\ell+5-s}(p_{I})+ \sum_{s=1}^{k+3}\tbinom{k+\ell+4-s}{k+3-s}f_s(p_1)E_{k+\ell+5-s}(p_{I^c})\bigg) \nonumber\\
    &\qquad\quad\quad\times [p^kp'^{\ell}] \, \hat{\mathsf{s}}_{h,1+|I|}(p,\boldsymbol{p}_{I})\hat{\mathsf{s}}_{g-h,1+|I^c|}(-p',\boldsymbol{p}_{I^c})\, ,
\end{align}
where again $[p^kp'^\ell]$ extracts the coefficient of the corresponding powers of the momenta while regarding the fractional parts as constant.

Finally, the first term of the recursion corresponds to the non-separating degeneration. The computation of the residue proceeds in exactly the same way as for the separating case, since the amplitude $\hat{\mathsf{s}}_{g-1,n+1}(r,-r',\boldsymbol{p})$ is again polynomial in $r$ and $r'$ with coefficients independent of $r,r'$. The result is obtained from \eqref{eq:term2 simplified} by the substitutions $p_I \to -q$ and $p_{I^c} \to q-p_1$, reflecting the momentum assignments in the non-separating channel.
Unlike the separating case, however, the fractional parts $\{q\}$ and $\{q-p_1\}$ now depend on the integration variable~$q$, and cannot be pulled out of the integral. This leads to the final contribution
\begin{align}\label{eq:term1 simplified}
    &p_1\hat{\mathsf{s}}_{g,n}(p_1,\boldsymbol{p})^{(1)}
    \nonumber\\
    &\qquad= \int_0^1 \d q\sum_{k,\ell\geq 0}(k+1)!(\ell+1)!\bigg(f_{k+\ell+5}(p_1) + 2p_1 E_{k+2}(q)E_{\ell+2}(q-p_1) \nonumber\\
    &\qquad\quad + \sum_{s=1}^{\ell+3}\tbinom{k+\ell+4-s}{\ell+3-s}f_s(p_1)E_{k+\ell+5-s}(q)+ \sum_{s=1}^{k+3}\tbinom{k+\ell+4-s}{k+3-s}f_s(p_1)E_{k+\ell+5-s}(q-p_1)\bigg) \nonumber\\
    &\qquad\quad\quad\times [q^kq'^{\ell}] \, \hat{\mathsf{s}}_{g-1,n+1}(q,-q',\boldsymbol{p})\, .
\end{align}
In this simplification we are still required to integrate factors involving Bernoulli polynomials depending on the fractional parts of partial sums of the momenta with respect to an internal momentum. In appendix \ref{app:Bernoulli integration} we explain how to carry out these integrals in practice, see (\ref{eq:first term recursion implemented}).

Combining \eqref{eq:term1 simplified}, \eqref{eq:term2 simplified} and \eqref{eq:term3 simplified} gives then a closed form recursion of the string amplitudes.

\section{Discussion} \label{sec:discussion}

In this paper, we derived an intersection number description of $c=1$ string amplitudes. The resulting expressions take the form of Feynman rules, sums over stable graphs with each graph contributing a VMS quantum volume, given in \eqref{eq:Feynman rules integer shifts}. These Feynman rules naturally compute discretized amplitudes in which momentum is conserved only modulo an integer. The physical $c=1$ amplitudes are recovered by restricting to the first Brillouin zone and analytically continuing to Lorentzian kinematics. We showed that these amplitudes satisfy perturbative spacetime unitarity and are piecewise polynomial, with the physical amplitudes exhibiting a reduced polynomial degree that is highly nontrivial from the worldsheet perspective. We further identified a spectral curve and topological recursion that govern the discretized amplitudes, together with a Mirzakhani-type recursion relation, thereby establishing a triality between the $c=1$ worldsheet string theory, matrix quantum mechanics, and a matrix integral. We conclude with a few remarks and open directions.

\paragraph{Generalization to $b \ne 1$.} Much of the analysis of this paper carries over to the case $b \ne 1$, i.e.\ the worldsheet theory where a Liouville theory of central charge $c=1+6(b+b^{-1})^2$ is coupled to a free boson with background charge $\hat{Q}=ib-ib^{-1}$ (sometimes referred to as a linear dilaton). Such a worldsheet theory can be viewed as problematic because a free boson with a background charge does not have a well-defined perturbative expansion. The dilaton varies linearly in time and there is no potential that keeps strings out of the strong coupling regime. Nonetheless, one can still compute the perturbative amplitudes with the result \eqref{eq:position space Feynman rules b ne 1}. These perturbative amplitudes do not have any obvious pathology. However, they exhibit the peculiar feature that momentum conservation depends on the genus, see eq.~\eqref{eq:anomalous momentum conservation}. Thus, for a given choice of external momenta, there is at most one perturbative amplitude that contributes to it, which then also equals the non-perturbative amplitude.
The corresponding momentum space Feynman rules read
\begin{multline}
    \hat{\mathsf{s}}^{(b)}_{g,n}(\boldsymbol{p})=\sum_{\Gamma \in \mathcal{G}_{g,n}}\frac{1}{|\text{Aut}(\Gamma)|}\int_{\RR/\ZZ} \d^L (b\boldsymbol{q})\int_{\bM_\Gamma} \prod_{v \in \mathcal{V}_\Gamma}\ex^{\frac{b^2+b^{-2}}{4} \kappa_1-\sum_k\frac{B_{2k}\kappa_{2k}}{2k \, (2k)!}}\prod_{i=1}^n \ex^{-p_i^2 \psi_i} \\
    \times\!\prod_{e=(\bullet,\circ) \in \mathcal{E}_\Gamma} \sum_{d=0}^\infty \frac{B_{2d+2}(\{b p_e\})(-\psi_\bullet-\psi_\circ)^d}{b^{2d+2}(d+1)!}\ . \label{eq:c=1 momentum space intersection numbers b ne 1}
\end{multline}
Here, momentum conservation at every vertex is anomalous, i.e.\ satisfies $\sum_i b p_i=\frac{1}{2}(b^2-1)(2g_v-2+n_v) \bmod 1$. There is also a corresponding dual topological recursion whose spectral curve is\footnote{The topological recursion depends on $\yy(z)$ only through its odd part $\frac{1}{2}(\yy(z)-\yy(\sigma(z)))$ under the local involution $\sigma(z)=2\pi m b-z$ at each branch point. For $b=1$, one can therefore replace $\yy(z) = -i\,\ex^{iz}$ by $\frac{1}{2}(-i\,\ex^{iz}+i\,\ex^{-iz})=\sin(z)$, which reproduces the spectral curve \eqref{eq:c1 spectral curve}. For $b \ne 1$, this reduction does not work independently of $m$.}
\be
\xx (z)=2\sqrt{2}\cos(b^{-1} z)\ , \qquad \yy(z)=-i\, \ex^{i b z}\ ,
\ee
whose set of branch points are located at $z=\pi m b$ with $m \in \ZZ$. This may be viewed as a family of deformations of the $c=1$ spectral curve, but its physical interpretation remains unclear to us.

A somewhat analogous situation appears in $\mathrm{AdS}_3$ holography with pure NS-NS background. For simplicity, we will restrict to bosonic strings. The $\mathrm{AdS}_3$ sigma model is described by the $\SL(2,\RR)_k$ WZW model on the worldsheet. It was shown in \cite{Eberhardt:2025sbi} that for $k=3$, an alternative localizing model exists that describes a localizing version of bosonic strings on $\mathrm{AdS}_3' \times X$ (where we view $X$ as the internal CFT). Its holographic dual is given by the CFT $\text{Sym}^N(\RR \times X)$, see also \cite{Eberhardt:2021vsx}. This model is in some ways similar to the $c=1$ string as it also has an exact momentum conservation corresponding to the translation symmetry along $\RR$ in the dual CFT. As in the $c=1$ string, one can turn on a linear dilaton slope on both sides. This modifies the dual CFT to $\text{Sym}^N(\RR_Q \times X)$, where $\RR_Q$ denotes a free boson with dilaton slope $Q$. Similar to the deformed $c=1$ string discussed above, only a single term contributes in the $N^{-1}$ expansion for fixed external momenta. On the worldsheet, one can also define a localizing model $\SL(2,\RR)_k'$ at generic $k$ (the slope of the linear dilaton is related to $k$ by $Q=\frac{k-3}{\sqrt{k-2}}$). The worldsheet theory is consistent with the only caveat that the OPE of physical vertex operators doesn't formally close because of the anomalous momentum conservation. This is also completely analogous to what happens in the linear dilaton theory where for a three point function, it is impossible to satisfy momentum conservation $\alpha_1+\alpha_2+\alpha_3=Q$ for physical vertex operators with $\alpha_i \in \frac{Q}{2}+i \RR$. Nonetheless, the worldsheet theory exists in the sense that it produces crossing symmetric correlators that can be consistently integrated over moduli space. It has also been discussed in \cite{Knighton:2024qxd}, where it was viewed as the leading term in a near-boundary expansion.

\paragraph{Chain of matrices.} The analysis in this paper raises the question: where does the matrix integral come from? We have phrased everything in terms of abstract topological recursion. Such a recursion can be realized inside an actual matrix integral in various different ways. While a single matrix integral can only accommodate hyperelliptic spectral curves for which $\yy(z)=z^2$, the $c=1$ spectral curve \eqref{eq:c1 spec param} is not of this form. It necessitates at least a \emph{two}-matrix integral \cite{Chekhov:2006vd}.
The natural arena for the matrix integral is precisely as a discretized matrix quantum mechanics integral, which takes the form
\be
\int \prod_i [D^{N\times N} \Phi_i]\, \exp\bigg[-\beta \sum_{i} \Big(\tfrac{1}{2 \varepsilon} (\Phi_{i+1}-\Phi_i)^2+\varepsilon \tr W(\Phi_i)\Big)\bigg]\ . \label{eq:c1 discretized MQM}
\ee
This is a matrix integral with potentially infinitely many matrices $\Phi_i$. The first term is a discretization of the kinetic term of matrix quantum mechanics and the second a potential. We expect that the appropriate choice is $W(\Phi)=-\frac{1}{2}\Phi^2$, which leads to the inverted harmonic oscillator potential of MQM. In eq.~\eqref{eq:c1 discretized MQM} we followed the notation of Gross and Klebanov \cite{Gross:1990ub}, who already conjectured that this matrix integral should be \emph{exactly} equivalent to MQM for small enough discretization length. One can view the triality discussed in this paper as a confirmation of Gross and Klebanov's conjecture.

This matrix integral has precisely the `chain of matrices' form studied in \cite{Eynard:2003kf, Eynard:2009zz} (but in a limit in which the length of the chain goes to infinity), where it was shown that observables in such a model obey topological recursion. Since the model is Gaussian for $W(\Phi)=-\frac{1}{2}\Phi^2$, the spectral curve is expected to be genus 0, consistent with the fact that a genus-0 curve has no moduli and the Gaussian model has no tunable couplings. This is precisely realized by the spectral curve \eqref{eq:c1 spec param}, the only surprise being that a particular cover of the spectral curve appears since infinitely many branch points participate in the topological recursion. We leave the explicit verification that the chain-of-matrices topological recursion reproduces our proposed recursion to future work.

\paragraph{Spectral curve from MQM.}
The $c=1$ spectral curve was previously motivated in \cite{Alexandrov:2004ks}, but to our knowledge it was never taken seriously as a spectral curve of a matrix integral.
Following the logic developed for the A-series minimal string \cite{Seiberg:2003nm}, one may define the spectral curve in terms of the boundary cosmological constant $\mu_\text{B}$ and the marked disk partition function with FZZT boundary conditions.
By analyzing the $b\to1$ limit, \cite{Alexandrov:2004ks} argued that the relevant curve arises from the subleading terms of the disk partition function, yielding $\xx(s)=\sqrt{\mu}\cosh(2\pi s)$ and $\yy(s)=\sqrt{\mu}\,2\pi s\sinh(2\pi s)$, where $\mu$ is the Liouville bulk cosmological constant.
To make contact with the hyperboloid in phase space describing the semiclassical ground state of the free-fermion system underlying $c=1$ MQM, one introduces the momentum variable $\mathsf{p}(s)$, related to the eigenvalue density obtained from the jump of the resolvent between two sheets.
This leads to the MQM spectral curve
\begin{align}
\xx(s)=\sqrt{\mu}\cosh(2\pi s)\,, \qquad
\mathsf{p}(s)=\frac{1}{2\pi i}\big(\yy(s+i)-\yy(s)\big)=\sqrt{\mu}\sinh(2\pi s)\, ,
\end{align}
which satisfies $\mathsf{p}^2-\xx^2=-\mu$, matching the Fermi sea profile in the semiclassical ground state of the MQM.
This coincides with our spectral curve \eqref{eq:c1 spectral curve}, up to reparametrization and setting $\mu=1$.
Notice that from standard genus counting, we have that $\omega_{0,1}\propto g_{\text{s}}^{-1}$. With the parametrization $\mathsf{p}^2-\mathsf{x}^2=-\mu$, $\omega_{0,1}=-\mathsf{p}(z)\d\xx(z) \propto \mu$ yielding the correct dictionary $\mu \propto g_{\text{s}}^{-1}$ between $c=1$ string theory and the MQM.

\paragraph{The chiral formalism.} A more direct perspective on the spectral curve in MQM arises in the chiral formalism developed in \cite{Alexandrov:2002fh,Alexandrov:2003ut}.
In this approach one studies the quantization of the single-fermion Hamiltonian $H=\tfrac12(p^2-x^2)$ in chiral variables $x_{\pm}=(x\pm p)/\sqrt2$.
The exact eigenfunctions $\psi_E^\pm(x_\pm)$ are related by a Fourier-transform integral, which in the string theoretic interpretation corresponds to the $S$-matrix relating asymptotic in and out closed string states.

The chiral formalism also provides a natural framework to study Euclidean sine-Liouville-type tachyon deformations, each associated with a deformation of the $c=1$ spectral curve.
It would be interesting to explore the topological recursion and intersection theory associated with these curves, as they may provide a window into time-dependent string backgrounds in two-dimensional string theory.

\paragraph{No recursion for physical amplitudes?}
The Mirzakhani-type recursion relation of section~\ref{subsec:recursion relation} acts on the discretized amplitudes $\hat{\mathsf{s}}_{g,n}$, which conserve momentum only modulo an integer. It does not close on the physical amplitudes $\mathsf{S}_{g,n}$, since intermediate states necessarily involve momentum conservation only modulo an integer. A natural question is whether there exists a refined recursion relation that closes directly on the physical amplitudes.

The main obstruction is that the physical amplitudes $\mathsf{S}_{g,n}(\omega_1,\dots,\omega_n)$ do not naturally originate from topological recursion, only the discretized amplitudes do. Of course, this is only suggestive and does not exclude the existence of a recursive formula of a different kind.

\paragraph{ZZ-instantons.}
It was observed in \cite{Alexandrov:2023fvb} that, in the presence of tachyon (sine-Liouville-type) deformations, ZZ-instantons arise as saddle points of the Fourier-transform integral relating in- and out-states in the chiral formalism.
Promoting the corresponding saddle-point equations to complex variables naturally leads to complex curves with double-point singularities, demonstrating a highly nontrivial agreement with ZZ-instanton effects computed using string field theory.
In the undeformed $c=1$ theory the relevant saddles are more degenerate, but they can nevertheless be analyzed directly at the level of saddle points of the Fourier-transform integral \cite{Alexandrov:2025pzs}.

The computation of ZZ-instanton contributions in matrix integrals is also well-understood \cite{Daul:1993bg, Bertola:2003rp, Ishibashi:2005zf, Eniceicu:2022dru}. They arise by performing a saddle point approximation of the effective potential felt by a single eigenvalue. However, for the spectral curve of the $c=1$ string, the formulas in the literature break down, because the Hessian around the saddle point turns out to have a zero eigenvalue. The analogous phenomenon was observed for the Virasoro minimal string at $b=1$, whose non-perturbative structure differs drastically from the $b \ne 1$ model \cite{Collier:2023cyw}. Physically, the zero eigenvalue seems to be a manifestation of the instanton zero mode associated with translations in target space. At present, it is not known how to correctly decouple this zero mode in the matrix model description and extend the triality described in this paper to the non-perturbative level.

\paragraph{Long strings, the adjoint sector of the MQM, and the gluing formula.}
Another future direction is the study of long strings in $c=1$ string theory \cite{Maldacena:2005hi,Karczmarek:2008sc,Balthazar:2018qdv}, which arise in a particular high-energy limit of long folded open strings on receding FZZT branes that extend back into the interacting region of spacetime near the Liouville wall.
These asymptotic states were conjectured to be dual to states in the adjoint sector of the MQM, where the fermion Hamiltonian acquires a highly nontrivial interaction among the fermions.
It would be interesting to understand their counterparts on the matrix integral side of the triality.

Furthermore, in view of the gluing formula discovered in \cite{Collier:2024mlg, DinstantonsGluing}, the worldsheet computation of long string scattering amplitudes appears to be only modestly more involved than the perturbative amplitudes described in this paper.
Long strings and their conjectured relation to a (Lorentzian) two-dimensional black hole were revisited in \cite{Witten:1991yr,Kazakov:2000pm,Maldacena:2005hi,Betzios:2022pji,Ahmadain:2022gfw}, as well as recent developments in the adjoint sector of the MQM in \cite{Klebanov:2026vmk}.

\paragraph{A new 3d TQFT?}
In the Virasoro minimal string \cite{Collier:2023cyw}, the string amplitudes (the quantum volumes $\mathsf{V}_{g,n}^{(b)}(\boldsymbol{p})$) admitted an elegant interpretation in terms of counting conformal blocks. The quantum volumes were realized as the trace of the identity in the (infinite-dimensional) Hilbert space of Virasoro TQFT associated to the worldsheet surface, gauged by the mapping class group of the surface. The origin of this phenomenon is that the conformal block inner product in VTQFT (the ``Verlinde inner product'' \cite{Verlinde:1989ua, Collier:2023fwi}), which is defined by integrating conformal blocks weighted by the correlation function of timelike Liouville CFT with central charge $26-c$ over Teichm\"uller space, is structurally similar to the worldsheet path integral of VMS. The latter arises from the former by taking the trace and gauging by the mapping class group. One might wonder whether $c=1$ string theory could analogously be used to define an alternative inner product on the vector space of $c=25$ Virasoro blocks with conformal weights $h_j=1-p_j^2$ and $\sum_j p_j=0$, in which the free boson path integral on the worldsheet replaces the timelike Liouville correlator as the measure. For $\Psi$ and $\Psi'$ conformal blocks with $c=25$ we can set
\be
\langle \Psi \mid \Psi' \rangle_{c=25}:= \int_{\mathcal{T}_{g,n}} \Psi^* \Psi' \,\langle V_{p_1}(z_1) \cdots V_{p_n}(z_n) \rangle'\ , \label{eq:modified inner product}
\ee
where $\langle V_{p_1}(z_1) \cdots V_{p_n}(z_n) \rangle'$ denotes the free boson correlator where we stripped off the momentum-conserving delta function. The integrand in \eqref{eq:modified inner product} also involves $b$-ghost insertions as in \eqref{eq:c=1 S-matrix worldsheet definition}. The inner product is positive definite because the free boson correlator is positive for fixed external momenta (either all real or all imaginary) for all moduli.
Normalizability with respect to this inner product is different from the notion of normalizability in VTQFT and thus this leads to an a priori different topology on the space of conformal blocks.
Crossing transformations act unitarily with respect to this inner product, since crossing transformations of $\Psi$ and $\Psi'$ can be absorbed by a coordinate change and using that the free boson correlator itself is crossing symmetric. Thus, we get a unitary representation of the set of crossing transformations on the resulting Hilbert space. Modulo possible functional analytic issues, such data defines a 3d TQFT built out of Virasoro conformal blocks of central charge $c=25$. It would be interesting to explore this TQFT further.

\paragraph{Generalization to $\mathcal{N}=1$ 2d strings.}
We expect the $c=1$ triality to extend to the type 0A and 0B two-dimensional string theories, whose worldsheet CFTs have $\mathcal{N}=(1,1)$ superconformal symmetry \cite{Douglas:2003up,Takayanagi:2003sm}. 
Unlike the 2d versions, the super-$\CC$LS \cite{Du:2025lya} and super-VMS \cite{Muhlmann:2025ngz,Rangamani:2025wfa,Johnson:2024fkm} worldsheet theories admit two inequivalent $\mathcal{N}=(1,1)$ algebras on the worldsheet for each type 0A/B and the choice of which one should be gauged leads to different string theories. 
In particular, one of the super-$\CC$LS type 0B theories is expected to recover the 2d type 0B string through the same procedure of extracting the relevant residue from the leading pole of the super-$\CC$LS string. 
Furthermore, we expect the resulting spectral curve, or more precisely the perturbative topological recursion, to coincide with the $c=1$ recursion discussed in this paper. 
While this is manifest from the MQM side of the triality --- the MQM state dual to the type 0B closed string vacuum corresponds to a Fermi sea filled on both sides of the potential, leading to two decoupled degrees of freedom, each with perturbative S-matrix elements identical to the $c=1$ case --- it is obscured from the worldsheet point of view \cite{Balthazar:2022atu}. 
This MQM expectation provides further evidence that this 0B super-$\CC$LS spectral curve coincides with the bosonic one, as proposed in \cite{Du:2025lya}. 
It would be interesting to understand whether this perturbative decoupling, and the resulting equivalence to the $c=1$ amplitudes, becomes manifest at the level of topological recursion or the associated intersection theory formula. 
Of course, non-perturbatively the two theories differ significantly; for instance, the spectrum of nodal singularities is expected to differ, reflecting the distinct D-instanton spectra of the type 0B and $c=1$ strings, and certain sectors of the type 0B S-matrix are IR divergent \cite{DeWolfe:2003qf,Balthazar:2022apu,Sen:2022clw}.

\section*{Acknowledgements}
We would like to thank Aleksandr Artemev, Xi Dong, Kelian H\"aring, Igor Klebanov, Clifford V. Johnson, Juan Maldacena, Beatrix M\"uhlmann, Nikita Nekrasov, Cumrun Vafa and Xi Yin for fruitful discussions and comments. We especially thank Beatrix M\"uhlmann for collaboration on related topics.
LE is supported by the European Research Council (ERC) under the European Union’s Horizon 2020 research and innovation programme (grant agreement No 101115511).
VAR is supported by the University of California President’s Postdoctoral Fellowship.

\appendix

\section{Amplitudes from free fermions}\label{app:MPR comparison}

In this appendix we briefly review the computation of S-matrix elements in $c=1$ string theory from the free-fermion perspective, in order to compare with the S-matrix elements obtained from the $c=1$ matrix integral. In \cite{Moore:1991zv}, Moore, Plesser, and Ramgoolam (MPR) developed a diagrammatic formalism that computes the exact S-matrix of the $c=1$ matrix quantum mechanics (MQM). In the double-scaling limit, the MQM reduces to a system of free fermions in an inverted harmonic potential.
Their construction implements an LSZ-type prescription that extracts scattering amplitudes from the exact single-particle eigenfunctions of the inverted oscillator.
The resulting diagrammatic rules yield the full non-perturbative S-matrix, including instanton corrections in the string coupling.

The state dual to the closed-string vacuum of the $c=1$ string is the one in which all scattering eigenstates with no incoming flux from the left side of the potential are occupied up to a Fermi energy level $\mu$ \cite{Balthazar:2019rnh}. The dictionary relating the MQM to the string theory is $g_s \propto \mu^{-1}$.\footnote{In the semi-classical limit of large $\mu$
this reduces to the usual picture in which the fermions fill the right Fermi sea of the inverted harmonic oscillator.}

An important ingredient in the MPR formalism is the reflection coefficient of a single scattering eigenstate. For an eigenstate $|E\rangle_{R}$ with no incoming flux from the left, the exact reflection coefficient takes the form
\begin{align}
R(\mu,E) = i\mu^{iE} \left[ \frac{1}{1+\e^{2\pi E}} \frac{\Gamma(\frac{1}{2}-iE)}{\Gamma(\frac{1}{2}+iE)} \right]^{\frac{1}{2}} \,.
\end{align}
As an example, the MPR diagrammatics yields the following simple formula for the $n-1 \to 1$ scattering amplitude, which reads after translating to our conventions (assuming $\omega_1,\dots,\omega_{n-1} \ge 0$),
\begin{multline}
\sum_{g=0}^\infty \hat{\mathsf{S}}_{g,n}(\omega_1,\ldots,\omega_{n}) g_\text{s}^{2g-2+n}\\
=-\frac{i}{2} \biggr(\prod_{j=1}^{n} \frac{\sqrt{2}}{\omega_j}\biggr) \sum_{I \subset [n-1]} (-1)^{|I|} \int_0^{\omega_I} \d x\, \frac{R(\mu,-\mu+\omega_{n}-x)}{R(\mu,-\mu -x)}\bigg|_{\mu=\frac{\sqrt{2}}{g_\text{s}}} \ , \label{eq:MPR n to 1}
\end{multline}
with $\omega_I=\sum_{i \in I} \omega_i$.
Expanding \eqref{eq:MPR n to 1} at small $g_\text{s}$ lets one extract arbitrary loop orders.\footnote{In our conventions, \eqref{eq:MPR n to 1} differs from the formula in \cite{Balthazar:2019rnh} by an overall factor $\frac{i}{2}$, which can be absorbed into the normalization $\mathcal{N}$ in \eqref{eq:S delta function strip off normalization}. In addition, our conventions interchange incoming and outgoing states relative to \cite{Balthazar:2019rnh}, a difference that can be traced back to the Euclidean rotation \eqref{eq:p omega relation}.}
The overall factor reflects our normalization of states: $\ket{\omega}^{\text{ours}}=\frac{\sqrt{2}}{\omega}\ket{\omega}^{\text{MPR}}$.
The amplitudes computed from \eqref{eq:MPR n to 1} agree with \eqref{eq:Sgn direct computation} for all cases displayed there.

\section{Some algebraic geometry} \label{app:algebraic geometry}
In this appendix, we collect the algebraic geometry needed for section~\ref{subsec:tree level rewriting}. We review the excess intersection formula on $\bM_{g,n}$, derive a graph expansion identity for exponentials of boundary classes, and extend it to the momentum-dependent setting relevant for $c=1$ string amplitudes. See \cite{JPPZ, Bae_Schmitt_Skowera_2022} for related mathematical accounts.

\subsection{Excess intersection formula}
Consider a stable graph $\Gamma$ with gluing morphism $\xi_\Gamma:\bM_\Gamma=\prod_{v \in \mathcal{V}_\Gamma} \bM_{g_v,n_v} \longrightarrow \bM_{g,n}$. For a class $\alpha \in \mathrm{H}^\bullet(\bM_\Gamma,\QQ)$ (we will suppress the coefficient ring in the following), we write $[\Gamma,\alpha]\equiv (\xi_\Gamma)_*(\alpha) \in \mathrm{H}^\bullet(\bM_{g,n})$. Following the conventions of the main text, we do not mod out by $\text{Aut}(\Gamma)$ in the definition of $\bM_\Gamma$ and instead track automorphism factors explicitly.

The main computational tool is the excess intersection formula \cite[Section~6.3]{Fulton:intersection}. For a boundary divisor $\mathscr{D}$ with inclusion $\xi_\mathscr{D}:\mathscr{D} \hookrightarrow \bM_{g,n}$, the normal bundle $\mathscr{N}$ of $\mathscr{D}$ has first Chern class $c_1(\mathscr{N})=-\psi_\bullet-\psi_\circ$, where $\psi_\bullet$ and $\psi_\circ$ are the cotangent line classes at the two branches of the node. The self-intersection formula then reads
\be
\xi_\mathscr{D}^*\,(\xi_\mathscr{D})_*(\alpha)=(-\psi_\bullet-\psi_\circ)\,\alpha\ . \label{eq:excess intersection}
\ee
From this and the projection formula, one derives inductively that for $k$ classes $\alpha_1,\dots,\alpha_k \in \mathrm{H}^\bullet(\mathscr{D})$,
\be
\prod_{i=1}^k (\xi_\mathscr{D})_*(\alpha_i)=(\xi_\mathscr{D})_* \bigg[(-\psi_\bullet-\psi_\circ)^{k-1}\prod_{i=1}^k \alpha_i\bigg]\ . \label{eq:pushforward product}
\ee
More generally, when two boundary strata $\bM_{\Gamma_1}$ and $\bM_{\Gamma_2}$ intersect, their set-theoretic intersection decomposes as
\be
\bM_{\Gamma_1} \cap \bM_{\Gamma_2}=\bigsqcup_{(\Gamma,\varphi_1,\varphi_2) \in \mathcal{G}_{\Gamma_1,\Gamma_2}} \bM_{\Gamma}\ .
\ee
Here $\mathcal{G}_{\Gamma_1,\Gamma_2}$ is the set of stable graphs $\Gamma$ equipped with graph morphisms $\varphi_i:\Gamma \to \Gamma_i$ (meaning $\Gamma$ is a degeneration of both $\Gamma_i$), subject to the constraint that every edge of $\Gamma$ lies in the image of at least one $\varphi_i$. The cup product of the corresponding tautological classes is given by \cite[Theorem~2.25]{Schmitt2020},
\be
[\Gamma_1,\alpha_1]\,[\Gamma_2,\alpha_2]=\!\!\sum_{(\Gamma,\varphi_1,\varphi_2) \in \mathcal{G}_{\Gamma_1,\Gamma_2}} \!\!\bigg[\Gamma, \xi_{\varphi_1}^*(\alpha_1)\,\xi_{\varphi_2}^*(\alpha_2) \!\!\prod_{e \in \varphi_{1}(\mathcal{E}_{\Gamma_1}) \cap \varphi_{2}(\mathcal{E}_{\Gamma_2})} \!\!(-\psi_\bullet-\psi_\circ)\bigg]\ , \label{eq:product tautological classes}
\ee
where $\xi_{\varphi_i}:\bM_\Gamma \to \bM_{\Gamma_i}$ is the forgetful map induced by $\varphi_i$ and the product over shared edges is the excess contribution. By induction, the $k$-fold product generalizes to
\be
    \prod_{i=1}^k [\Gamma_i,\alpha_i]=\sum_{(\Gamma,\varphi_1,\dots,\varphi_k)} \bigg[\Gamma,\prod_i \xi_{\varphi_i}^*(\alpha_i) \prod_{e \in \mathcal{E}_\Gamma} (-\psi_\bullet-\psi_\circ)^{\#\{i \, |\, e \in \varphi_i(\mathcal{E}_{\Gamma_i})\}-1}\bigg] \ , \label{eq:arbitrary product tautological classes}
\ee
where the sum runs over stable graphs $\Gamma$ such that each edge of $\Gamma$ appears in at least one $\varphi_i(\mathcal{E}_{\Gamma_i})$.

\subsection{Graph expansion}
We now specialize to the case where all $\Gamma_i$ are boundary divisors (stable graphs with a single edge). Let $\alpha \in \QQ[[\psi_\bullet,\psi_\circ]]$ be a formal power series in the nodal $\psi$-classes, taken to be the same for every divisor. We claim:
\begin{align}
    \exp\bigg(\sum_{\mathscr{D}} \frac{[\mathscr{D},\alpha]}{|\text{Aut}(\mathscr{D})|}\bigg)=\sum_{\Gamma \in \mathcal{G}_{g,n}}\frac{1}{|\text{Aut}(\Gamma)|} \bigg[\Gamma,\prod_{e \in \mathcal{E}_\Gamma}\frac{\ex^{(-\psi_\bullet-\psi_\circ)\alpha}-1}{-\psi_\bullet-\psi_\circ} \bigg]\ . \label{eq:graph expansion}
\end{align}
Here, the sum on the left runs over all boundary divisors of $\bM_{g,n}$. The automorphism group $\text{Aut}(\mathscr{D})$ is $\ZZ_2$ for $\mathscr{D}_\text{irr}$ and $\mathscr{D}_{g/2,\emptyset}$ (on $\bM_{g,0}$ with $g$ even), and trivial otherwise. On the right, the sum runs over all stable graphs of type $(g,n)$, including the trivial graph (which contributes the $k=0$ term, i.e.\ the identity $1$).

To prove \eqref{eq:graph expansion}, we expand the exponential as follows,
\be
\exp\bigg(\sum_{\mathscr{D}} \frac{[\mathscr{D},\alpha]}{|\text{Aut}(\mathscr{D})|}\bigg)=\sum_{(k_{\mathscr{D}})_\mathscr{D} \ge 0} \prod_\mathscr{D} \frac{[\mathscr{D},\alpha]^{k_{\mathscr{D}}}}{|\text{Aut}(\mathscr{D})|^{k_{\mathscr{D}}}\, k_\mathscr{D}!}\ . \label{eq:exponential expansion}
\ee
We evaluate each term using \eqref{eq:arbitrary product tautological classes}. For a given stable graph $\Gamma$, each morphism from one of the $k_\mathscr{D}$ copies of the divisor $\mathscr{D}$ sends its unique edge to some edge $e \in \mathcal{E}_\Gamma(\mathscr{D})$, where $\mathcal{E}_\Gamma(\mathscr{D})=\{e \in \mathcal{E}_\Gamma \mid \mathscr{D}_e=\mathscr{D}\}$. Let $k_e \ge 1$ denote the number of copies mapping to edge $e$, so that $k_\mathscr{D}=\sum_{e \in \mathcal{E}_\Gamma(\mathscr{D})} k_e$. Since each copy of $\mathscr{D}$ can be oriented in $|\text{Aut}(\mathscr{D})|$ ways, the number of ordered morphism tuples with prescribed $(k_e)$ is $\prod_\mathscr{D} |\text{Aut}(\mathscr{D})|^{k_\mathscr{D}} \frac{k_\mathscr{D}!}{\prod_{e \in \mathcal{E}_\Gamma(\mathscr{D})} k_e!}$. This factor precisely cancels the $|\text{Aut}(\mathscr{D})|^{k_\mathscr{D}} k_\mathscr{D}!$ in \eqref{eq:exponential expansion}, and summing over $(k_\mathscr{D})$ subject to $k_\mathscr{D}=\sum_e k_e$ becomes unrestricted. The remaining overcounting from graph automorphisms permuting edges of the same type contributes $\frac{1}{|\text{Aut}(\Gamma)|}$. Combining with \eqref{eq:exponential expansion}, we obtain
\begin{align}
    \exp\bigg(\sum_{\mathscr{D}} \frac{[\mathscr{D},\alpha]}{|\text{Aut}(\mathscr{D})|}\bigg)&=\sum_{\Gamma \in \mathcal{G}_{g,n}} \frac{1}{|\text{Aut}(\Gamma)|} \!\sum_{\substack{(k_e)_{e \in \mathcal{E}_\Gamma} \\ k_e\ge 1}} \!\bigg[\Gamma,\prod_{e \in \mathcal{E}_\Gamma }\frac{\alpha^{k_e}(-\psi_\bullet-\psi_\circ)^{k_e-1}}{k_e!}\bigg] \ .
\end{align}
Summing over $k_e$ independently for each edge yields the claimed formula \eqref{eq:graph expansion}.

\subsection{Tree-level graph expansion with momenta} \label{subsec:tree level graph expansion}
For the application in section~\ref{subsec:tree level rewriting}, we need a variant of \eqref{eq:graph expansion} where the class $\alpha$ depends on the edge momentum.

At tree level ($g=0$), all edges are separating and the edge momentum $p_e$ is determined by the external momenta: a separating edge of type $(0,I)$ carries momentum $p_I=\sum_{i \in I} p_i$. We assume that $\alpha$ is an even function of $p_I$, since the sign of $p_I$ depends on the choice of which side of the node is labelled by $I$ vs.\ $I^c$. Since distinct separating divisors $\mathscr{D}_{0,I}$ and $\mathscr{D}_{0,J}$ (with $\{I,I^c\} \ne \{J,J^c\}$) intersect transversally whenever they are compatible, the combinatorial argument of the previous subsection applies divisor by divisor. More precisely, because distinct genus-0 divisors have no shared edges, the excess factors $(-\psi_\bullet-\psi_\circ)$ in \eqref{eq:arbitrary product tautological classes} only arise from multiple copies of the same divisor, so the proof of \eqref{eq:graph expansion} carries through with $\alpha$ replaced by $\alpha_{p_I}$ for each divisor $\mathscr{D}_{0,I}$ independently. We obtain
\be
\exp\bigg(\frac{1}{2}\sideset{}{'}\sum_{I \subset [n]} (\xi_I)_*\,\alpha_{p_I}\bigg)=\sum_{\Gamma \in \mathcal{G}_{0,n}}\frac{1}{|\text{Aut}(\Gamma)|} \bigg[\Gamma,\prod_{e \in \mathcal{E}_\Gamma}\frac{\ex^{(-\psi_\bullet-\psi_\circ)\alpha_{p_e}}-1}{-\psi_\bullet-\psi_\circ} \bigg]\ . \label{eq:tree level graph expansion with momenta}
\ee
Here, $\xi_I \equiv \xi_{0,I}$ and the factor $\frac{1}{2}$ compensates the double counting from summing over both $I$ and $I^c$. The primed sum restricts to $2 \le |I| \le n-2$ (stability). On the right-hand side, only tree graphs appear since all genus-0 edges are separating, and $p_e$ denotes the momentum flowing through the edge $e$. This identity is used in \eqref{eq:tree level sum over graphs identity} of the main text, with $\alpha_{p}=\mathcal{F}(\{p\},-\psi_\bullet-\psi_\circ)$ as defined in \eqref{eq:mathcal F definition}.

\section{Integrating Bernoulli polynomials with fractional parts} \label{app:Bernoulli integration}

Throughout the main text, we often encountered Bernoulli polynomials featuring arguments that depend on the fractional parts of partial sums of the Euclidean momenta. These played an important role in the discretized amplitudes and in the analytic continuation to the physical sheet, they lead to the discontinuities responsible for perturbative spacetime unitarity. In both the stable graph expansion Feynman rules (\ref{eq:c=1 momentum space intersection numbers}) and the Mirzakhani recursion (\ref{eq:Mirzarec c1}), we have to integrate products of Bernoulli polynomials involving fractional parts of partial sums of momenta that share some internal momentum. Here we describe how to deal with these sums in practice.

\paragraph{Loop integrals in momentum space Feynman rules.}
In the momentum space Feynman rules for the discretized $c=1$ string amplitudes (\ref{eq:c=1 momentum space intersection numbers}), we have to integrate loop momenta over the compact space $\mathbb{R}/\mathbb{Z}$. Each stable graph carries the edge factors $B_{2d+2}(\{p_e\})$, depending on the fractional part of the edge momentum. Thus we have to integrate products of Bernoulli polynomials of the following form\footnote{For the application to the loop integrals the Bernoulli polynomials are always of even degree $n_j = 2d_j+2$, but for other purposes we will find it useful to consider the more general case.} 
\begin{equation}
    \int_{\mathbb{R}/\mathbb{Z}}\d q\, \prod_{j=1}^N B_{n_j}(\{q+q_j\})\, ,
\end{equation}
where $q_j$ correspond to partial sums of the external and internal momenta. These integrals can be tedious to evaluate in general because of the many case distinctions depending on the hierarchy of the fractional parts of the $q_j$. To deal with these integrals, we appeal to the Fourier expansion of the Bernoulli polynomials
\begin{align}\label{eq:loop momentum integral as funny sum}
    \int_{\mathbb{R}/\mathbb{Z}}\d q\, \prod_{j=1}^N B_{n_j}(\{q+q_j\}) &= \prod_{j=1}^N(-n_j!)\int_{\mathbb{R}/\mathbb{Z}}\d q\sum_{m_j\ne 0,\, j=1,\ldots, N}\frac{\e^{2\pi i \sum_{j=1}^Nm_j(q+ q_j)}}{\prod_{j=1}^N(2\pi i m_j)^{n_j}}\\
    &= \prod_{j=1}^N (-n_j!)\sum_{\substack{m_j\ne 0,\, j=1,\ldots,N\\\sum_{j=1}^N m_j =0}}\frac{\e^{2\pi i \sum_{j=1}^Nm_j q_j}}{\prod_{j=1}^N(2\pi i m_j)^{n_j}}\, .
\end{align}
To proceed, we will analyze the following closely-related constrained sums
\begin{equation}\label{eq:constrained sum}
    \mathcal{D}^{(N)}_a(n_1,\ldots,n_N;q_1,\ldots,q_N) \equiv \sum_{\substack{m_j\ne0,\,j=1,\ldots,N\\\sum_{j=1}^Nm_j\ne 0}}\frac{\e^{2\pi i(\sum_{j=1}^N m_j q_j)}}{\prod_{j=1}^N (2\pi i m_j)^{n_j}(\sum_{j=1}^N 2\pi i m_j)^{a}}\, .
\end{equation}
This sum admits a relatively straightforward recursion relation, which can be derived by making use of the partial fraction expansion to expand one of the terms in the denominator. Defining $\mathcal{N}\equiv n_N+a$, we have
\begin{align}
    &\mathcal{D}_a^{(N)}(n_1\ldots,n_N;q_1,\ldots,q_N)\nonumber\\
    &\quad= (-1)^{n_1}E_{\mathcal{N}}(q_N)\mathcal{D}_{n_1}^{(N-2)}(n_2,\ldots,n_{N-1};q_2-q_1,\ldots q_{N-1}-q_1)\nonumber\\
    &\quad\quad + \sum_{s=1}^{n_N}(-1)^{n_N-s}\begin{pmatrix}\mathcal{N}-s-1 \\ n_N-s\end{pmatrix}E_s(q_N)\mathcal{D}_{\mathcal{N}-s}^{(N-1)}(n_1,\ldots,n_{N-1};q_1,\ldots,q_{N-1})\nonumber\\
    &\quad\quad - (-1)^{n_N}\begin{pmatrix}\mathcal{N}-1\\ n_N-1\end{pmatrix}\mathcal{D}^{(N-1)}_{\mathcal{N}}(n_1,\ldots,n_{N-1};q_1-q_N,\ldots,q_{N-1}-q_N)\nonumber\\
    &\quad\quad + \sum_{s=1}^{a}(-1)^{n_N}\!\begin{pmatrix}\mathcal{N}-s-1\\ a-s\end{pmatrix}E_s(q_N)\mathcal{D}_{\mathcal{N}-s}^{(N-1)}(n_1,\ldots,n_{N-1};q_1-q_N,\ldots, q_{N-1}-q_N)\nonumber\\
    &\quad\quad - (-1)^{n_N}\begin{pmatrix}\mathcal{N}-1\\ a-1\end{pmatrix} \mathcal{D}_{\mathcal{N}}^{(N-1)}(n_1,\ldots, n_{N-1};q_1,\ldots, q_{N-1})\, .\label{eq:DN recursion}
\end{align}
Recall that $E_n(q)$ is the modified Bernoulli polynomial depending on the fractional part of its argument introduced in (\ref{eq:modified bernoulli}). We notice that at each step of the recursion we ``peel off'' an additional Bernoulli polynomial (and have to subtract some diagonal terms). The result is that the sums can be expressed in terms of a sum of terms involving products of at most $N$ Bernoulli polynomials with fractional arguments.
The base cases are
\begin{equation}
    \mathcal{D}_a^{(0)} = 0 \, (a>0),\quad \mathcal{D}_0^{(0)}=1,\quad \mathcal{D}_a^{(1)}(n;q) = E_{a+n}(q)\, .
\end{equation}

The loop momentum integral (\ref{eq:loop momentum integral as funny sum}) can be simply expressed in terms of these sums. By writing $m_N = -\sum_{j=1}^{N-1}m_j$, we have
\begin{multline}
    \int_{\mathbb{R}/\mathbb{Z}}\d q\, \prod_{j=1}^N B_{n_j}(\{q+q_j\}) \\ = (-1)^{n_N}\prod_{j=1}^N(-n_j!)\mathcal{D}^{(N-1)}_{2d_{N}+2}(n_1,\ldots,n_{N-1};q_1-q_N,\ldots,q_{N-1}-q_N)\, .\label{eq:loop momentum integral result}
\end{multline}
Thus the loop integral can be evaluated in terms of the constrained sum (\ref{eq:constrained sum}) that itself can be expressed as a sum over terms involving products of at most $N-1$ Bernoulli polynomials through the recursion (\ref{eq:DN recursion}).

\paragraph{Simplifying the first term in the Mirzakhani recursion relation.}
In (\ref{eq:Mirzarec c1}), we presented a recursive representation of the discretized $c=1$ string amplitudes akin to Mirzakhani's recursion relation for the Weil-Petersson volumes. We then presented a simplification that facilitates efficient implementation of the recursion. A substantial advantage of this simplification is that it allowed us to treat the fractional terms in the amplitude as effectively constant. But for the term in the recursion relation corresponding to the non-separating degeneration, the simplified implementation \eqref{eq:term1 simplified}, while compact and useful, still requires the evaluation of an integral against Bernoulli polynomials with arguments depending on the fractional parts of partial sums of momenta. For completeness, we present here a way to compute the sort of integral that appears. It is structurally very similar to the loop momentum integrals discussed above.

For example we might be interested in the following generic contribution to the sub-amplitude $\hat{\mathsf{s}}_{g-1,n+1}$ that appears in (\ref{eq:term1 simplified})
\begin{equation}\label{eq:Bernoulli contribution to handle term}
    [q^k q'^{\ell}]\hat{\mathsf{s}}_{g-1,n+1}(q,-q',\boldsymbol{p}) \supset \prod_{j=1}^N B_{n_j}(\{q+q_j\})\, ,
\end{equation}
where $k$ and $\ell$ are even and the $q_j$ represent partial sums of the external momenta. It is sufficient to consider the case that the Bernoulli polynomials depend only on one of the internal momenta because of momentum conservation. Thus we are tasked with computing the following integral
\begin{align}
    p_1\hat{\mathsf{s}}_{g,n}(p_1,\boldsymbol{p})^{(1)}
    &\supset \int_{\RR/\ZZ} \d q\, (k+1)!(\ell+1)!\bigg(f_{k+\ell+5}(p_1) + 2p_1 E_{k+2}(q)E_{\ell+2}(q-p_1) \nonumber\\
    &\quad + \sum_{s=1}^{\ell+3}\tbinom{k+\ell+4-s}{\ell+3-s}f_s(p_1)E_{k+\ell+5-s}(q)\nonumber\\
    &\quad+ \sum_{s=1}^{k+3}\tbinom{k+\ell+4-s}{k+3-s}f_s(p_1)E_{k+\ell+5-s}(q-p_1)\bigg) \prod_{j=1}^N B_{n_j}(\{q+q_j\})\, .
\end{align}
For each term on the right-hand side, we can straightforwardly apply the result we derived for the loop momentum integrals (\ref{eq:loop momentum integral result}). Thus we can express the contribution to the first term of the recursion relation from (\ref{eq:Bernoulli contribution to handle term}) in terms of the constrained sums (\ref{eq:constrained sum}) that we showed could themselves be computed recursively in (\ref{eq:DN recursion}). All together, this gives the following contribution to the non-separating term in the recursion relation
\begin{align}
    &p_1\hat{\mathsf{s}}_{g,n}(p_1,\boldsymbol{p})^{(1)}\nonumber\\
    &\quad\supset  (k+1)!(\ell+1)!\prod_{j=1}^N(-n_j!)\bigg((-1)^{n_N}f_{k+\ell+5}(p_1)\nonumber\\
    &\quad\qquad\times \mathcal{D}^{(N-1)}_{n_N}(n_1,\ldots,n_{N-1};q_1-q_N,\ldots,q_{N-1}-q_N)\nonumber\\
    &\quad\quad+ 2p_1\mathcal{D}^{(N+1)}_{k+2}(\ell+2,n_1,\ldots,n_N;-p_1,q_1,\ldots,q_N)\nonumber\\
    &\quad\quad+ \sum_{s=1}^{\ell+3}\tbinom{k+\ell+4-s}{\ell+3-s}f_s(p_1)\mathcal{D}^{(N)}_{k+\ell+5-s}(n_1,\ldots,n_N;q_1,\ldots,q_N)\nonumber\\
    &\quad\quad +  \sum_{s=1}^{k+3}\tbinom{k+\ell+4-s}{k+3-s}f_s(p_1)\mathcal{D}^{(N)}_{k+\ell+5-s}(n_1,\ldots,n_N;q_1+p_1,\ldots,q_N+p_1)\bigg)\, . \label{eq:first term recursion implemented}
\end{align}

This implementation makes it straightforward to evaluate $c=1$ amplitudes for which the non-separating degeneration term of the recursion involves integrating nontrivial amplitudes which themselves depend piecewise polynomially on the momenta.  For example, the one-loop three-point amplitude is given by
\begin{align}
    \hat{\mathsf{s}}_{1,3}(p_1,p_2,p_3) &= \mathsf{V}_{1,3}^{(b=1)}(p_1,p_2,p_3) + \big(F(p_1,p_2) + 5~\text{perms}\big)\ , \\
    F(p_1,p_2) &= \frac{167}{362880} - \frac{p_1^2}{480} + E_2(p_1)\bigg(\frac{-13+30p_1^2-15p_1^4}{720} - 2E_4(p_1) - \frac{8}{3}E_4(p_2)\bigg)\nonumber\\
    &\quad+ E_4(p_1)\bigg(\frac{3}{4} - \frac{p_1^2}{4} - 2p_2^2\bigg) - \frac{16}{3}E_5(p_1)E_1(p_2) - \frac{13}{2}E_6(p_1)\ .
\end{align}
One can verify that this analytically continues to the correct physical amplitude (\ref{eq:S1n result}) upon restriction to the first Brillouin zone. This example makes it clear that iterating the recursion leads to an explosion in the complexity of the discretized amplitudes as the terms involving Bernoulli polynomials with fractional arguments proliferate wildly, even as the physical $c=1$ amplitudes obtained by analytic continuation to the physical sheet remain comparatively tame.

\bibliographystyle{JHEP}
\bibliography{bib}
\end{document}